\documentclass[10pt,a4paper]{article}
\usepackage[margin=25mm]{geometry}
\usepackage{graphicx,xcolor}
\usepackage{etoolbox}
\AtBeginEnvironment{table}{\small}
\usepackage[colorlinks=true,allcolors=blue]{hyperref}
\usepackage[T1]{fontenc}
\usepackage[utf8]{inputenc}
\usepackage{textcomp}
\usepackage{amsmath,amssymb,bm}
\usepackage{booktabs}
\usepackage[labelfont=bf,labelsep=period,font=small]{caption}
\usepackage[caption=false,font=footnotesize,labelfont=sf,textfont=sf]{subfig}
\usepackage{tikz}
\usetikzlibrary{arrows.meta,positioning,calc,fit,backgrounds,shapes.misc}
\usepackage{cite}
\usepackage{multirow}
\usepackage{url}
\graphicspath{{figs/}}

\newcommand{\muV}{\textmu V}
\newcommand{\muVsq}{\textmu V$^2$}
\newcommand{\ccdan}{CC-DAN}
\newsavebox{\overviewbox}
\newcommand{\nohyph}{\hyphenpenalty=10000 \exhyphenpenalty=10000 \relax}

\newcommand{\minb}{\min(\mathrm{NN},\mathrm{SSI})}

\begin{document}

\title{Coordinate-conditioned detector-atom network for montage-agnostic EEG channel completion: zero-shot transfer and a regime map}

\author{Hiroshi Higashi\\[2pt]
\small Faculty of Engineering Science, Kansai University, 3-3-35 Yamate-cho, Suita, Osaka 564-8680, Japan\\
\small E-mail: \href{mailto:hgshrs@kansai-u.ac.jp}{hgshrs@kansai-u.ac.jp}, ORCID: \href{https://orcid.org/0000-0001-8880-3411}{0000-0001-8880-3411}}
\date{}
\maketitle

\begin{abstract}
\textit{Objective.}
Electroencephalography (EEG) electrode layouts differ across devices, from consumer headsets to 128-channel nets, and learned methods for completing missing electrodes assume a fixed layout and within-dataset training, so they cannot be applied to unseen devices.
We aimed at a single pretrained model that completes EEG at arbitrary positions from an arbitrary electrode set without calibration on the target data, and at charting where such learned completion beats classical interpolation.
\textit{Approach.}
We propose the coordinate-conditioned detector-atom network (CC-DAN), a self-supervised convolutional dictionary model in which the atoms are spherical-harmonic functions of electrode coordinates and the detector pools observed electrodes with coordinate-dependent weights, making it independent of electrode number, order, and names.
In a plug-and-play protocol, a single model pretrained on five public datasets (132 subjects) was applied zero-shot to nine held-out sets with 8 to 128 channels and compared with nearest-neighbor and spherical spline interpolation (SSI), a within-subject oracle, and learned baselines on signal error, band-wise error, and downstream tasks.
\textit{Main results.}
With five or fewer observed electrodes, sparse real layouts, and hemispheric dropout, CC-DAN beat both interpolations in nearly all subjects and approached the oracle; on the EGI net and 8- to 16-channel devices it kept this advantage up to 15 electrodes, whereas SSI was superior with 15 or more electrodes on dense 10-05 montages.
CC-DAN was also the most robust to electrode-position errors.
Mean-squared-error (MSE) gains did not carry over to downstream accuracy, and the amplitude shrinkage of MSE-optimal prediction degraded power-based measures, which a variance-matching loss and gain calibration mitigate.
\textit{Significance.}
The resulting regime map tells practitioners where learned completion pays off and where classical interpolation remains the method of choice, and the plug-and-play protocol lets one pretrained model connect an unseen device to a fixed-montage pipeline without retraining.
\end{abstract}


\section{Introduction}
The spatial sampling of EEG varies widely across devices: clinical 10-20 systems with 19 channels, consumer headsets with 4--16 channels, and research nets with 62--128 channels (10-10 systems or the EGI Geodesic Sensor Net) coexist, and electrodes drop out or fail routinely.
Completing missing electrodes (interpolation, completion, spatial super-resolution) is used to replace bad electrodes, to adapt recordings to pipelines that assume a fixed montage, and to pool datasets.
An attractive use of the second kind is to carry a model trained on a high-density montage, such as an EEG-based brain-age or pathology classifier, to a sparse consumer or bedside device without new labels or retraining.
The standard training-free method is spherical spline interpolation (SSI) \cite{perrin1989}, but its accuracy collapses where observed electrodes are sparse or extrapolation is required.

Recent learned methods---deep super-resolution with CNN/GAN \cite{han2019,corley2018}, Transformer \cite{estformer2025}, or diffusion models \cite{stad2024,srgdiff2026,tgsd2026,cafe2026}, graph-convolutional harmonization \cite{km26}, and large single-dataset training \cite{svantesson2021}---assume a fixed high-density layout with a fixed split and are trained and evaluated per dataset.
In practice the electrode set of the device at hand does not match the training set, and electrode names often cannot even be mapped (e.g., EGI nets versus 10-10 systems).
Coordinate-queryable implicit representations \cite{scalpinr2026} and mask-adaptive attention \cite{amacw2023} remain within-dataset, the foundation models REVE \cite{reve2025} and LUNA \cite{luna2025} do not report signal-level reconstruction of missing electrodes, and harmonization methods \cite{km26,imac2025,regionpooling2024} assume a fixed target montage.
Low-order spherical bases represent scalp fields efficiently \cite{park2026}, which supports our atom parameterization.
Common gaps are the fixed-layout assumption, the absence of cross-dataset zero-shot evaluation, inadequate classical baselines, reference and scale harmonization not being treated as part of the method, and missing validation on downstream tasks, so it is unknown how far an improvement in mean squared error (MSE) propagates to actual analyses.

This paper asks: \emph{Can a pretrained model complete EEG at arbitrary positions from an arbitrary electrode set without calibration on the target data?
Where does it beat classical interpolation, and do MSE gains propagate to downstream tasks?}
Our contributions are as follows.
\begin{enumerate}
\item \textbf{Multichannel DAN and \ccdan{}.}
We extend the single-channel detector-atom network \cite{higashi2025} to multichannel atoms with shared activations (Section~\ref{sec:mdan}), then parameterize the spatial part of each atom as a spherical-harmonic expansion of electrode coordinates and replace the detector by coordinate-dependent pooling that is invariant to the number of electrodes (Section~\ref{sec:ccdan}), so that no electrode-name correspondence is needed.
\item \textbf{Plug-and-play protocol.}
With a reference (observed-set average) and scale determined by the target's observed electrodes only, we present the first systematic zero-shot evaluation of a model pretrained on multi-laboratory, multi-paradigm, multi-net public data on unseen datasets (Sections~\ref{sec:setup} and \ref{sec:zeroshot}).
\item \textbf{Regime map.}
We chart, as a function of electrode count, layout, and frequency band, where learning beats classical interpolation, together with the ceiling given by a within-subject oracle (Section~\ref{sec:regime}).
\item \textbf{Downstream tasks and amplitude shrinkage.}
We evaluate completed signals on motor-imagery classification, steady-state visual evoked potential (SSVEP) detection, and $\alpha$ topography, and show that the amplitude shrinkage inherent to MSE-optimal prediction degrades power-based measures (Section~\ref{sec:downstream}).
\item \textbf{Comparison methods and evaluation protocol.}
We compare in-dataset fixed-layout super-resolution (SRGDiff), the foundation model LUNA (REVE was not evaluable), name-based DAN, and graph convolutional network (GCN) harmonization \cite{km26} on identical metrics, and identify artificial subjects and weak SSI baselines in prior evaluations (Sections~\ref{sec:comparison} and \ref{sec:protocol}).
\end{enumerate}

\section{Methods}\label{sec:method}
Fig.~\ref{fig:overview} gives an overview of the method.

\begin{figure}[tbp]
\centering
\sbox{\overviewbox}{%
\begin{tikzpicture}[
  font=\small,
  >={Stealth[length=2mm]},
  every node/.append style={execute at begin node=\nohyph},
  box/.style={draw,rounded corners=2pt,align=center,inner sep=3.5pt,minimum height=9mm,fill=white},
  proc/.style={box,fill=blue!6},
  data/.style={box,fill=orange!10},
  lab/.style={font=\normalsize\bfseries,anchor=west},
  note/.style={font=\small,align=center},
  side/.style={font=\small,align=left,anchor=west,inner sep=1pt},
  pics/head/.style={code={
    \draw[line width=0.5pt] (0,0) circle (0.5);
    \draw[line width=0.5pt] (-0.08,0.49) -- (0,0.6) -- (0.08,0.49);
    \draw[line width=0.5pt] (-0.5,0.07) .. controls (-0.6,0.1) and (-0.6,-0.1) .. (-0.5,-0.07);
    \draw[line width=0.5pt] (0.5,0.07) .. controls (0.6,0.1) and (0.6,-0.1) .. (0.5,-0.07);
  }},
  pics/wave/.style={code={
    \draw[line width=0.5pt] (0,0) -- (1.4,0);
    \draw[#1,line width=0.8pt] plot[domain=0:1.4,samples=80,smooth] (\x,{0.13*sin(12*\x r)+0.07*sin(31*\x r+1)});
  }},
  pics/spikes/.style={code={
    \draw[line width=0.5pt] (0,0) -- (1.4,0);
    \foreach \x/\h in {0.10/0.26,0.36/0.40,0.45/0.14,0.78/0.36,1.12/0.22,1.28/0.30}
      \draw[red!80!black,line width=0.9pt] (\x,0) -- (\x,\h);
  }},
]
\node[lab] at (-0.3,1.2) {(a) Problem setting};
\pic at (1.0,0) {head};
\foreach \x/\y in {0/0.33,-0.3/0.12,0.3/0.12,-0.14/-0.26,0.17/-0.28,0.36/-0.06,-0.36/-0.12,0/-0.06,0.06/0.14} \fill[red] ($(1.0,0)+(\x,\y)$) circle (1.6pt);
\node[note] at (1.0,-1.1) {observed set $\mathcal O$\\ $\{\bm p_i\}$, $x(\bm p_i,t)$};
\draw[->,thick] (1.8,0) -- node[above,note]{completion} node[below,note]{no calibration} (4.6,0);
\pic at (5.4,0) {head};
\foreach \x/\y in {0/0.33,-0.3/0.12,0.3/0.12,-0.14/-0.26,0.17/-0.28,0.36/-0.06,-0.36/-0.12,0/-0.06,0.06/0.14} \fill[red] ($(5.4,0)+(\x,\y)$) circle (1.6pt);
\foreach \x/\y in {-0.18/0.36,0.2/0.36,-0.38/0.3,0.4/0.26,-0.24/-0.06,0.21/-0.1,-0.06/-0.38,0.33/-0.36,-0.36/-0.34,0.12/0.02,-0.12/0.18} \draw[blue!70!black,line width=0.7pt] ($(5.4,0)+(\x,\y)$) circle (1.6pt);
\node[note] at (5.4,-1.1) {queries $\{\bm q_j\}$\\ estimates $\hat x(\bm q_j,t)$};
\node[side,text width=8.6cm] at (7.0,-0.3) {\textbf{observed-set average reference}: $x(\bm p_i,t)-\bar x_{\mathcal O}(t)$ for input, output, and target\\[2pt] \textbf{canonical coordinates}: electrodes are handled by position only, no electrode names};
\node[lab] at (-0.3,-2.0) {(b) \ccdan{}};
\node[data,text width=2.4cm] (in) at (0.95,-4.2) {observed signals $\bm X_{\mathcal O}$\\ coordinates $\{\bm p_i\}$};
\node[proc,right=3mm of in,text width=3.6cm] (det) {\textbf{Detector D-3}\\ SH embedding $Y(\bm p)\in\mathbb R^{81}$\\ shared filter bank with coordinate-dependent mixing $\sum_r\alpha_r(\bm p)F_r$\\ coordinate-dependent pooling $w_h(\bm p)$ over $\mathcal O$\\ temporal convolutions $\times3$};
\node[data,right=3mm of det,text width=2.1cm] (act) {activations\\ $z_n(t)\ge0$\\ $N$=32, sparse, shared};
\node[proc,right=3mm of act,text width=3.2cm] (gen) {\textbf{Atom generation}\\ $u_k=\sum_n c_{n,k}\star z_n$\\ ($K$=81 virtual ch.)\\ $a_n(\bm q,\tau)=\sum_k c_{n,k}(\tau)Y_k(\bm q)$};
\node[data,right=3mm of gen,text width=2.1cm] (out) {projection $Y(\bm q)$\\ $\hat x(\bm q,t)$ at any position};
\draw[->] (in) -- (det);
\draw[->] (det) -- (act);
\draw[->] (act) -- (gen);
\draw[->] (gen) -- (out);
\pic at ($(in.south)+(-0.7,-0.45)$) {wave={black}};
\pic at ($(in.south)+(-0.7,-0.95)$) {wave={black}};
\pic at ($(act.south)+(-0.7,-0.5)$) {spikes};
\pic at ($(out.south)+(-0.7,-0.5)$) {wave={red!80!black}};
\node[lab] at (-0.3,-7.1) {(c) Masks and plug-and-play};
\pic at (1.0,-8.4) {head};
\foreach \x/\y in {-0.24/0.24,0.18/0.33,0.36/-0.12,-0.33/-0.18,0.02/-0.36} \fill[red] ($(1.0,-8.4)+(\x,\y)$) circle (1.5pt);
\node[note] at (1.0,-9.5) {random $k$\\ (log-uniform)};
\pic at (3.7,-8.4) {head};
\foreach \x/\y in {0/0.36,-0.36/0.14,0.36/0.14,-0.18/0,0.18/0,-0.36/-0.18,0.36/-0.18,0/-0.36,-0.18/-0.22,0.18/-0.22} \fill[red] ($(3.7,-8.4)+(\x,\y)$) circle (1.5pt);
\node[note] at (3.7,-9.5) {real layouts\\ (coord.-matched)};
\pic at (6.4,-8.4) {head};
\foreach \x/\y in {0.12/0.36,0.36/0.18,0.12/0.12,0.36/-0.12,0.12/-0.12,0.33/-0.33,0.12/-0.36,0.24/0} \fill[red] ($(6.4,-8.4)+(\x,\y)$) circle (1.5pt);
\draw[gray,dashed] (6.4,-8.9) -- (6.4,-7.9);
\node[note] at (6.4,-9.5) {region dropout\\ (hemisphere etc.)};
\node[side,text width=7.8cm] at (7.8,-8.8) {\textbf{pretraining}: masks applied to complete recordings (5 montages, 132 subjects)\\[2pt] \textbf{inference}: preprocess $\to$ coordinates $\to$ observed-set reference $\to$ scale by observed std $\to$ complete; no training-data statistics};
\end{tikzpicture}}%
\resizebox{\textwidth}{!}{\usebox{\overviewbox}}
\caption{Overview.
(a) EEG at arbitrary query positions is estimated from an arbitrary set of observed electrodes.
Input, output, and evaluation target are expressed in the observed-set average reference, and electrodes are handled by their coordinates in a canonical frame only.
(b) \ccdan{}.
A coordinate-dependent detector estimates non-negative sparse activations $z_n(t)$ shared by all electrodes, and convolving them with atoms whose spatial part is a spherical-harmonic (SH) expansion generates the signal at any position, \eqref{eq:model}; training minimizes \eqref{eq:loss} on complete recordings under the masks in (c).
(c) During training, diverse observed sets (random $k$, coordinate-matched real layouts, region dropout) are applied to complete recordings for self-supervised learning; at inference the model is applied without calibration using a reference and scale determined by the target's observed electrodes only.}
\label{fig:overview}
\end{figure}

\subsection{Problem Setting and Reference}\label{sec:reference}
Given signals $x(\bm p_i,t)$ of an observed electrode set $\mathcal{O}$ with coordinates $\{\bm p_i\}$, we estimate signals $\hat x(\bm q_j,t)$ at query positions $\{\bm q_j\}$.
Because EEG depends on the reference electrode, we adopt the \textbf{observed-set average reference}: the input is $x(\bm p_i,t)-\bar x_{\mathcal O}(t)$, where $\bar x_{\mathcal O}$ is the mean over the observed electrodes, and the output and evaluation target are expressed in the same reference.
This reference is computable from the observed electrodes alone, so the target side contains no unknown quantity and the evaluation is reference-invariant and leakage-free.
Using the common average reference (CAR) of all electrodes as the target would require estimating the offset of an average that includes unobserved electrodes, which is consistently worse (Section~\ref{sec:ablation}).

\subsection{Canonical Coordinates and Spatial Basis}\label{sec:coords}
All montages are expressed in the coordinate frame of the \texttt{standard\_1005} template \cite{oostenveld2001} of MNE-Python \cite{mne2013} (the canonical frame); other templates (e.g., EGI GSN-HydroCel) are placed in head coordinates via the fiducials and then mapped to the canonical frame.
The GSN-to-10-20 correspondence table (19 pairs) in EGI's technical documentation deviates by $7.2^\circ$ on average even after fiducial alignment (from the coordinate origin: $12^\circ$ at Fz, $20$--$21^\circ$ at T7/T8; Table~S1), because the GSN has no electrodes at the 10-20 positions; electrodes must therefore be handled by coordinates rather than by names.
As the spatial basis we use real spherical harmonics $\{Y_k(\bm q)\}_{k=1}^{K}$ of degree $L=8$ ($K=81$).

\subsection{Multichannel Detector-Atom Network}\label{sec:mdan}
Fig.~\ref{fig:concept} contrasts the three models of this and the next subsection.
The detector-atom network (DAN) \cite{higashi2025} is a neural implementation of convolutional dictionary learning that represents a single-channel signal $x[i]$ as a superposition of time-shifted, amplitude-modulated short waveforms (atoms) $a_n[j]$, $j=0,\dots,M-1$:
\begin{equation}
x[i]\approx\sum_{n=1}^{N}\sum_{j=0}^{M-1} a_n[j]\,z_n[i-j],\qquad z_n[i]\ge 0.
\label{eq:dan1}
\end{equation}
A \emph{detector} (convolutional layers with ReLU) estimates the non-negative activations $z_n[i]$ from the input, an \emph{atom network} generates component signals by convolving $z_n$ with $a_n$, and both are trained jointly with a reconstruction loss and a sparsity loss; atoms are short (about 0.5 s) and pretrained models transfer to other data without calibration \cite{higashi2025}.

\begin{figure}[tbp]
\centering
\resizebox{\textwidth}{!}{%
\begin{tikzpicture}[
  font=\footnotesize,
  lab/.style={font=\footnotesize\bfseries,anchor=west},
  note/.style={font=\footnotesize,align=left,anchor=west,text width=6.0cm,inner sep=1pt},
  pics/sig/.style={code={\draw[gray!60,line width=0.3pt] (0,0) -- (1.9,0);
    \draw[black,line width=0.7pt] plot[domain=0:1.9,samples=90,smooth] (\x,{#1*(0.16*sin(11*\x r)+0.09*sin(29*\x r+1)+0.05*sin(53*\x r+2))});}},
  pics/sigm/.style={code={\draw[gray!60,line width=0.3pt] (0,0) -- (1.9,0);
    \draw[gray!70,dashed,line width=0.7pt] plot[domain=0:1.9,samples=90,smooth] (\x,{#1*(0.16*sin(11*\x r)+0.09*sin(29*\x r+1)+0.05*sin(53*\x r+2))});}},
  pics/atom/.style={code={\draw[gray!60,line width=0.3pt] (0,0) -- (0.8,0);
    \draw[blue!70!black,line width=0.8pt] plot[domain=0:0.8,samples=70,smooth] (\x,{#1*exp(-((\x-0.4)^2)/0.05)*sin(24*(\x-0.4) r)});}},
  pics/spk/.style={code={\draw[gray!60,line width=0.3pt] (0,0) -- (1.3,0);
    \foreach \x/\h in {0.10/0.22,0.34/0.34,0.45/0.11,0.74/0.30,1.02/0.18,1.18/0.26}
      \draw[red!80!black,line width=0.9pt] (\x,0) -- (\x,\h);}},
  pics/head/.style={code={\draw[line width=0.5pt] (0,0) circle (0.5);
    \draw[line width=0.5pt] (-0.08,0.49) -- (0,0.6) -- (0.08,0.49);
    \draw[line width=0.5pt] (-0.5,0.07) .. controls (-0.6,0.1) and (-0.6,-0.1) .. (-0.5,-0.07);
    \draw[line width=0.5pt] (0.5,0.07) .. controls (0.6,0.1) and (0.6,-0.1) .. (0.5,-0.07);}},
]
\node[lab] at (-0.55,0.75) {(a) Single-channel DAN};
\pic at (0,0.15) {sig=1};
\node at (0.95,-0.25) {$x[i]$};
\node at (2.2,0.15) {$\approx$};
\node at (2.7,0.15) {$\sum_n$};
\pic at (3.1,0.15) {atom=0.32};
\node at (3.5,-0.25) {$a_n[j]$};
\node at (4.1,0.15) {$\star$};
\pic at (4.3,0.15) {spk};
\node at (4.95,-0.25) {$z_n[i]\ge 0$};
\node[note] at (5.75,0.1) {short atoms and sparse non-negative activations};
\node[lab] at (-0.55,-0.85) {(b) Multichannel DAN (name-based)};
\foreach \k/\y in {1/-1.25,2/-1.6,4/-2.3} {\pic at (0,\y) {sig=0.55}; \node[anchor=east,inner sep=1pt] at (-0.02,\y) {$c$=\k};}
\pic at (0,-1.95) {sigm=0.55};
\node[anchor=east,inner sep=1pt] at (-0.02,-1.95) {$c$=3};
\node[gray!60!black] at (0.95,-2.72) {$\hat x_c[i]$ ($c$=3 missing)};
\node at (2.2,-1.78) {$\approx$};
\node at (2.7,-1.78) {$\sum_n$};
\foreach \s/\y in {0.17/-1.25,0.10/-1.6,-0.13/-1.95,-0.07/-2.3} {\pic at (3.1,\y) {atom=\s};}
\draw[rounded corners=2pt,gray!70] (3.02,-1.02) rectangle (3.98,-2.5);
\node at (3.5,-2.72) {$a_{n,c}[j]$};
\node at (4.1,-1.78) {$\star$};
\pic at (4.3,-1.78) {spk};
\node at (4.95,-2.72) {$z_n[i]$};
\node[note] at (5.75,-1.78) {one activation shared by all channels, so the missing channel is generated as $\sum_n a_{n,3}\star z_n$};
\node[lab] at (-0.55,-3.15) {(c) \ccdan{} (coordinate-conditioned)};
\pic at (0.9,-4.05) {head};
\foreach \x/\y in {0/0.33,-0.3/0.12,0.3/0.12,-0.14/-0.26,0.17/-0.28,0.36/-0.06,-0.36/-0.12,0/-0.06} \fill[red] ($(0.9,-4.05)+(\x,\y)$) circle (1.4pt);
\foreach \x/\y in {-0.18/0.36,0.2/0.36,-0.38/0.3,0.4/0.26,-0.24/-0.06,0.21/-0.1,-0.06/-0.38,0.33/-0.36,-0.36/-0.34,0.12/0.02,-0.12/0.18} \draw[blue!70!black,line width=0.6pt] ($(0.9,-4.05)+(\x,\y)$) circle (1.4pt);
\node[align=center] at (0.9,-4.95) {observed $\bm p_i$ (red)\\ queries $\bm q_j$ (blue)};
\node at (2.2,-4.05) {$\approx$};
\node at (2.7,-4.05) {$\sum_n$};
\begin{scope}
\clip (3.5,-4.05) circle (0.5);
\shade[inner color=red!75,outer color=white] (3.27,-3.82) circle (0.5);
\shade[inner color=blue!75,outer color=white,opacity=0.85] (3.73,-4.28) circle (0.5);
\end{scope}
\pic at (3.5,-4.05) {head};
\node at (3.5,-4.75) {$a_n(\bm q,\tau)$};
\node at (4.1,-4.05) {$\star$};
\pic at (4.3,-4.05) {spk};
\node at (4.95,-4.75) {$z_n(t)$};
\node[note] at (5.75,-4.05) {the spatial part $\sum_k c_{n,k}(\tau)Y_k(\bm q)$ is a continuous field, so the output exists at any $\bm q_j$ without electrode names};
\end{tikzpicture}}
\caption{From the single-channel DAN to \ccdan{}.
(a) A single-channel signal is a sum of short atoms $a_n$ convolved with sparse non-negative activations $z_n$, \eqref{eq:dan1}.
(b) The multichannel DAN gives each atom a channel-specific component $a_{n,c}$ while all channels share one activation, \eqref{eq:mdan}; a channel that is not observed (dashed, $c$=3) is still generated from the shared $z_n$ and its atom components, which is the principle of completion, but the components are tied to the training electrode set.
(c) \ccdan{} replaces the channel-specific components by a spherical-harmonic field $a_n(\bm q,\tau)$ over the scalp, \eqref{eq:model}, so the signal can be generated at any query position from any observed set.}
\label{fig:concept}
\end{figure}

We extend the DAN to multichannel signals $\bm X\in\mathbb R^{C\times T}$: each atom becomes a multichannel waveform with channel components $a_{n,c}[j]$, and the activation $z_n[i]$ is \emph{shared by all channels},
\begin{equation}
\hat x_c[i]=\sum_{n=1}^{N}\sum_{j=0}^{M-1} a_{n,c}[j]\,z_n[i-j],\qquad z_n[i]\ge 0 .
\label{eq:mdan}
\end{equation}
Each atom is thus a spatiotemporal template with a fixed spatial pattern, matching the instantaneous-mixing structure of the EEG forward problem.
Because activations are shared, unobserved channels are generated from the $z_n$ estimated on observed channels and the unobserved components $a_{n,c}$ of the atoms; this is the principle of channel completion.

The detector must estimate $z_n$ under an observation mask $\bm m\in\{0,1\}^C$.
Instead of feeding the zero-filled $\bm X\odot\bm m$ concatenated with $\bm m$ to the convolutional layers (design D-1), design D-2 computes per-channel temporal filter outputs $h_{c,h}[i]$ and pools them over the observed channels with learnable non-negative weights $w_{h,c}$,
\begin{equation}
g_h[i]=\frac{\sum_{c} m_c\,w_{h,c}\,h_{c,h}[i]}{\sum_c m_c\,w_{h,c}},
\end{equation}
before temporal convolutions output $z_n$.
D-2 is invariant to the number of observed channels and consistently outperformed D-1 in preliminary experiments (BCI IV-1, 44 observed electrodes: 22.5 vs 28.8 \muVsq{}).

Training is self-supervised from complete recordings.
For each batch we draw a random mask $\bm m$ (Section~\ref{sec:masks}) and reconstruct all channels from $\bm X\odot\bm m$ by minimizing
\begin{equation}
\mathcal L=\frac{\sum_{c,i}(1-m_c)\,(\hat x_c[i]-x_c[i])^2}{\sum_c(1-m_c)\,T}
+\lambda_{\mathrm{obs}}\frac{\sum_{c,i}m_c\,(\hat x_c[i]-x_c[i])^2}{\sum_c m_c\,T}
+\lambda_{\mathrm{sp}}\frac{1}{NT}\sum_{n,i} z_n[i]
\label{eq:loss}
\end{equation}
with $\lambda_{\mathrm{obs}}=1$ and $\lambda_{\mathrm{sp}}=0.05$.
The missing-channel term drives completion accuracy, the observed-channel term the fidelity of the decomposition, and the sparsity term makes atoms oscillatory and activations sparse without hurting MSE (Section~\ref{sec:interp}).
Because the atoms in \eqref{eq:mdan} have independent coefficients per channel, this name-based model presupposes the training electrode set and cannot be applied to other nets, custom grids, or mismatched coordinates.

\subsection{Coordinate-Conditioned DAN (\ccdan{})}\label{sec:ccdan}
\ccdan{} replaces the spatial part of the atoms by a continuous function of coordinates (Fig.~\ref{fig:concept}(c)):
\begin{equation}
\begin{split}
\hat x(\bm q,t)&=\sum_{n=1}^{N}\sum_{\tau} a_n(\bm q,\tau)\, z_n(t-\tau),\\
a_n(\bm q,\tau)&=\sum_{k=1}^{K} c_{n,k}(\tau)\,Y_k(\bm q),
\end{split}
\label{eq:model}
\end{equation}
where $z_n(t)\ge 0$ is the activation of atom $n$ (shared by all electrodes) and $c_{n,k}(\tau)$ are learned coefficients ($N=32$, atom length 0.5 s $=125$ samples, $K=81$).
Instead of $C\times N\times M$ free parameters, \ccdan{} represents atoms at any position with $K\times N\times M$ coefficients, and the low-order expansion builds in smoothness on the sphere as a prior.
Reconstruction is done in basis space, $u_k=\sum_n c_{n,k}\star z_n$ ($K$ virtual channels), and projected with $Y(\bm q)$, so the cost does not grow when coordinates differ from trial to trial.

\emph{Detector (D-3).}
A shared temporal filter bank (64 channels, kernel length 25 samples) is applied to each observed electrode, and a coordinate-dependent low-rank mixture $F(\bm p)=\sum_{r=1}^{R}\alpha_r(\bm p)F_r$ ($R=4$, $\alpha_r(\bm p)=Y(\bm p)^\top\bm\beta_r$) yields position-specific features.
Coordinate-dependent non-negative pooling weights $w_h(\bm p)=\mathrm{softplus}(Y(\bm p)^\top\bm\gamma_h)$ then form a weighted mean over the observed set (invariant to permutation and electrode count), and three temporal convolution layers output $z_n(t)$.
The model has 510{,}016 parameters, comparable to the name-based DAN (about 488k).
The loss is \eqref{eq:loss} with the target in the observed-set average reference; a Laplace--Beltrami smoothness penalty $\sum_k l_k(l_k+1)\|c_k\|^2$ can be added but is not used in the main experiments.
At inference the model takes $(\bm X_{\mathcal O},\{\bm p_i\})$ and query coordinates $\{\bm q_j\}$, neither of which needs to have appeared in training.

\subsection{Training Mask Distribution}\label{sec:masks}
The observed sets seen in training are diversified:
(a) random sets with log-uniform electrode count $k\in[3,\min(64,C-1)]$ (50\% of training steps);
(b) coordinate-matched real layouts (11 templates including 10-20 with 19 ch, Emotiv EPOC 14 ch, OpenBCI 8/16 ch, Muse 4 ch, motor 9/3 ch, and BCI IV-2a 21 ch; training electrodes within $8^\circ$ of a template electrode are observed, with an additional partial cap in 30\% of these cases; 30\% of steps; Table~S2);
(c) region dropout (hemisphere, frontal, posterior, temporal, and random caps of radius $20$--$60^\circ$; 20\% of steps).
Each step draws a batch of 32 trials from one dataset (probability proportional to the square root of its trial count) and trains with that dataset's electrode set and coordinates.

\subsection{Plug-and-Play Protocol}\label{sec:pnp}
On the target side, only common preprocessing (0.5--40 Hz, 250 Hz, 4-s windows), electrode coordinates (template names or measured positions), re-referencing to the observed-set average, and scaling by the median standard deviation of the observed electrodes are required; no statistics of the training data and no electrode-name correspondence are needed.

\subsection{Amplitude Calibration}\label{sec:calib}
The MSE-minimizing prediction is a conditional expectation whose temporal variance is smaller than that of the truth (amplitude shrinkage; Section~\ref{sec:downstream}), which matters when completed signals are used for power analysis.
We study two remedies.
(a) \emph{Post-hoc gain table}: on the pretraining corpus, predictions are made under the training mask distribution and the amplitude ratio $g=\mathrm{std}(x)/\mathrm{std}(\hat x)$ of each missing electrode is tabulated as the median over bins of the angular distance $d$ to the nearest observed electrode (seven bins from 0--5 to $>45^\circ$) and the number of observed electrodes $k$ (eight bins from 3 to $\ge64$); 181{,}665 samples, overall median 1.28 (Table~S3).
At inference the prediction of each missing electrode is multiplied by the table value; no ground truth on the target data is needed.
(b) \emph{Variance-matching loss}: we add to \eqref{eq:loss} the mean over missing electrodes of $(\log\mathrm{std}(\hat x_c)-\log\mathrm{std}(x_c))^2$ with weight $\lambda_{\mathrm{var}}\in\{1,3\}$ (models vm1 and vm3), pretrained with the same corpus, settings, and epochs as the main model.
The gain table can also be applied to vm1 (vm1+cal).

\section{Experimental Setup}\label{sec:setup}
\subsection{Data}
\emph{Pretraining corpus (Tier-1, 132 subjects).}
Schirrmeister2017 (128 ch, 127 used as Cz was the reference; 14 subjects, motor execution \cite{schirrmeister2017}), GrosseWentrup2009 (128 ch, 10 subjects, motor imagery \cite{grossewentrup2009}), and the motor-imagery (MI), SSVEP, and event-related potential (ERP) paradigms of Lee2019 (62 ch, 54 subjects \cite{lee2019}): five montage configurations and about 60k 4-s trials, obtained with MOABB \cite{moabb2018} and MNE-Python \cite{mne2013}.

\emph{Evaluation sets (not used in training).}
BCI Competition IV-1 (59-ch BrainAmp, four real subjects \cite{blankertz2007,tangermann2012}), BCI Competition III-IIIa (60-ch custom grid, three subjects \cite{schlogl2005}), BCI Competition IV-2a (22 ch, nine subjects \cite{tangermann2012}), HBN-EEG (HBN; EGI GSN-128, children aged 5--21, 40 subjects, resting state \cite{hbn2024}), Nakanishi2015 (8-ch SSVEP, nine subjects \cite{nakanishi2015}), and BrainInvaders 2013a (BI2013a; 16-ch P300, 24 subjects \cite{bi2013a}).
In addition, each Lee2019 paradigm is evaluated with a model trained without it (cross-paradigm transfer within the same subjects and device; Section~\ref{sec:transfer}).
Layouts are shown in Fig.~S1 and dataset details are given in Table~\ref{tab:datasets}.

\begin{table}[tbp]
\centering\footnotesize\setlength{\tabcolsep}{3pt}
\caption{Datasets.
``Trials'' Is the Number of 4-s Windows After Preprocessing (All Subjects).
Of the Seven Subjects of BCI IV-1, the Four Real Subjects (800 Trials) Were Used for Evaluation.}
\label{tab:datasets}
\resizebox{\textwidth}{!}{%
\begin{tabular}{lp{1.8cm}p{3.6cm}p{2.6cm}rrrrp{2.9cm}}
\toprule
Dataset & Role & Device / montage & Recording reference & fs [Hz] & ch & Subj. & Trials & Source\\
\midrule
Schirrmeister2017 \cite{schirrmeister2017} & pretraining & 128-ch 10-05 (BrainAmp) & Cz (excluded; 127 ch) & 500 & 127 & 14 & 13{,}484 & MOABB (NEMAR)\\
GrosseWentrup2009 \cite{grossewentrup2009} & pretraining & 128-ch extended 10-20 & CAR (as distributed) & 500 & 128 & 10 & 3{,}000 & MOABB\\
Lee2019 MI \cite{lee2019} & pretraining / held-out & 62-ch 10-05 (BrainAmp) & nasion & 1000 & 62 & 54 & 10{,}800 & MOABB (NEMAR)\\
Lee2019 SSVEP \cite{lee2019} & pretraining / held-out & same & nasion & 1000 & 62 & 54 & 10{,}800 & MOABB (NEMAR)\\
Lee2019 ERP \cite{lee2019} & pretraining / held-out & same (window per 20 stimuli) & nasion & 1000 & 62 & 54 & 22{,}356 & MOABB (NEMAR)\\
\midrule
BCI IV-1 \cite{blankertz2007,tangermann2012} & evaluation & 59-ch 10-05 (BrainAmp) & nose & 1000 & 59 & 7 (4 real) & 1{,}400 & BCI Competition IV\\
BCI III-IIIa \cite{schlogl2005} & evaluation & 60-ch custom grid (Neuroscan) & mastoid & 250 & 60 & 3 & 840 & BCI Competition III\\
BCI IV-2a \cite{tangermann2012} & evaluation & 22-ch extended 10-20 & left mastoid & 250 & 22 & 9 & 5{,}184 & MOABB\\
HBN-EEG \cite{hbn2024} & evaluation & EGI GSN-HydroCel-129, resting & Cz (excluded; 128 ch) & 500 & 128 & 40 & 2{,}062 & AWS S3 (fcp-indi), Release 1\\
Nakanishi2015 \cite{nakanishi2015} & evaluation & 8-ch occipital (SSVEP) & unknown & 256 & 8 & 9 & 1{,}620 & MOABB\\
BI2013a \cite{bi2013a} & evaluation & 16-ch 10-20 (P300; window per 15 stimuli) & unknown & 512 & 16 & 24 & 2{,}338 & MOABB (Zenodo)\\
\bottomrule
\end{tabular}}
\end{table}

\emph{Preprocessing and quality control.}
Preprocessing was a fourth-order zero-phase Butterworth band-pass at 0.5--40 Hz, resampling to 250 Hz, removal of dead channels (median standard deviation $<1$ \muV{}), within-dataset CAR, and a global scale (median channel standard deviation).
For HBN, bad channels (standard deviation above three times the median) were replaced by spherical splines per subject, and up to 60 clean 4-s windows per subject were kept (2062 in total).
The artificial subjects c, d, and e of BCI IV-1 \cite{tangermann2012} were excluded (Section~\ref{sec:protocol}).

\subsection{Training Details}
Tier-1 pretraining used Adam (learning rate $10^{-3}$, no weight decay) for 40 epochs with cosine annealing to zero, batches of 32 trials from one dataset (about 1900 steps per epoch), and gradient-norm clipping at 1.0; the Laplace--Beltrami penalty and coordinate jitter were not used (Table~\ref{tab:hparams}).
No model selection was performed; the final-epoch weights were evaluated.
One run took 39 min on an RTX 5080 and was repeated with three seeds.
The regression test (Section~\ref{sec:regression}) used leave-one-subject-out (LOSO) over the four real subjects of BCI IV-1 with the artificial subjects included on the training side.
Table~\ref{tab:hparams} summarizes the Tier-1 pretraining configuration.
The variance-matched models (vm1, vm3) and the seed replicates were trained with identical settings except for $\lambda_{\mathrm{var}}$ and the seed.

\begin{table}[tbp]
\centering\footnotesize\setlength{\tabcolsep}{3pt}
\caption{Tier-1 Pretraining Configuration.}
\label{tab:hparams}
\begin{tabular}{p{3.0cm}p{11.0cm}}
\toprule
Item & Value\\
\midrule
Training data & Schirrmeister2017, Lee2019 MI/SSVEP/ERP, GrosseWentrup2009 (5 sets, 132 subjects, 60{,}440 trials)\\
Model & spherical harmonics $L$=8 ($K$=81), rank-4 coordinate-dependent filters, $N$=32 atoms, atom length 125 samples, 510{,}016 parameters\\
Optimization & Adam, learning rate $10^{-3}$, cosine annealing (all steps), weight decay 0, gradient clipping 1.0\\
Epochs and batches & 40 epochs, batches of 32 trials (from one dataset), 1 epoch $=$ total trials / 32 steps\\
Dataset sampling & probability proportional to (number of trials)$^{0.5}$\\
Mask distribution & random $k\in[3,\min(64,C-1)]$ (log-uniform) 50\%, real layouts 30\%, region dropout 20\%\\
Loss & $\lambda_{\mathrm{obs}}=1$, $\lambda_{\mathrm{sp}}=0.05$, Laplace--Beltrami penalty 0, coordinate jitter 0 mm, target $=$ observed-set average reference\\
Variance matching (vm1 / vm3) & add $\lambda_{\mathrm{var}}=1$ / $3$ (otherwise identical)\\
Seeds & 0 (main results), 1, 2 (replicates)\\
\bottomrule
\end{tabular}
\end{table}

\subsection{Comparison Methods}
Training-free: nearest-neighbor interpolation (NN) and SSI \cite{perrin1989} (equivalent to MNE, $m=4$, seven series terms, regularization $10^{-5}$).
Ceiling: a within-subject ridge-regression oracle (a linear map from observed to missing electrodes learned on other trials of the same subject).
Learned: the name-based multichannel DAN (D-2), a reimplementation of the GCN harmonization of K~M \emph{et al} \cite{km26} (KM26), SRGDiff \cite{srgdiff2026} (public code, patched and trained within the target data), and LUNA \cite{luna2025} (public weights, zero-shot reconstruction).
REVE \cite{reve2025} could not be evaluated because its masked-autoencoder (MAE) decoder is not released, and the public code of TGSD \cite{tgsd2026} could not be run (Appendix~\ref{app:comparison}).

\subsection{Evaluation Conditions and Metrics}\label{sec:metrics}
For each evaluation set we evaluate random $k$ electrodes ($k\in\{3,5,9,15,21,29,44\}$, redrawn per trial), coordinate-matched real layouts (10-20 with 19 ch, Emotiv 14 ch, OpenBCI 8 ch, motor 9/3 ch, IV-2a-type 21 ch; the number of matched electrodes is given in parentheses), and region dropout (left hemisphere, frontal, posterior).
Metrics are the MSE on missing electrodes (\muVsq{}, observed-set average reference, 200 trials per subject), the band-wise relative error (error power divided by true power; $\delta$ 0.5--4, $\theta$ 4--8, $\alpha$ 8--13, $\beta$ 13--30, $\gamma$ 30--40 Hz), and the amplitude gain (median over missing electrodes of predicted over true standard deviation).
Statistics are paired per subject against $\minb$ (number of subjects won and Wilcoxon signed-rank test).

\subsection{Downstream Tasks}\label{sec:downstream_setup}
Completed outputs keep the observed electrodes at their true values, fill only the missing electrodes, and are re-referenced to the all-electrode CAR before the downstream model.
(i) \emph{Motor-imagery classification}: EEGNet \cite{lawhern2018} trained on real full-montage data is applied to completed trials (BCI IV-2a: LOSO over nine subjects; Lee2019 MI: 44 training / 10 test subjects), with a classifier retrained on the observed electrodes only as a reference.
(ii) \emph{SSVEP detection}: canonical correlation analysis (CCA \cite{lin2006}) on 13 parieto-occipital electrodes of Lee2019 SSVEP (54 subjects) applied to the completed electrodes.
(iii) \emph{$\alpha$ topography}: on HBN resting 128 ch (40 subjects), the root-mean-square (RMS) error of $\log\alpha$ power at missing electrodes and the correlation with the true topography.
Each task compares NN, SSI, \ccdan{}, and gain-calibrated \ccdan{}; the $\alpha$ topography task also evaluates vm1 and its calibrated version.

\section{Results}\label{sec:results}
\subsection{Regression Test: Dropping Name Correspondence Does Not Hurt}\label{sec:regression}
On the benchmark of the name-based multichannel DAN (BCI IV-1, between-subject LOSO, CAR-59 target; Table~\ref{tab:regression}), \ccdan{} with spherical-harmonic atoms and the rank-4 coordinate-dependent detector outperformed the multichannel DAN D-2 in all conditions ($k$=9: 26.0 vs 27.3, $k$=3: 44.1 vs 46.9 \muVsq{}), whereas a radial-basis-function (RBF) basis was clearly worse ($k$=9: 35.4), and training time was about 1/2.3 of the name-based model.

\begin{table}[tbp]
\centering\footnotesize
\caption{Regression Test (BCI IV-1, LOSO Over the Four Real Subjects, Trained on BCI IV-1 Only, CAR-59 Target; MSE on Missing Electrodes in \muVsq{}).}
\label{tab:regression}
\begin{tabular}{lrrrrrrr}
\toprule
Method & $k$=44 & $k$=15 & $k$=9 & $k$=5 & $k$=3 & motor\_9 & motor\_3\\
\midrule
NN & 14.4 & 24.2 & 33.1 & 49.0 & 67.2 & 32.2 & 54.5\\
SSI & 3.9 & 11.4 & 21.2 & 53.9 & 69.6 & 34.4 & 63.1\\
Multichannel DAN D-2 (name-based) & 15.0 & 20.5 & 27.3 & 35.0 & 46.9 & 24.8 & 50.8\\
\ccdan{} SH $L$=8 & 16.1 & 21.8 & 27.5 & 35.7 & 46.0 & 25.6 & 51.1\\
\ccdan{} SH $L$=4 & 16.0 & 20.7 & 25.6 & 33.5 & 44.5 & 25.1 & 50.0\\
\ccdan{} SH $L$=8 + Laplace--Beltrami penalty & 15.2 & 20.5 & 26.5 & 34.5 & 46.3 & 23.8 & 50.2\\
\ccdan{} RBF (64 points) & 27.3 & 31.6 & 35.4 & 41.5 & 49.3 & 33.2 & 53.3\\
\ccdan{} SH $L$=8 + rank-4 & 14.6 & 20.2 & \textbf{26.0} & \textbf{33.9} & \textbf{44.1} & \textbf{23.8} & \textbf{50.0}\\
\bottomrule
\end{tabular}
\end{table}

\subsection{Zero-Shot Benchmark}\label{sec:zeroshot}
\begin{figure}[tbp]
\centering
\subfloat[BCI IV-1 (59 ch)]{\includegraphics[width=0.32\textwidth]{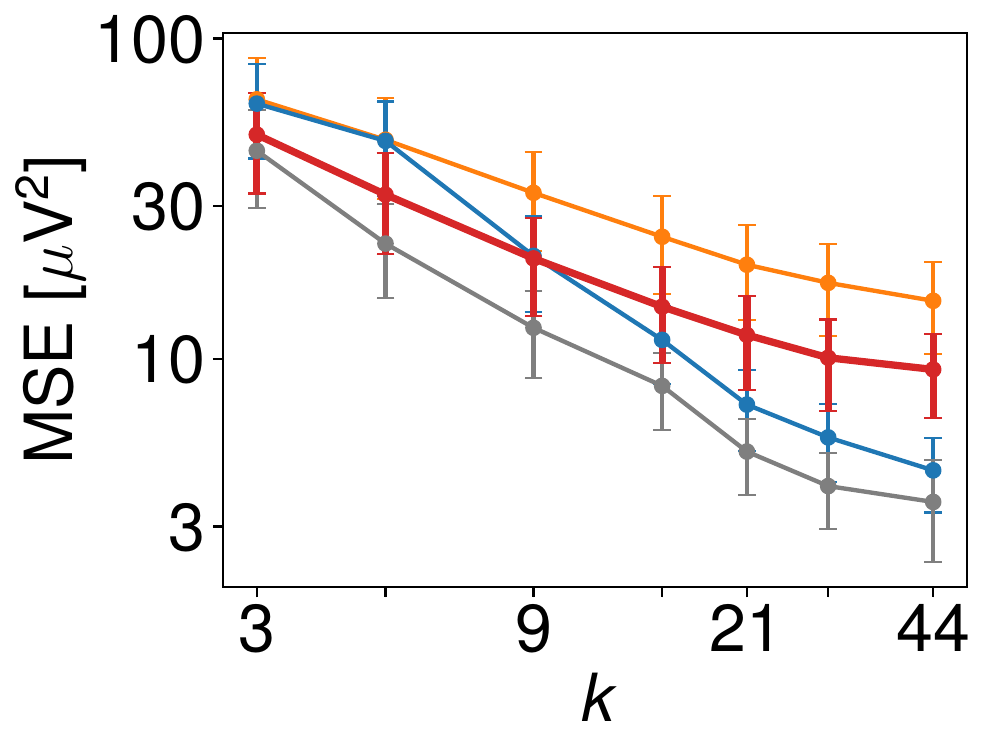}}\hfill
\subfloat[BCI III-IIIa grid (60 ch)]{\includegraphics[width=0.32\textwidth]{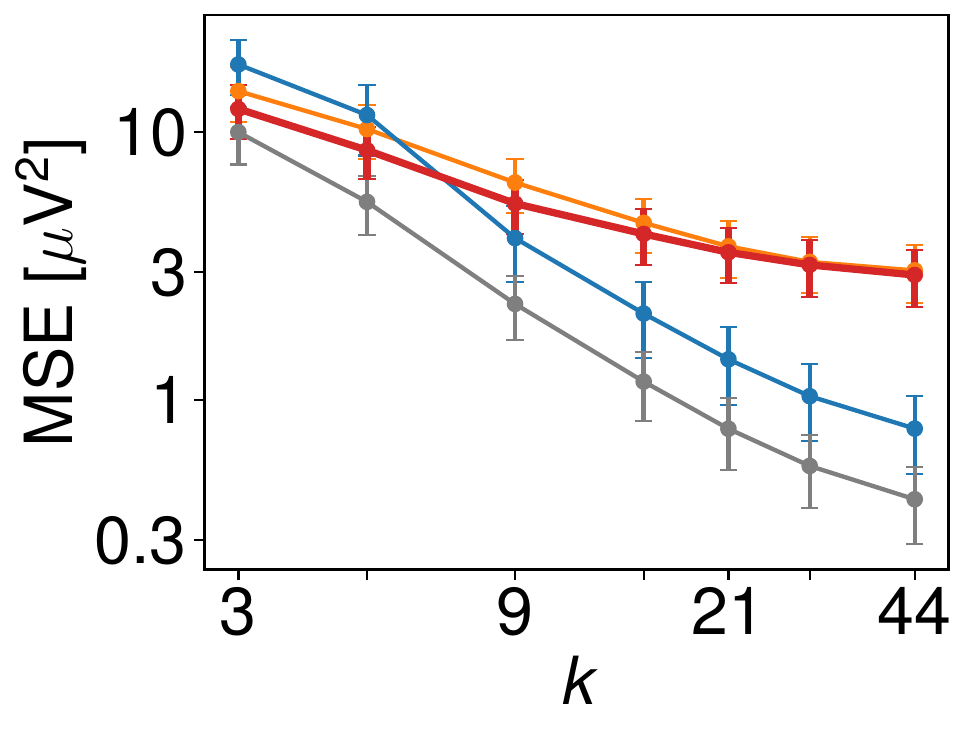}}\hfill
\subfloat[BCI IV-2a (22 ch)]{\includegraphics[width=0.32\textwidth]{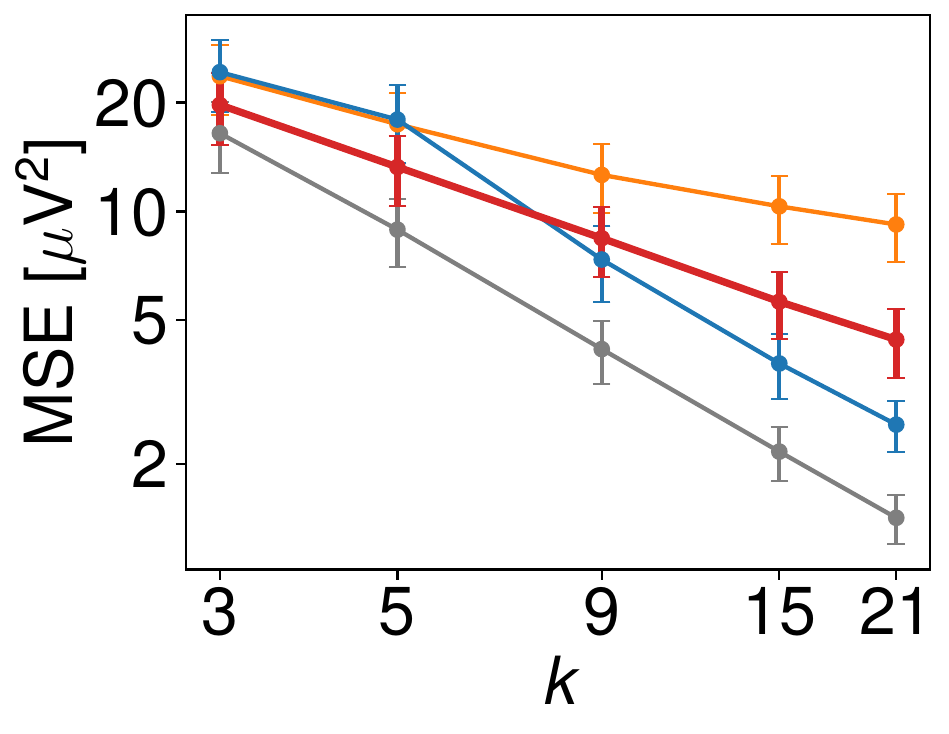}}\\[1mm]
\subfloat[HBN: EGI 128 ch, pediatric resting]{\includegraphics[width=0.32\textwidth]{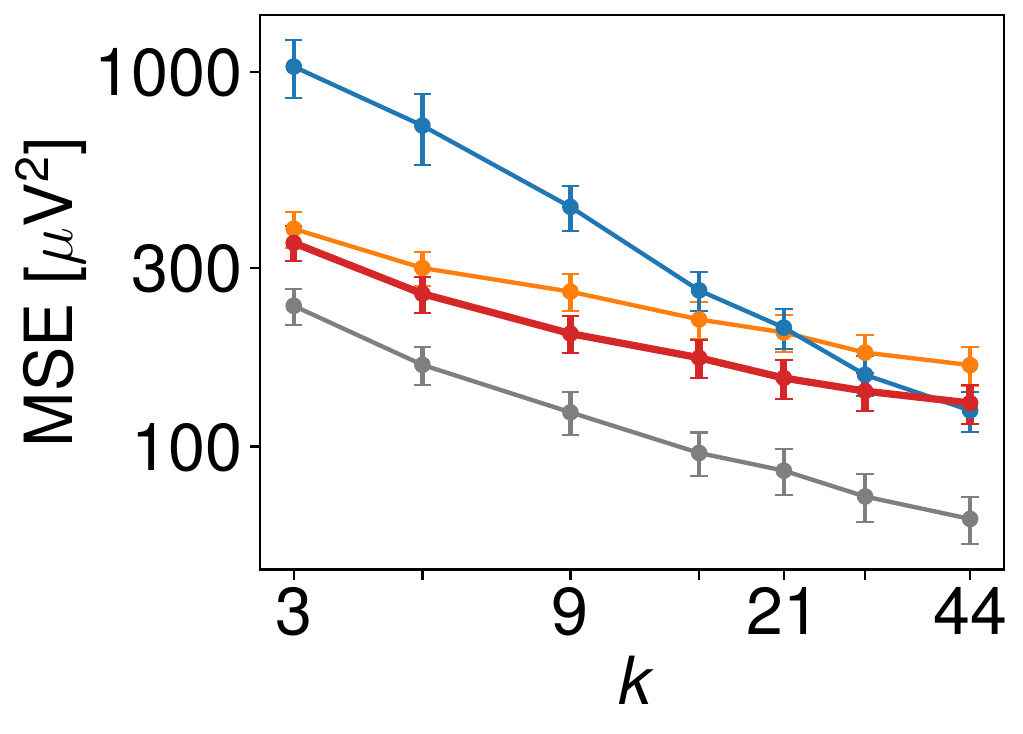}}\hfill
\subfloat[Nakanishi2015 (8 ch, SSVEP)]{\includegraphics[width=0.32\textwidth]{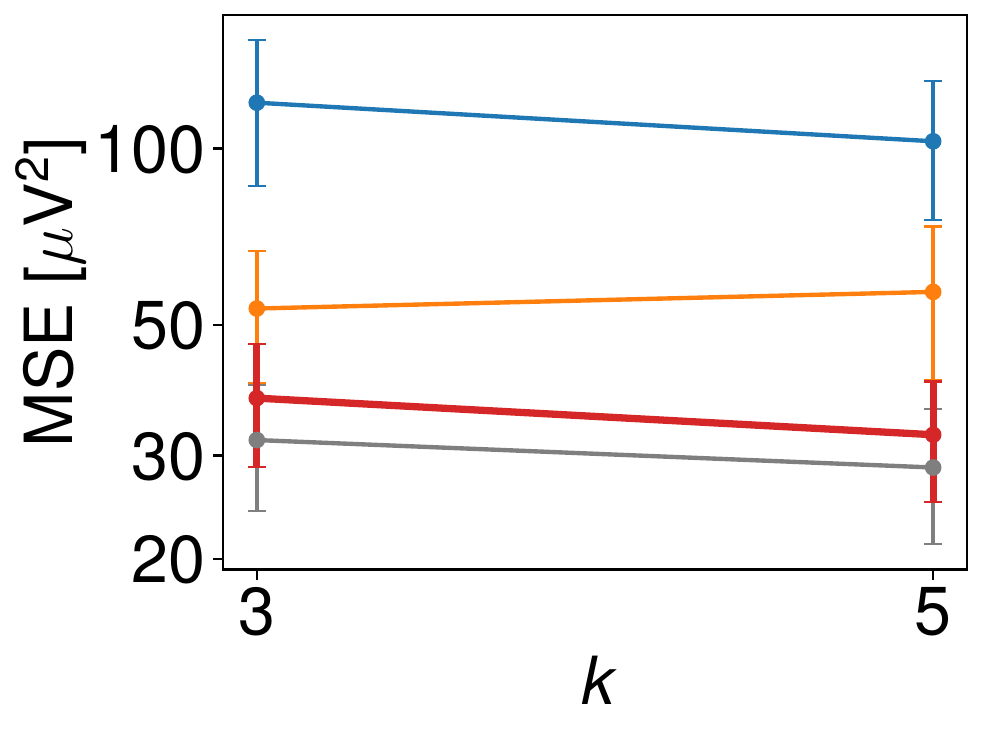}}\hfill
\subfloat[BrainInvaders 2013a (16 ch, P300)]{\includegraphics[width=0.32\textwidth]{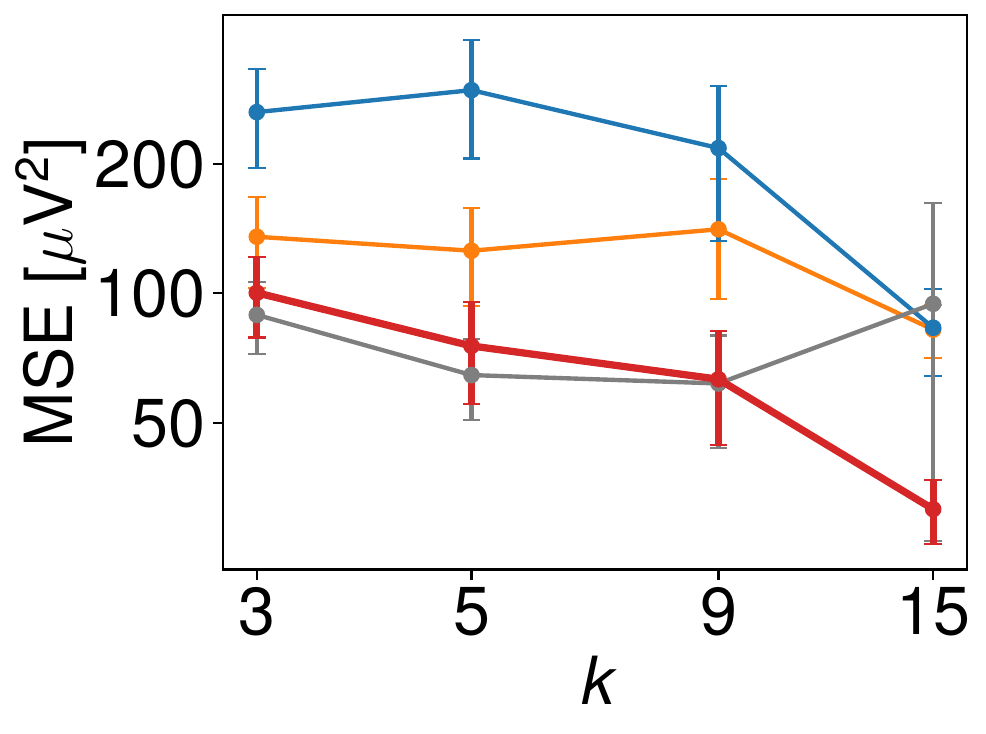}}\\[2mm]
\includegraphics[width=0.38\textwidth,trim=7 112 7 109,clip]{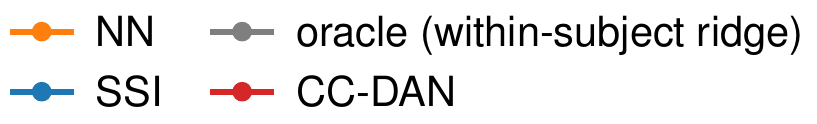}
\caption{Zero-shot completion: MSE on missing electrodes versus the number of observed electrodes $k$ (mean $\pm$ SEM over subjects, log-log) for NN, SSI, the within-subject ridge oracle (ceiling), and \ccdan{}, which never saw the evaluation data.
Results on real layouts and region dropout are shown in Fig.~\ref{fig:main_layouts} and cross-paradigm transfer on Lee2019 in Fig.~S2.}
\label{fig:main_k}
\end{figure}

\begin{figure}[tbp]
\centering
\subfloat[BCI IV-1 (59 ch)]{\includegraphics[width=0.33\textwidth]{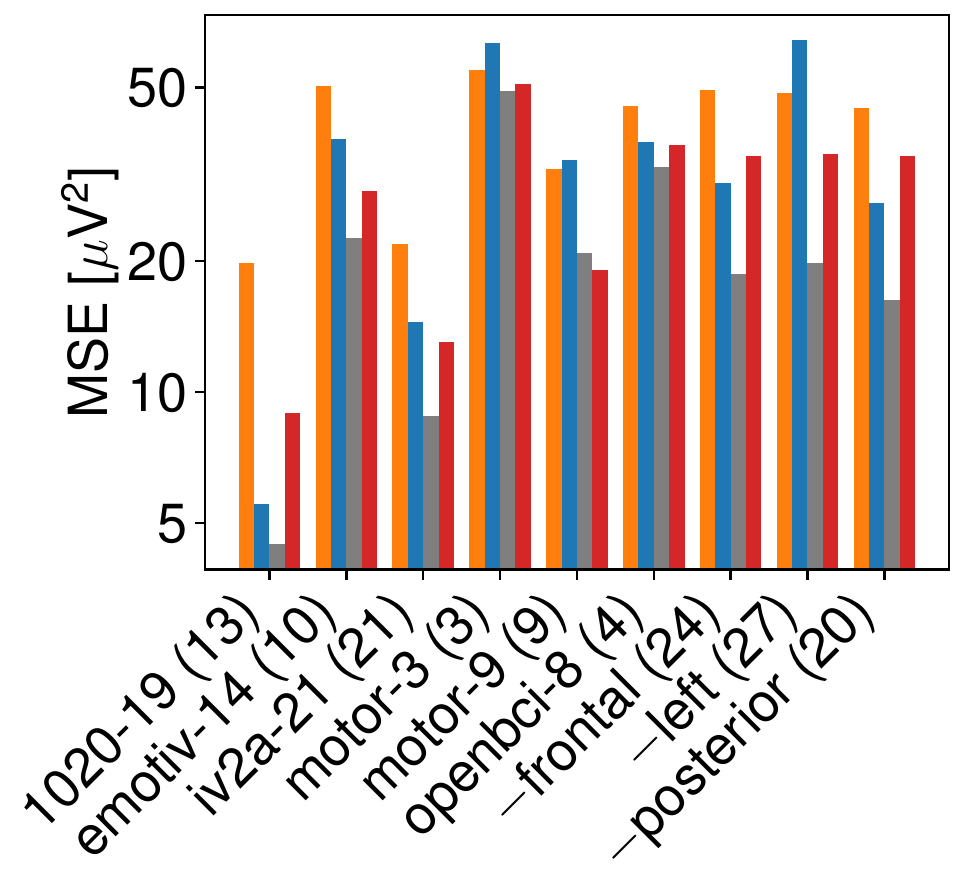}}\hfill
\subfloat[BCI III-IIIa grid (60 ch)]{\includegraphics[width=0.33\textwidth]{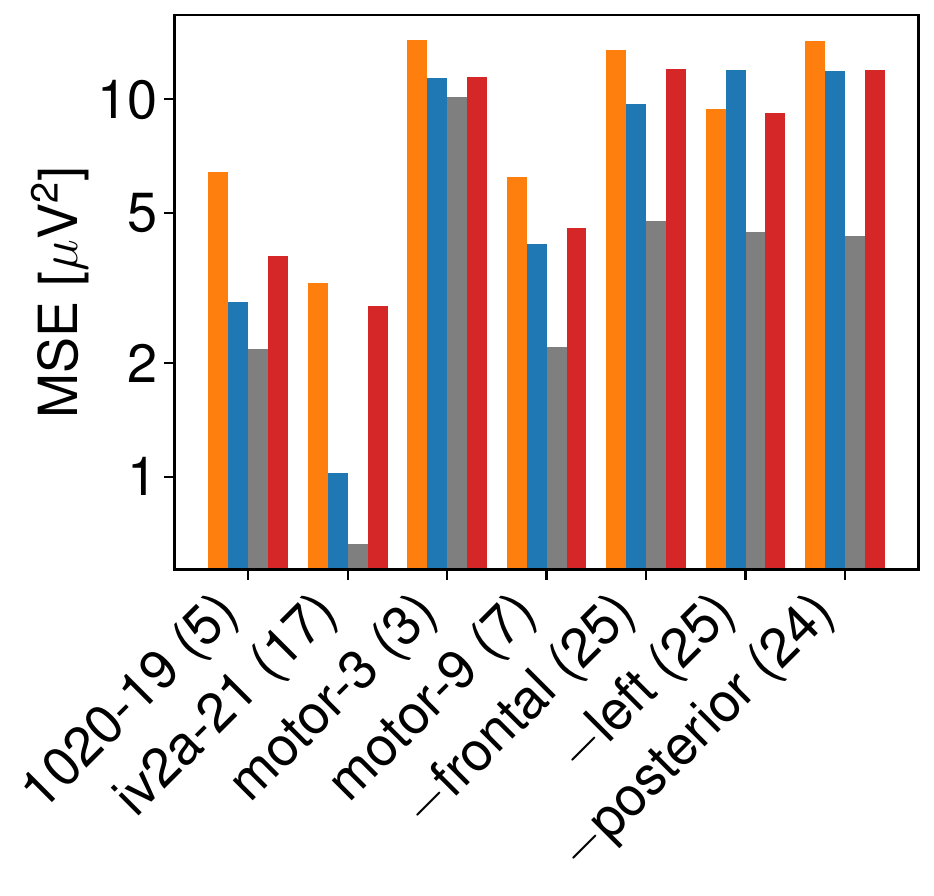}}\hfill
\subfloat[BCI IV-2a (22 ch)]{\includegraphics[width=0.33\textwidth]{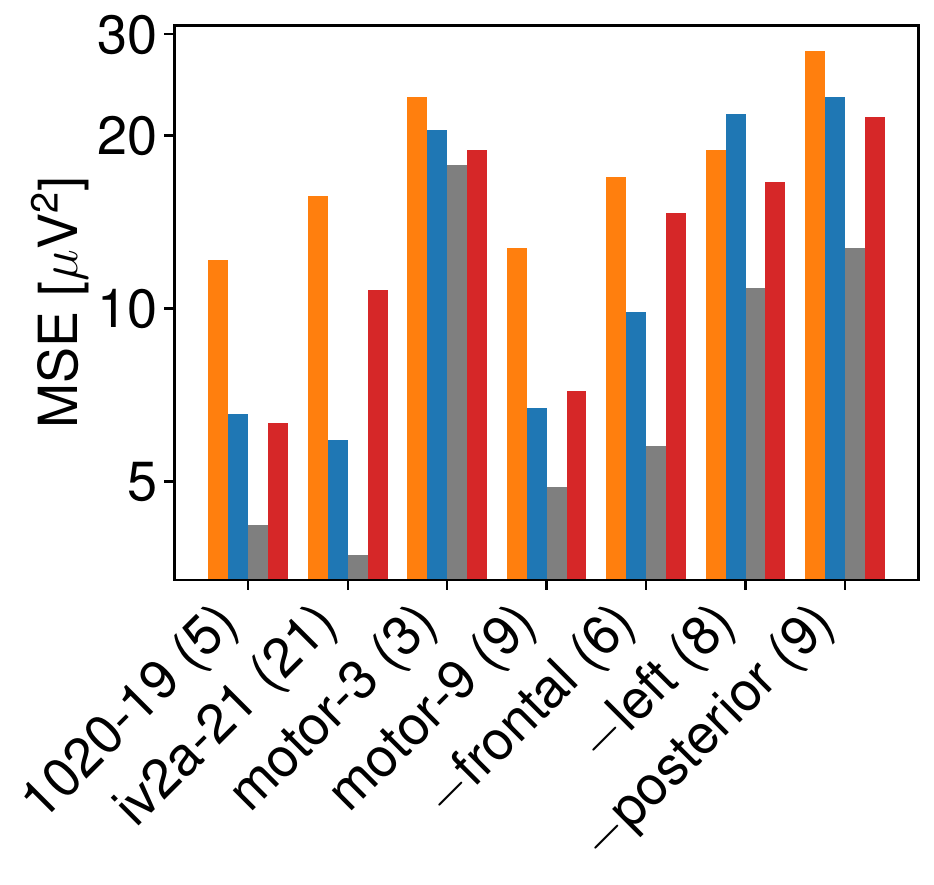}}\\[1mm]
\subfloat[HBN: EGI 128 ch, pediatric resting]{\includegraphics[width=0.33\textwidth]{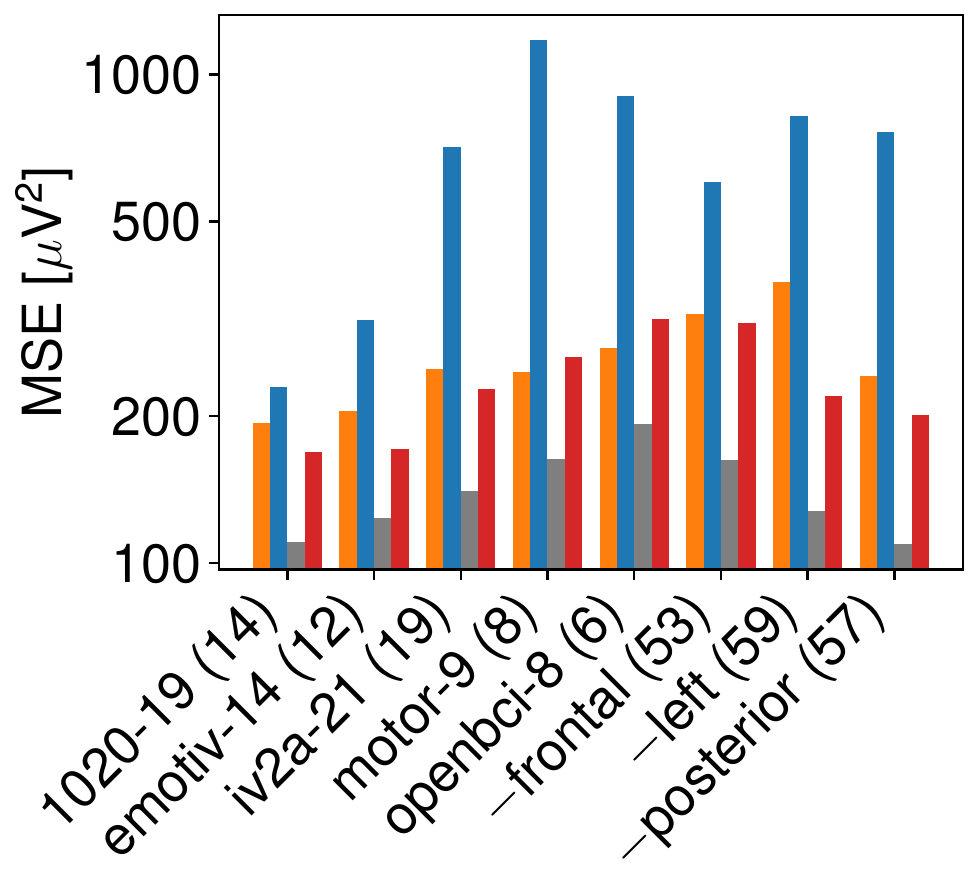}}\hfill
\subfloat[Nakanishi2015 (8 ch, SSVEP)]{\includegraphics[width=0.33\textwidth]{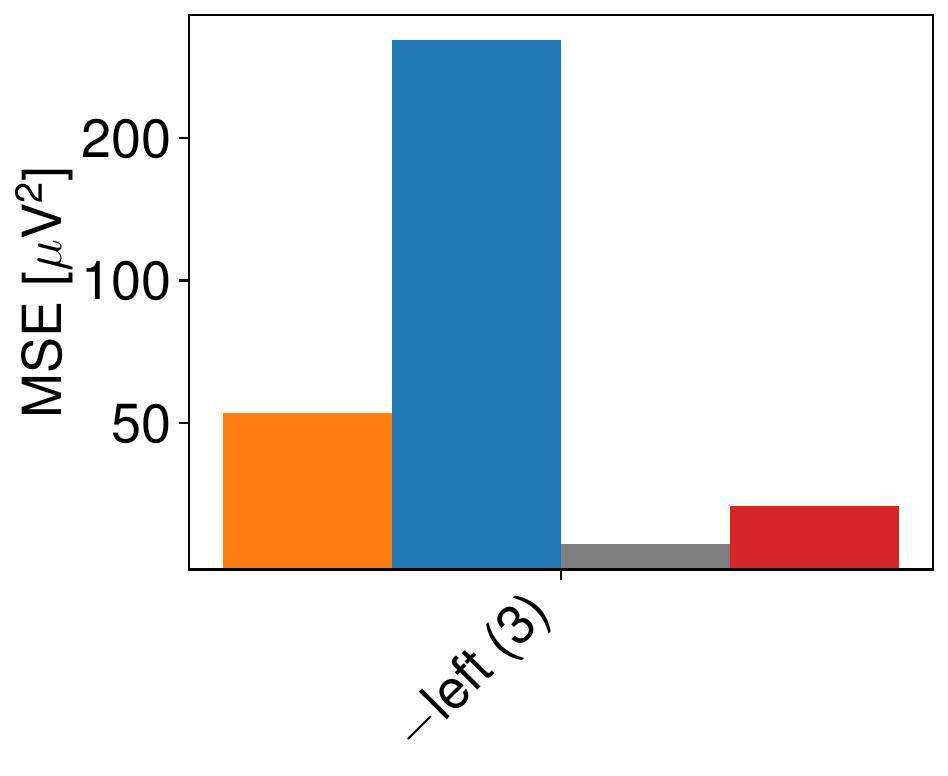}}\hfill
\subfloat[BrainInvaders 2013a (16 ch, P300)]{\includegraphics[width=0.33\textwidth]{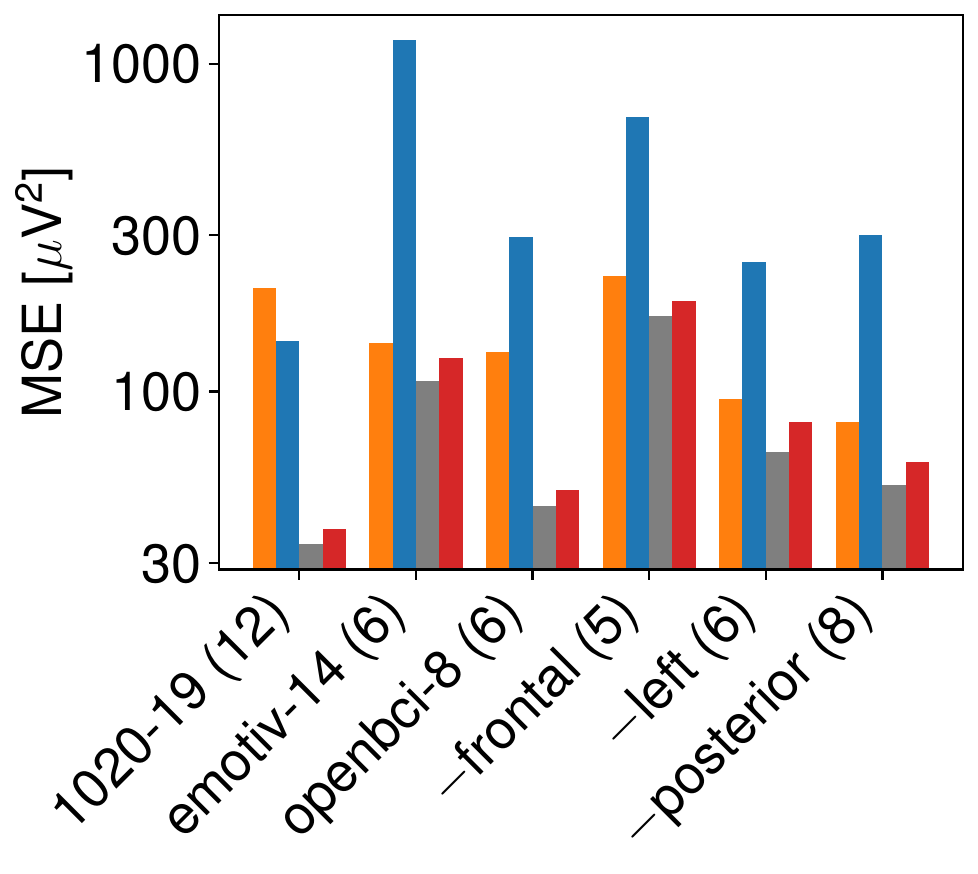}}\\[1mm]
\subfloat[Lee2019 MI (62 ch)*]{\includegraphics[width=0.33\textwidth]{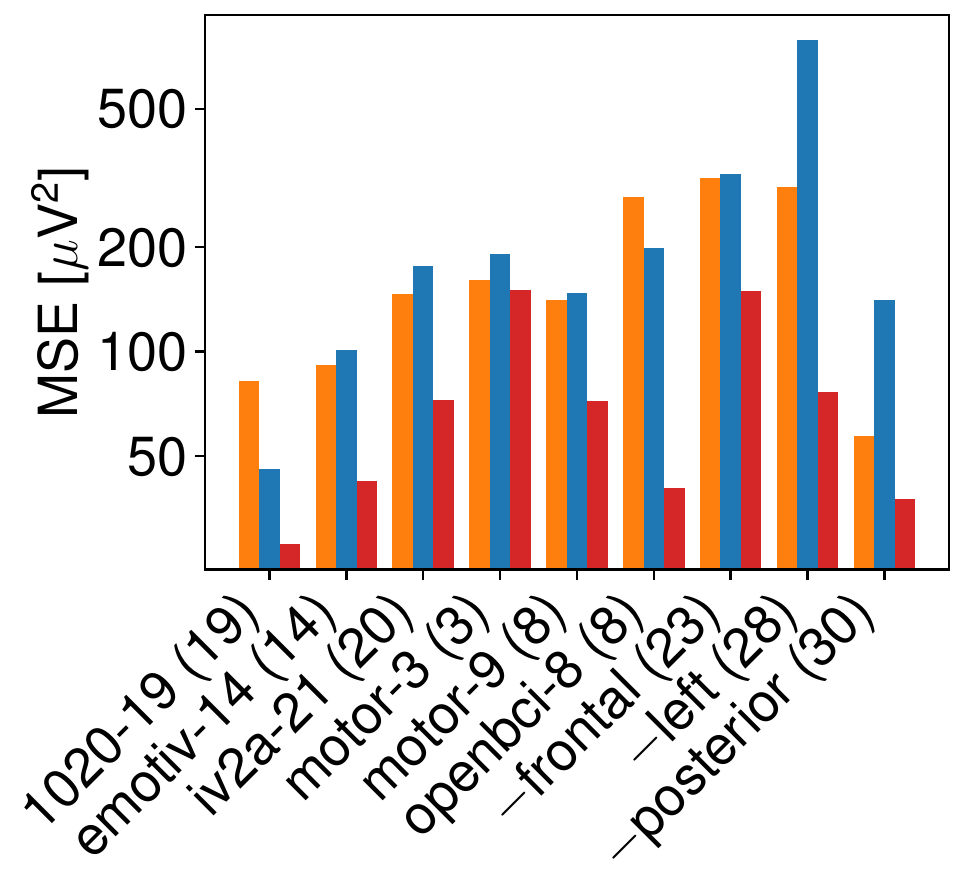}}\hfill
\subfloat[Lee2019 SSVEP (62 ch)*]{\includegraphics[width=0.33\textwidth]{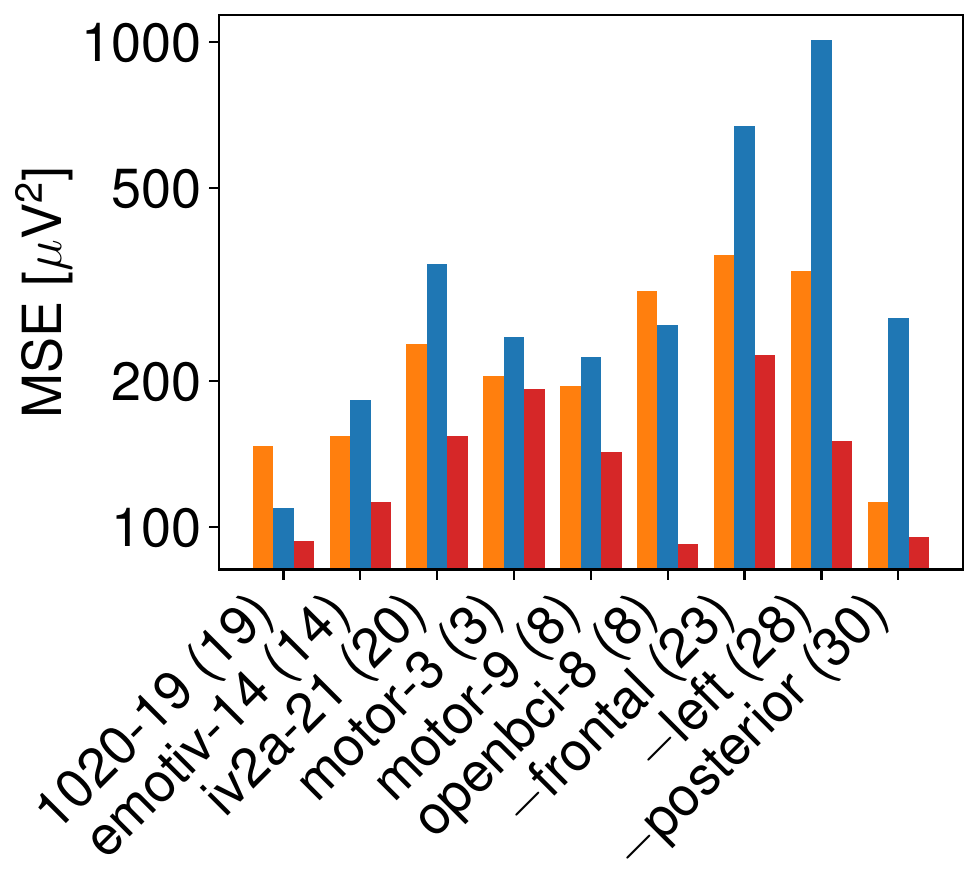}}\hfill
\subfloat[Lee2019 ERP (62 ch)*]{\includegraphics[width=0.33\textwidth]{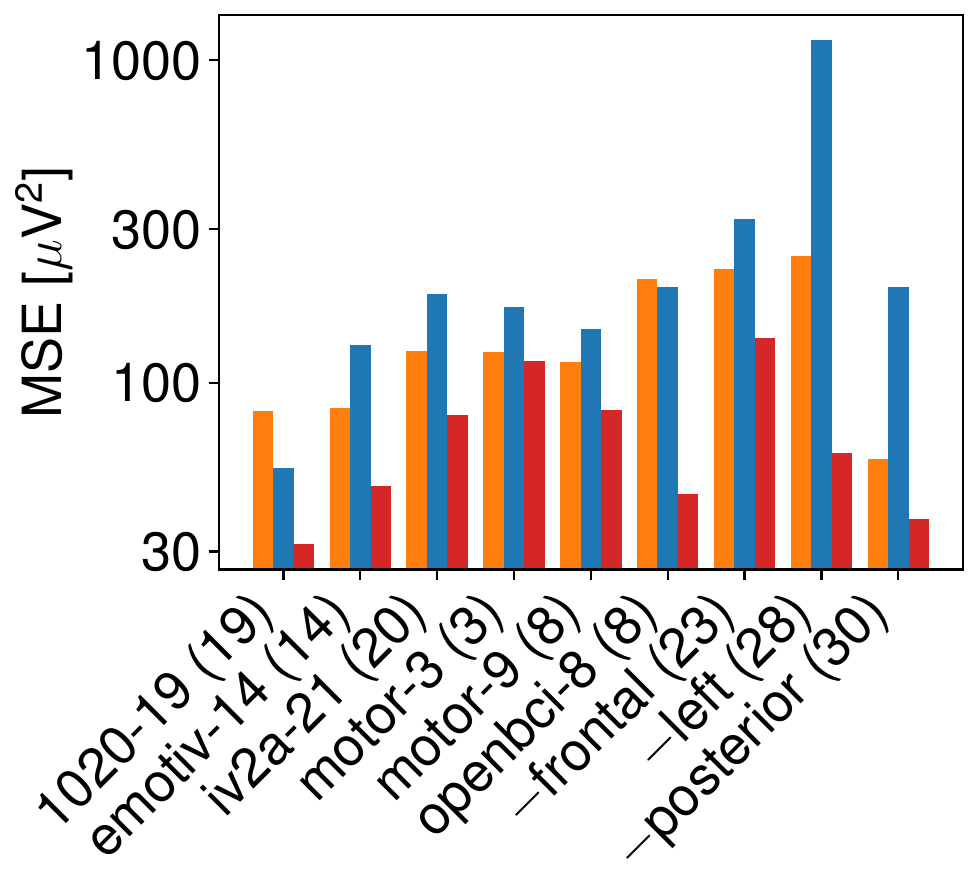}}\\[2mm]
\includegraphics[width=0.6\textwidth,trim=10 129 13 128,clip]{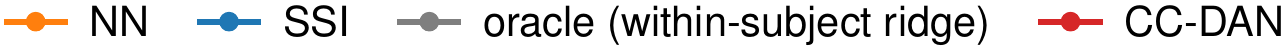}
\caption{Zero-shot completion on coordinate-matched real layouts and region dropout: MSE on missing electrodes (mean over subjects $\pm$ SEM) for NN, SSI, the within-subject ridge oracle, and \ccdan{}; the number of matched electrodes is given in parentheses, and $-$left etc.\ denote region dropout with the number of dropped electrodes in parentheses.
* Within-laboratory cross-paradigm transfer (Fig.~S2).}
\label{fig:main_layouts}
\end{figure}

Figs.~\ref{fig:main_k} and \ref{fig:main_layouts} and Table~\ref{tab:main} show the zero-shot performance of \ccdan{} pretrained on Tier-1.

\begin{table}[tbp]
\centering\footnotesize\setlength{\tabcolsep}{3pt}
\caption{Main Zero-Shot Results (MSE on Missing Electrodes in \muVsq{}, Mean Over Subjects).
Bold: Better Than $\minb$.}
\label{tab:main}
\resizebox{\textwidth}{!}{%
\begin{tabular}{llrrrr}
\toprule
Data & Condition & NN & SSI & Oracle & \ccdan{}\\
\midrule
BCI IV-1 59 ch & $k$=3\,/\,5\,/\,9 & 64.7\,/\,48.3\,/\,33.0 & 62.7\,/\,48.0\,/\,21.0 & 44.7\,/\,22.9\,/\,12.5 & \textbf{50.2\,/\,32.6\,/\,20.6}\\
 & $k$=15\,/\,44 & 24.1\,/\,15.2 & 11.5\,/\,4.5 & 8.2\,/\,3.6 & 14.6\,/\,9.3\\
 & motor\_9\,/\,emotiv\_14 (10) & 32.5\,/\,50.3 & 34.2\,/\,38.1 & 20.9\,/\,22.6 & \textbf{19.0\,/\,29.0}\\
 & left-hemisphere dropout (27) & 48.6 & 64.2 & 19.8 & \textbf{35.3}\\
BCI III-IIIa 60 ch & $k$=3\,/\,5\,/\,9 & 14.2\,/\,10.3\,/\,6.5 & 17.9\,/\,11.6\,/\,4.0 & 10.0\,/\,5.5\,/\,2.3 & \textbf{12.2\,/\,8.6}\,/\,5.4\\
BCI IV-2a 22 ch & $k$=3\,/\,5\,/\,9 & 23.7\,/\,17.4\,/\,12.6 & 24.3\,/\,18.0\,/\,7.4 & 16.5\,/\,8.9\,/\,4.2 & \textbf{19.7\,/\,13.3}\,/\,8.4\\
 & motor\_3\,/\,left-hemisphere dropout (8) & 23.3\,/\,18.9 & 20.4\,/\,21.8 & 17.7\,/\,10.8 & \textbf{18.8\,/\,16.6}\\
HBN EGI 128 ch & $k$=3\,/\,5\,/\,9 & 381\,/\,300\,/\,259 & 1035\,/\,720\,/\,437 & 238\,/\,165\,/\,123 & \textbf{350\,/\,256\,/\,200}\\
 & $k$=15\,/\,44 & 218\,/\,165 & 261\,/\,125 & 96\,/\,64 & \textbf{173}\,/\,131\\
 & 1020\_19 (14)\,/\,emotiv\_14 (12) & 193\,/\,205 & 229\,/\,315 & 110\,/\,124 & \textbf{169\,/\,172}\\
 & left-hemisphere dropout (59) & 376 & 822 & 128 & \textbf{220}\\
Nakanishi2015 8 ch & $k$=3\,/\,5 & 53.4\,/\,57.0 & 120\,/\,103 & 31.9\,/\,28.6 & \textbf{37.6\,/\,32.5}\\
BI2013a 16 ch & $k$=3\,/\,5\,/\,9 & 135\,/\,126\,/\,141 & 264\,/\,297\,/\,218 & 89\,/\,65\,/\,62 & \textbf{100\,/\,75\,/\,63}\\
 & 1020\_19 (12)\,/\,openbci\_8 (6) & 207\,/\,132 & 143\,/\,296 & 34\,/\,45 & \textbf{38\,/\,50}\\
\bottomrule
\end{tabular}}
\end{table}

(1) With at most five observed electrodes, the mean MSE was below $\minb$ on every evaluation set (BCI IV-1: $-18$ to $-32\%$; BCI IV-2a: $-15$ to $-21\%$; HBN: $-16\%$), and so were all subjects except at $k$=3 on HBN (29/40): BCI IV-1 4/4, BCI III-IIIa 3/3, BCI IV-2a 9/9 (Wilcoxon $p$=0.004), HBN $k$=5 39/40, Nakanishi2015 9/9 ($p$=0.004), and BI2013a 24/24 ($p<0.001$).
(2) On the sparse real layouts (motor 9 ch, Emotiv 14 ch) \ccdan{} was best on BCI IV-1 and BI2013a but not on BCI III-IIIa or on motor 9 ch of BCI IV-2a and HBN, and under hemispheric dropout it was best and SSI worst (BCI IV-1: 64 vs 35 \muVsq{}); per subject, 4/4 on BCI IV-1, 8/9 under left-hemisphere dropout on BCI IV-2a ($p$=0.008), 30--38/40 on HBN except OpenBCI 8 ch (8/40) and frontal dropout (19/40), and 23--24/24 on BI2013a were below $\minb$, and the motor 3-ch layout on BCI IV-2a fell to 6/9.
(3) On high-density 10-05 montages (BCI IV-1, III-IIIa, IV-2a) with 15 or more observed electrodes, SSI was superior; the boundary lies near $k$=9, where the per-subject wins split (3/4, 0/3, 2/9).
(4) On the EGI net (HBN) and the 8- to 16-channel devices (Nakanishi2015, BI2013a), SSI failed throughout (extrapolation to peripheral electrodes), and \ccdan{} beat NN for $k\le 21$ (HBN: 38--39/40 subjects for $k$=5--15, $p<0.001$; BI2013a: 23--24/24 in every condition) and approached the within-subject oracle (Nakanishi2015 $k$=5: 32.5 vs oracle 28.6, SSI 103).
(5) The seed-to-seed standard deviation was at most 1.7 \muVsq{}.

\subsection{Cross-Net, Cross-Paradigm, and Cross-Laboratory Transfer}\label{sec:transfer}
A model trained only on 10-05-system nets transferred to the EGI GSN-128 (HBN) through coordinates alone, a pair for which no electrode-name correspondence exists (Table~\ref{tab:main}).
The large absolute errors on HBN (130--350 \muVsq{}) reflect the amplitude of pediatric resting EEG (channel variance $\approx150$ \muVsq{}); relative errors are on par with BCI IV-1 and IV-2a.
On HBN at $k$=21/29/44, \ccdan{} gave 152/141/131 vs NN 202/178/165 and SSI 208/155/125 (37/23/14 subjects won), so the advantage is confined to $k\le 21$, and it lost to NN on the peripheral OpenBCI 8-ch layout (316 vs 275).

Evaluating each Lee2019 paradigm with Tier-1 trained without it, 50--54/54 subjects were below $\minb$ in every condition (MI $k$=9: 58 vs NN 127 / SSI 140 \muVsq{}; SSVEP $k$=9: 117 vs 228 / 311; ERP $k$=9: 58 vs 116 / 158; Fig.~S2).
Because the three paradigms share the same 54 subjects, device, and montage, this is within-laboratory cross-paradigm transfer.
True zero-shot corresponds to Nakanishi2015 and BI2013a, whose subjects, devices (g.USBamp, Nexus), and paradigms (SSVEP, P300) all differ from training; there too all subjects were below $\minb$, including region dropout on BI2013a (e.g., posterior dropout 61 vs NN 81 / SSI 299 \muVsq{}; 23--24/24 subjects).

\subsection{Regime Map and Band-Wise Error}\label{sec:regime}
\begin{figure}[tbp]
\centering
\begin{minipage}[c]{0.50\textwidth}\centering
\subfloat[Regime map]{\includegraphics[width=\linewidth]{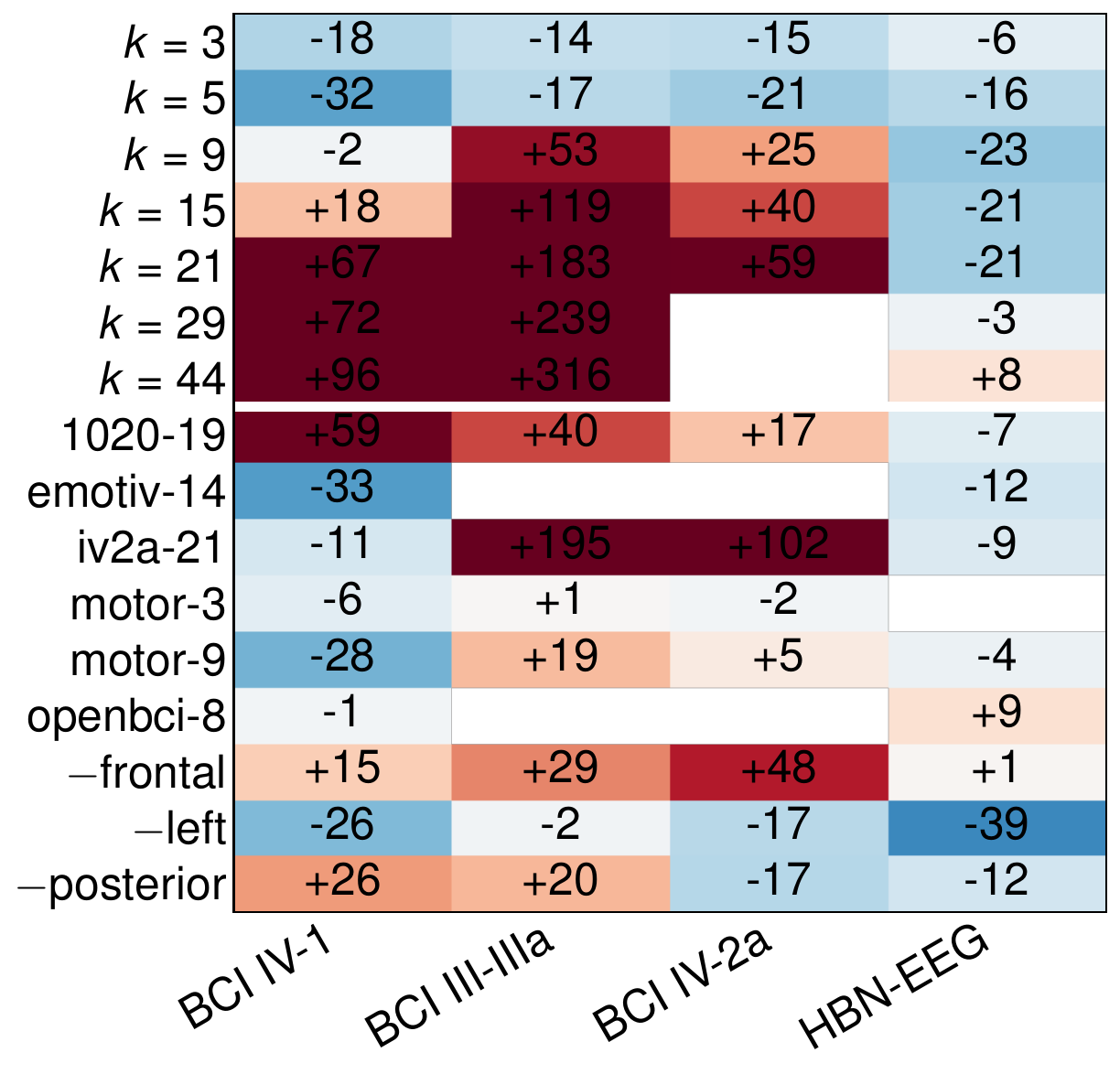}}
\end{minipage}\hfill
\begin{minipage}[c]{0.08\textwidth}\centering\includegraphics[width=\linewidth,trim=186 7 7 7,clip]{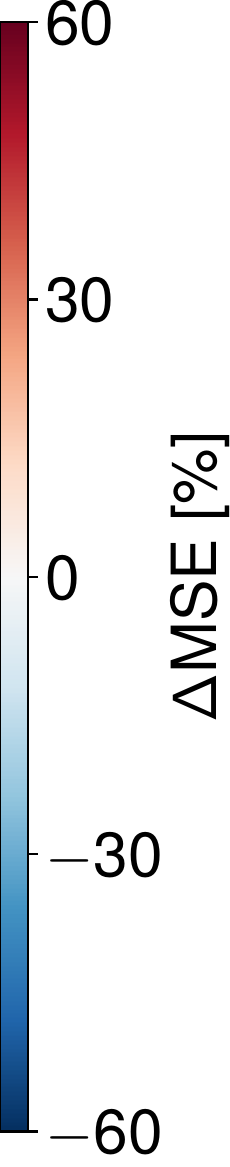}\end{minipage}\hfill
\begin{minipage}[c]{0.38\textwidth}\centering
\subfloat[Band-wise relative error]{\includegraphics[width=\linewidth]{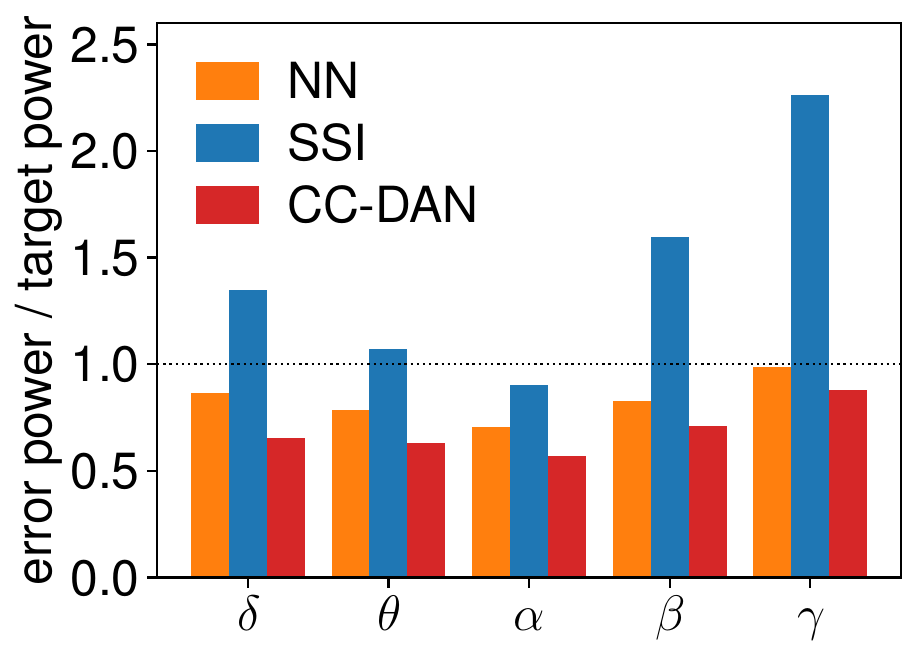}}
\end{minipage}
\caption{Regime map.
(a) Median over subjects of the relative change in MSE of \ccdan{} against $\minb$ ($\Delta$MSE) per condition (rows) and dataset (columns) (\%; blue: learning wins, red: training-free wins); the upper block covers random $k$ electrodes and the lower block real layouts and region dropout.
(b) Band-wise relative error at missing electrodes (error power over true power; $k$=5, $k$=9, and motor 9 ch pooled over all evaluation sets; dotted line: error as large as the signal).}
\label{fig:regime}
\end{figure}

Fig.~\ref{fig:regime} shows the relative change against $\minb$ per condition and dataset and the band-wise relative error.
Learning wins for $\le 5$--9 observed electrodes, sparse real layouts, hemispheric dropout, and nets outside the 10-10 system and low-channel devices, whereas SSI wins for $k\ge 15$ on 10-05 systems, dense grids, and frontal or posterior dropout (where SSI extrapolation works).
The gap to the within-subject oracle is small at low density (BCI IV-1 $k$=3: 50 vs 45) and large at high density ($k$=9: 20.6 vs 12.5).
By band, at $k$=5 \ccdan{} was best in every band ($\delta$ 0.72 / $\alpha$ 0.71 / $\gamma$ 0.97 vs SSI 1.99 / 1.44 / 3.44 and NN 1.02 / 0.90 / 1.20), the $\beta$/$\gamma$ error power of SSI exceeded the true power at low density (extrapolation adds noise), and at $k$=9--15 \ccdan{} remained best in the $\gamma$ band ($k$=15: 0.83 vs SSI 1.14), so the regime of learned methods widens toward higher frequencies.

\subsection{Scaling, Robustness to Coordinate Error, and Ablation}\label{sec:ablation}
\begin{figure}[tbp]
\centering
\subfloat[BCI IV-1: corpus size]{\includegraphics[width=0.29\textwidth]{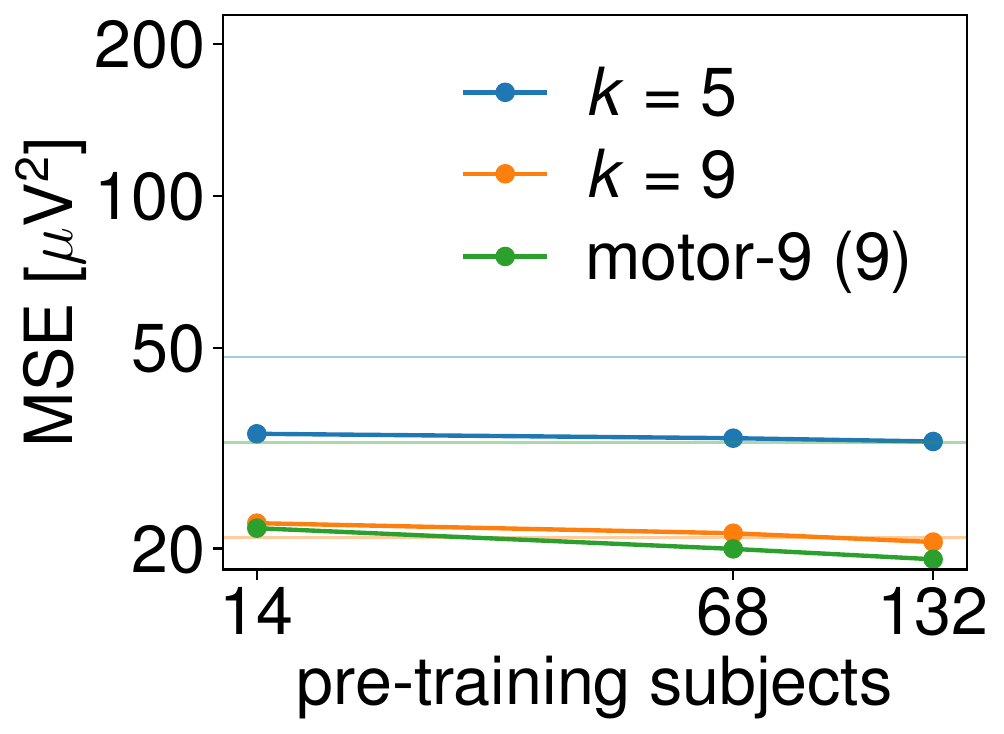}}\hfill
\subfloat[BCI IV-1: position error]{\includegraphics[width=0.29\textwidth]{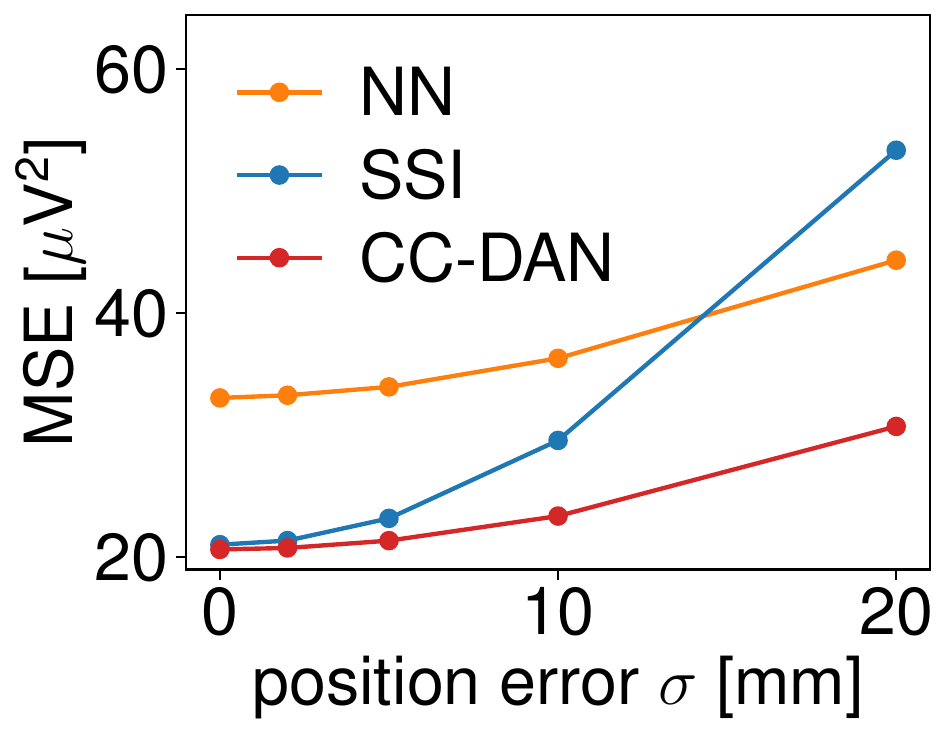}}\hfill
\subfloat[BCI IV-1: ablation]{\includegraphics[width=0.40\textwidth]{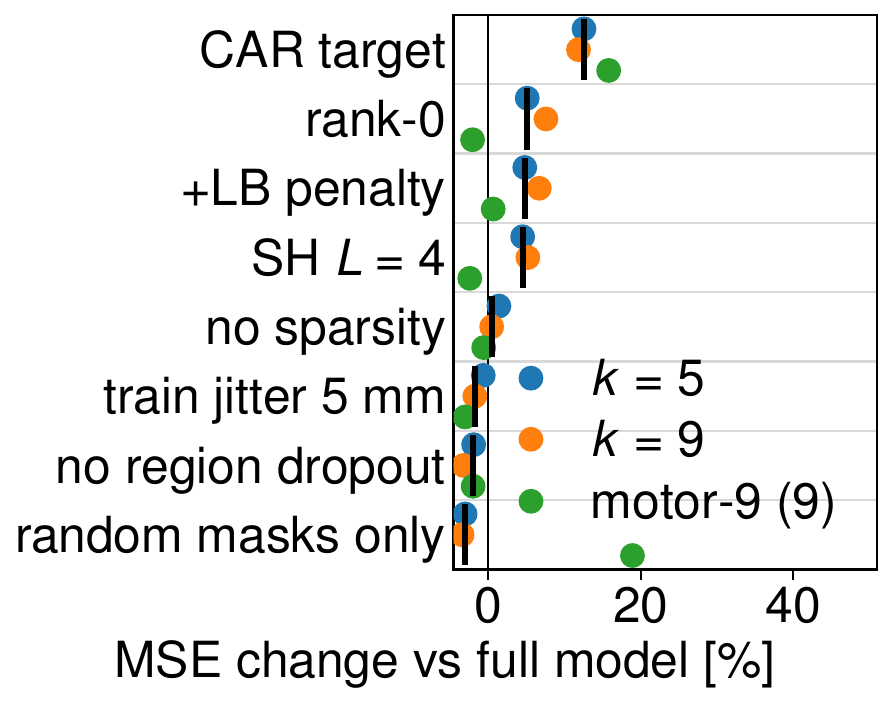}}\\[1mm]
\subfloat[BCI IV-2a: corpus size]{\includegraphics[width=0.29\textwidth]{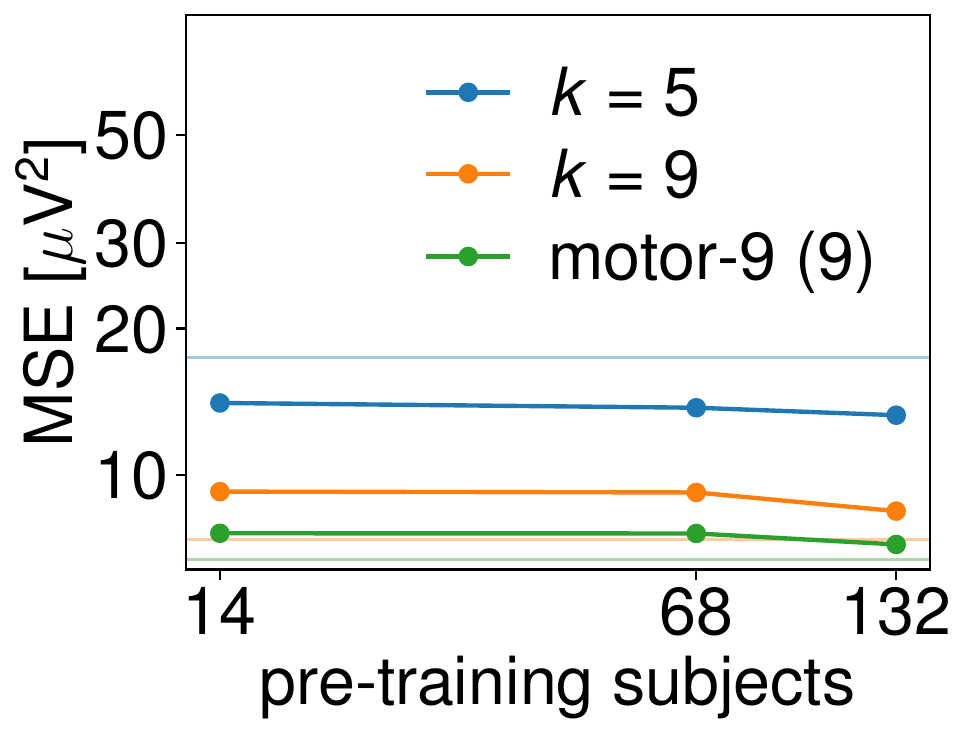}}\hfill
\subfloat[BCI IV-2a: position error]{\includegraphics[width=0.29\textwidth]{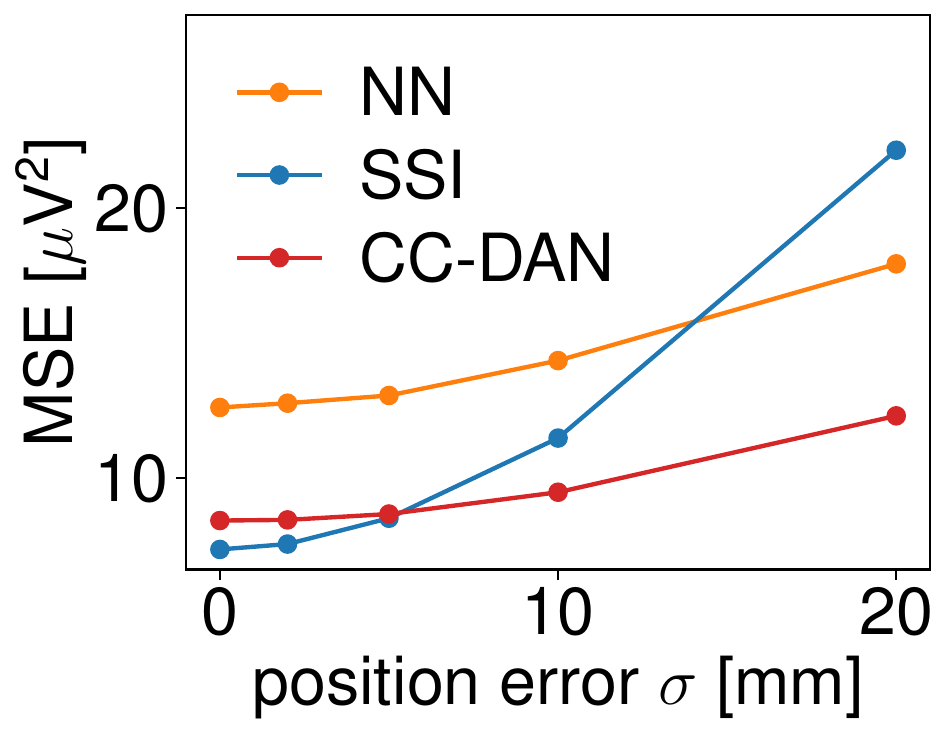}}\hfill
\subfloat[BCI IV-2a: ablation]{\includegraphics[width=0.40\textwidth]{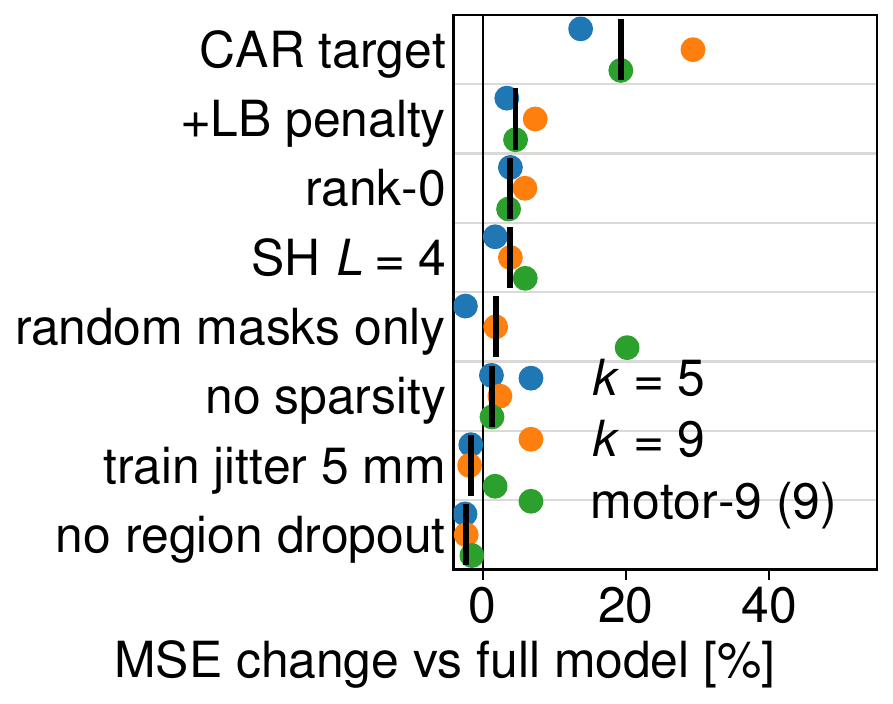}}
\caption{Sensitivity analyses on BCI IV-1 (top row) and BCI IV-2a (bottom row).
(a, d) Zero-shot MSE versus the number of pretraining subjects (14: Schirrmeister2017 only; 68: +Lee2019 MI; 132: Tier-1) at $k$=5, $k$=9, and motor 9 ch (log-log); thin horizontal lines: $\minb$ for the same condition.
(b, e) MSE when isotropic Gaussian error with standard deviation $\sigma$ is added to the electrode positions given to each method ($k$=9; Table~\ref{tab:perturb}).
(c, f) Ablation pretrained on Schirrmeister2017 only: relative change in zero-shot MSE against the full model (vertical line at 0) for the three conditions of (a); variants are sorted by the median over the conditions of each dataset (black tick); exact values in Table~\ref{tab:ablation}; +LB: Laplace--Beltrami smoothness penalty; SH $L$=4: degree-4 spherical harmonics.}
\label{fig:scaling}
\end{figure}

As shown in Fig.~\ref{fig:scaling} for BCI IV-1 (top row) and BCI IV-2a (bottom row), (a) increasing the number of pretraining subjects from 14 (Schirrmeister2017 only) to 68 (+Lee2019 MI) to 132 (Tier-1) reduced the zero-shot error monotonically in every condition (BCI IV-1 $k$=9: 22.4 $\to$ 21.4 $\to$ 20.6).
(b) Adding isotropic Gaussian error $\sigma$ to the electrode positions at test time (Table~\ref{tab:perturb}) degraded SSI at $\sigma$=20 mm from 21 to 53 and NN from 33 to 44 on BCI IV-1 $k$=9, whereas \ccdan{} went from 21 to 31; the same held on BCI IV-2a (SSI 7.4 $\to$ 22, \ccdan{} 8.4 $\to$ 12), so \ccdan{} was the most robust to coordinate error even without training-time jitter.
(c) In the ablation (trained on Schirrmeister2017 only, Table~\ref{tab:ablation}), a CAR target reference worsened MSE by $+12$ to $30\%$ and using random masks only worsened the real-layout conditions by $+19\%$, whereas removing region-dropout masks, training jitter, or the sparsity term had neutral to few-percent effects, rank-0 (shared filters only) cost $-2$ to $+8\%$, and $L=4$ and the smoothness penalty were slightly worse.

\begin{table}[tbp]
\centering\footnotesize
\caption{MSE on Missing Electrodes (\muVsq{}) Under Electrode-Position Error at Test Time (Isotropic Gaussian, Standard Deviation $\sigma$ in mm).}
\label{tab:perturb}
\resizebox{\textwidth}{!}{%
\begin{tabular}{llrrrrr}
\toprule
Data & Condition & Method & $\sigma$=0 & 5 & 10 & 20\\
\midrule
BCI IV-1 & $k$=9 & NN / SSI / \ccdan{} & 33.0 / 21.0 / 20.6 & 33.9 / 23.1 / 21.3 & 36.3 / 29.5 / 23.3 & 44.3 / 53.3 / 30.7\\
BCI IV-1 & motor\_9 & NN / SSI / \ccdan{} & 32.5 / 34.2 / 19.0 & 33.1 / 35.3 / 20.2 & 41.1 / 57.9 / 31.5 & 54.8 / 83.0 / 51.2\\
BCI IV-2a & $k$=9 & NN / SSI / \ccdan{} & 12.6 / 7.4 / 8.4 & 13.1 / 8.5 / 8.7 & 14.4 / 11.5 / 9.5 & 17.9 / 22.2 / 12.3\\
BCI IV-2a & $k$=5 & NN / SSI / \ccdan{} & 17.4 / 18.0 / 13.3 & 17.8 / 19.7 / 13.5 & 18.9 / 26.3 / 14.2 & 22.0 / 52.6 / 16.8\\
\bottomrule
\end{tabular}}
\end{table}

\begin{table}[tbp]
\centering\footnotesize
\caption{Ablation (Pretrained on Schirrmeister2017 Only, Zero-Shot, MSE on Missing Electrodes in \muVsq{}).}
\label{tab:ablation}
\begin{tabular}{lrrrrrr}
\toprule
 & \multicolumn{3}{c}{BCI IV-1} & \multicolumn{3}{c}{BCI IV-2a}\\
\cmidrule(lr){2-4}\cmidrule(lr){5-7}
Variant & $k$=5 & $k$=9 & motor\_9 & $k$=5 & $k$=9 & motor\_9\\
\midrule
Full (SH $L$=8, rank-4, observed-set average reference) & 33.8 & 22.4 & 22.0 & 14.1 & 9.2 & 7.6\\
rank-0 (shared filters only) & 35.5 & 24.2 & 21.5 & 14.6 & 9.8 & 7.9\\
SH $L$=4 & 35.3 & 23.6 & 21.4 & 14.3 & 9.6 & 8.0\\
+ Laplace--Beltrami penalty & 35.4 & 24.0 & 22.1 & 14.5 & 9.9 & 7.9\\
Target reference = CAR & 38.0 & 25.1 & 25.4 & 16.0 & 12.0 & 9.0\\
Random masks only & 32.7 & 21.7 & 26.1 & 13.7 & 9.4 & 9.1\\
No region-dropout masks & 33.1 & 21.7 & 21.5 & 13.7 & 9.0 & 7.5\\
No sparsity term & 34.2 & 22.6 & 21.8 & 14.2 & 9.5 & 7.7\\
Training jitter 5 mm & 33.5 & 22.1 & 21.3 & 13.8 & 9.1 & 7.7\\
\bottomrule
\end{tabular}
\end{table}

\subsection{Comparison Methods}\label{sec:comparison}
Table~\ref{tab:comparison} compares methods on BCI IV-1.
The fixed-layout super-resolution model SRGDiff \cite{srgdiff2026}, trained within BCI IV-1 after patching the public code (subject a held out, six subjects for training), reached 28.2 on motor 9 ch (subject a: SSI 36.0, NN 34.1) and 22.1 on the IV-2a-type 21 ch (subject a: SSI 14.6); even in-dataset training did not reach zero-shot \ccdan{} (19.0 / 13.0).
The reconstruction error of its variational autoencoder (VAE) with all electrodes as input (7.2 / 5.9 \muVsq{}) is small, so the bottleneck is the low-to-high-resolution latent mapping; the diffusion branch of the public code did not function because of a conditioning mismatch, so the model effectively reduced to a convolutional latent encoder (Appendix~\ref{app:comparison}).
Training on three real subjects and evaluating the fourth in four folds gave the same picture (Table~\ref{tab:comparison}, footnote d): only 1/4 subjects beat $\minb$ on each layout, and no fold reached zero-shot \ccdan{}.
Zero-shot reconstruction by the foundation model LUNA \cite{luna2025} gave an MSE equal to the signal power in every condition (equivalent to predicting zero) because its latent representation carries only frequencies below 4 Hz.
The reimplemented KM26 GCN \cite{km26} was impractical at low density ($\ge 96$ \muVsq{}) and reconstructed even the observed channels poorly (22--40 \muVsq{}).

\begin{table}[tbp]
\centering\footnotesize\setlength{\tabcolsep}{4pt}
\caption{Comparison Methods (BCI IV-1, MSE on Missing Electrodes in \muVsq{}).
Upper Block: Name-Based Protocol (CAR-59 Target, LOSO Within BCI IV-1); Lower Block: Plug-and-Play Protocol (Observed-Set Average Reference).}
\label{tab:comparison}
\resizebox{\textwidth}{!}{%
\begin{tabular}{llrrr}
\toprule
Method & Setting & iv2a\_21 & motor\_9 & $k$=9\\
\midrule
KM26 GCN reimplementation \cite{km26} & BCI IV-1 LOSO & 105 & 96 & 105\\
Multichannel DAN D-2 (name-based) & BCI IV-1 LOSO & 19.6 & 24.8 & 27.3\\
\ccdan{} rank-4$^{\mathrm{a}}$ & BCI IV-1 LOSO & 19.6 & 23.8 & 26.0\\
\midrule
SRGDiff \cite{srgdiff2026}, long setting$^{\mathrm{b}}$ & trained within BCI IV-1 (subject a held out) & 22.1 & 28.2 & --\\
SRGDiff, short setting (200 epochs) & same & 27.7 & 29.5 & --\\
SRGDiff VAE oracle (all electrodes as input)$^{\mathrm{c}}$ & same & 7.2 & 5.9 & --\\
SRGDiff, trained on three real subjects$^{\mathrm{d}}$ & trained within BCI IV-1 (mean over four folds) & 24.1 & 29.8 & --\\
LUNA \cite{luna2025}$^{\mathrm{e}}$ & zero-shot & -- & 58.7 & 53.2\\
REVE \cite{reve2025}$^{\mathrm{f}}$ & -- & -- & -- & --\\
SSI / NN & training-free & 14.5 / 21.9 & 34.2 / 32.5 & 21.0 / 33.0\\
\ccdan{} Tier-1 & zero-shot & 13.0 & \textbf{19.0} & \textbf{20.6}\\
\bottomrule
\end{tabular}}
\par\smallskip
\begin{minipage}{0.97\textwidth}\footnotesize\raggedright
$^{\mathrm{a}}$ Configuration of Table~\ref{tab:regression}.
$^{\mathrm{b}}$ Fixed-layout training; the diffusion branch of the public code is inactive (Appendix~\ref{app:comparison}).
$^{\mathrm{c}}$ The autoencoder is not the bottleneck.
$^{\mathrm{d}}$ Mean over folds a, b, f, g; SSI 15.1 / 34.4 and NN 21.7 / 32.2 on the same four subjects.
$^{\mathrm{e}}$ Error approximately equal to the signal power.
$^{\mathrm{f}}$ Not evaluable because the decoder is not released.
\end{minipage}
\end{table}

\subsection{Downstream Tasks and Amplitude Shrinkage}\label{sec:downstream}
\emph{Motor-imagery classification.}
Applying an EEGNet trained on the full montage to completed trials recovered 85--95\% of the full-montage accuracy from motor 9 ch (BCI IV-2a: \ccdan{} 0.78, SSI 0.80 vs full montage 0.83; Fig.~\ref{fig:downstream}(a, b), Table~S4).
\ccdan{} and SSI were on par: \ccdan{} was better with $\le 5$ observed electrodes and on the Emotiv layout (Lee2019 MI, motor 3 ch: 0.64 vs SSI 0.54; Emotiv 14 ch: 0.67 vs 0.64), SSI slightly better with 9--19 electrodes, and NN consistently worse.
A classifier retrained on the observed electrodes alone was best in every condition, so completion adds no information for classification.
Amplitude calibration was neutral on BCI IV-2a ($\pm0.03$) but hurt on Lee2019 motor 9 ch and $k$=9 ($-0.05$ to $-0.06$), because it also amplifies the noise component of the missing electrodes.

\emph{SSVEP detection.}
On Lee2019 SSVEP (full-montage CCA accuracy 0.954, chance 0.25; Fig.~\ref{fig:downstream}(c), Table~S5), when occipital electrodes were observed, CCA on the observed electrodes alone gave 0.92, NN/SSI completion was on par (0.91--0.92), and \ccdan{} completion was worse (0.84--0.87); without occipital electrodes (motor 9 ch, $k$=9), completion recovered 0.42--0.50 with NN/SSI (0.46--0.50) $\ge$ \ccdan{} (0.42--0.43).
CCA is invariant to per-channel scale, so the deficit of \ccdan{} is due to the spatial reproduction (smoothing) of the stimulus-frequency component rather than to amplitude.

\emph{$\alpha$ topography.}
On HBN resting 128 ch, the RMS error of $\log_{10}\alpha$ power at missing electrodes was 0.19--0.35 for NN and 0.18--0.67 for SSI, whereas \ccdan{} was worst at 0.29--0.67, and its correlation with the true topography was also low (Fig.~\ref{fig:downstream}(d), Table~S6; $k$=9: NN 0.56, SSI 0.41, \ccdan{} 0.20).

\emph{The cause is amplitude shrinkage.}
The amplitude gain of \ccdan{} at missing electrodes was 0.61--0.89 on BCI IV-1, IV-2a, and HBN for $k\ge5$ and fell to 0.18--0.54 at $k$=3 and on Nakanishi2015, whereas NN and SSI were mostly close to one (Table~S7); this shrinkage is $-1$ to $-4$ dB in $\alpha$ power (HBN $k$=9: $-3.1$ dB; BCI IV-2a $k$=9: $-3.9$ dB).
This shrinkage is inherent to MSE-optimal prediction: for truth $y$ and observation $\bm x$, the MSE-minimizing prediction is the conditional expectation $\hat y=\mathbb E[y\mid\bm x]$, and the law of total variance $\mathrm{Var}(y)=\mathrm{Var}(\hat y)+\mathbb E[\mathrm{Var}(y\mid\bm x)]$ shows that its variance is smaller than that of the truth by the component not determined by the observation.
With $\rho$ the correlation between prediction and truth, the amplitude gain of this prediction equals $\rho$ and its error is $(1-\rho^2)\sigma^2$, where $\sigma^2$ is the variance of the truth; fewer observed electrodes mean smaller $\rho$, so shrinkage is strongest exactly at the low densities where MSE gains are largest (Table~S7).
NN and SSI, by contrast, are fixed-weight linear combinations of observed signals that are not MSE-optimized, so they preserve variance (SSI even inflates amplitude in extrapolation regions), whereas the within-subject ridge oracle is MSE-optimal and shrinks likewise (amplitudes as low as \ccdan{} on Nakanishi2015).

\begin{figure}[tbp]
\centering
\subfloat[BCI IV-1]{\includegraphics[width=0.245\textwidth]{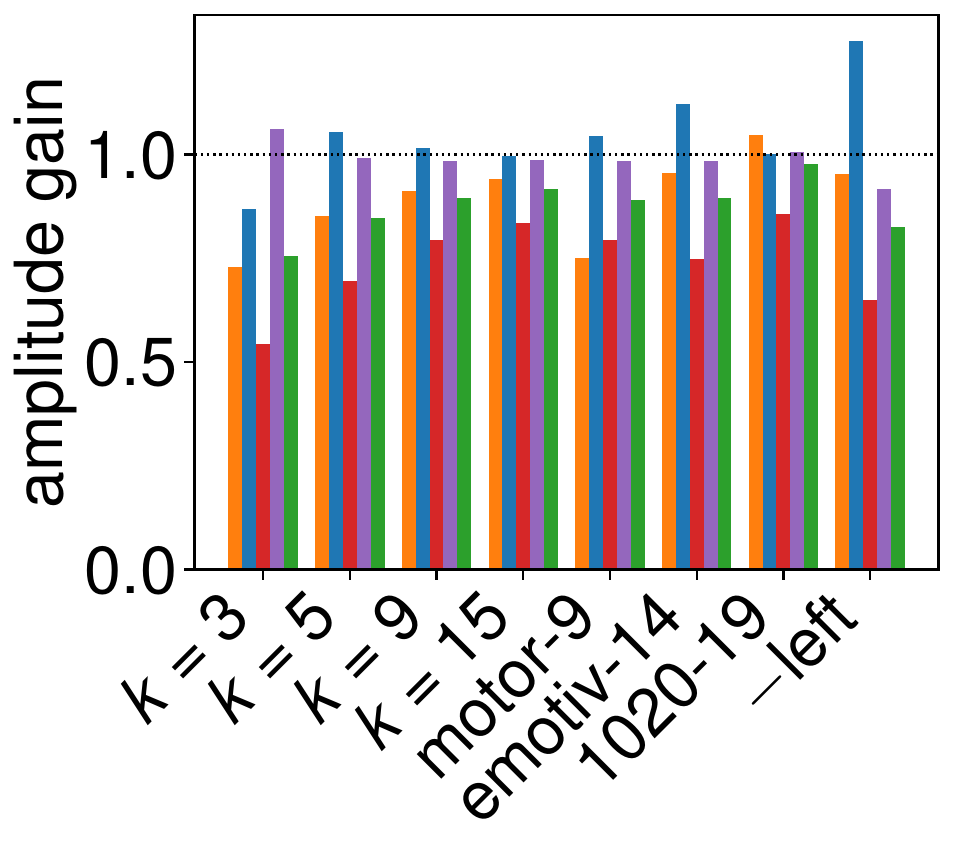}}\hfill
\subfloat[BCI IV-2a]{\includegraphics[width=0.245\textwidth]{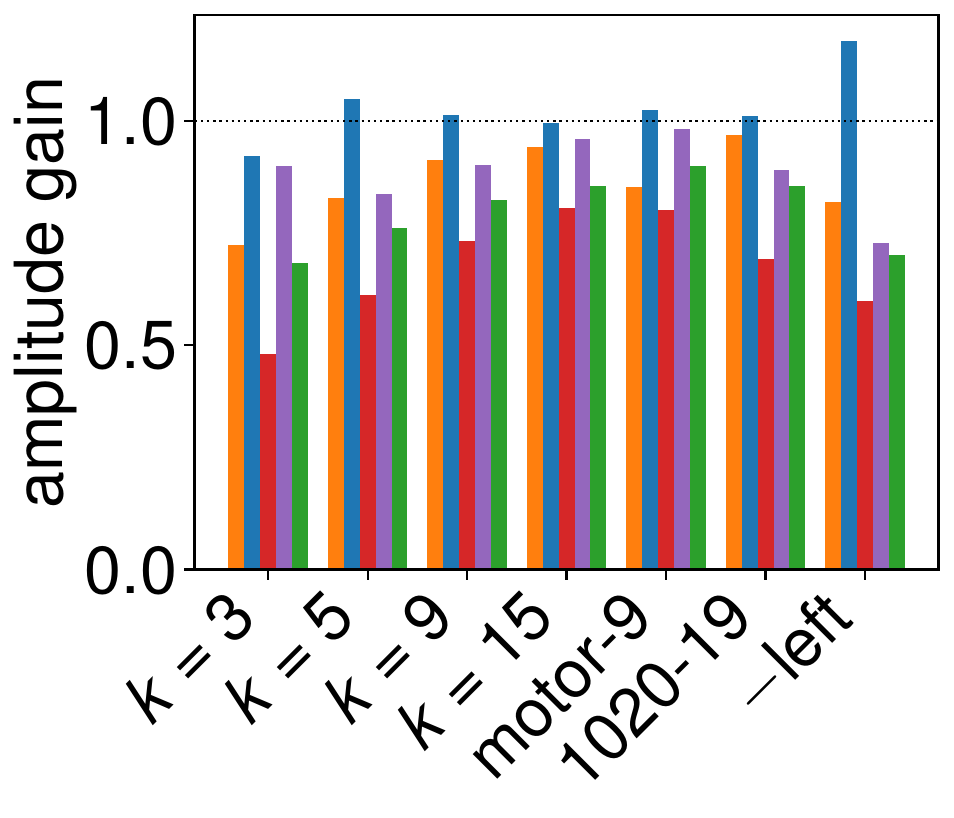}}\hfill
\subfloat[HBN]{\includegraphics[width=0.245\textwidth]{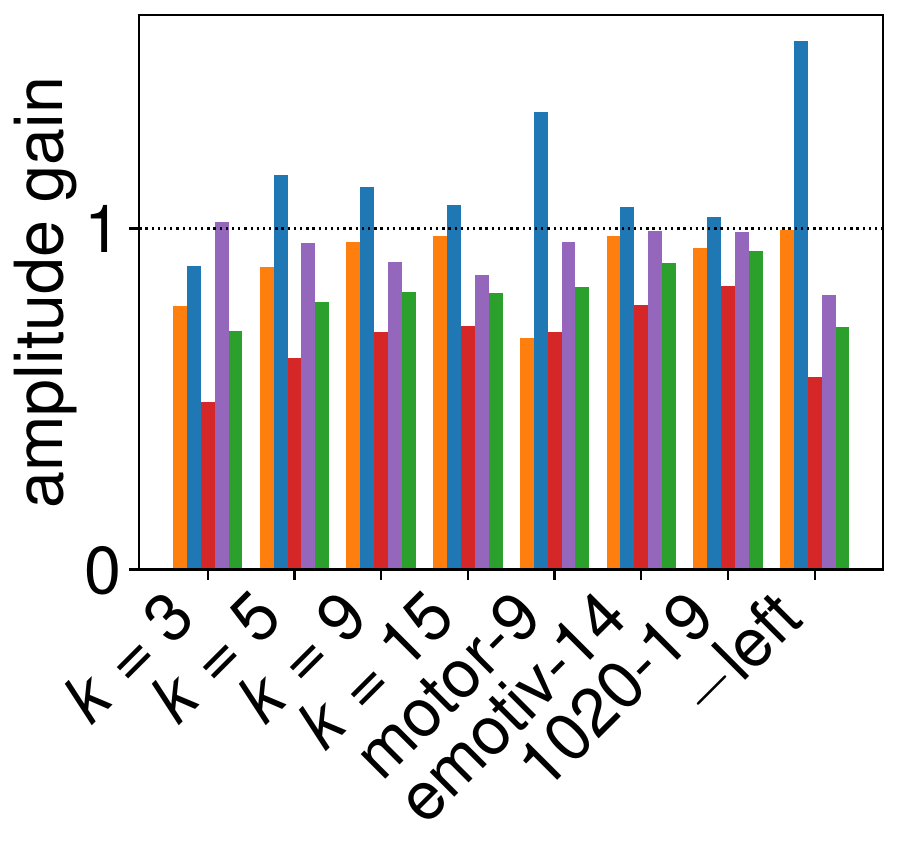}}\hfill
\subfloat[BI2013a]{\includegraphics[width=0.245\textwidth]{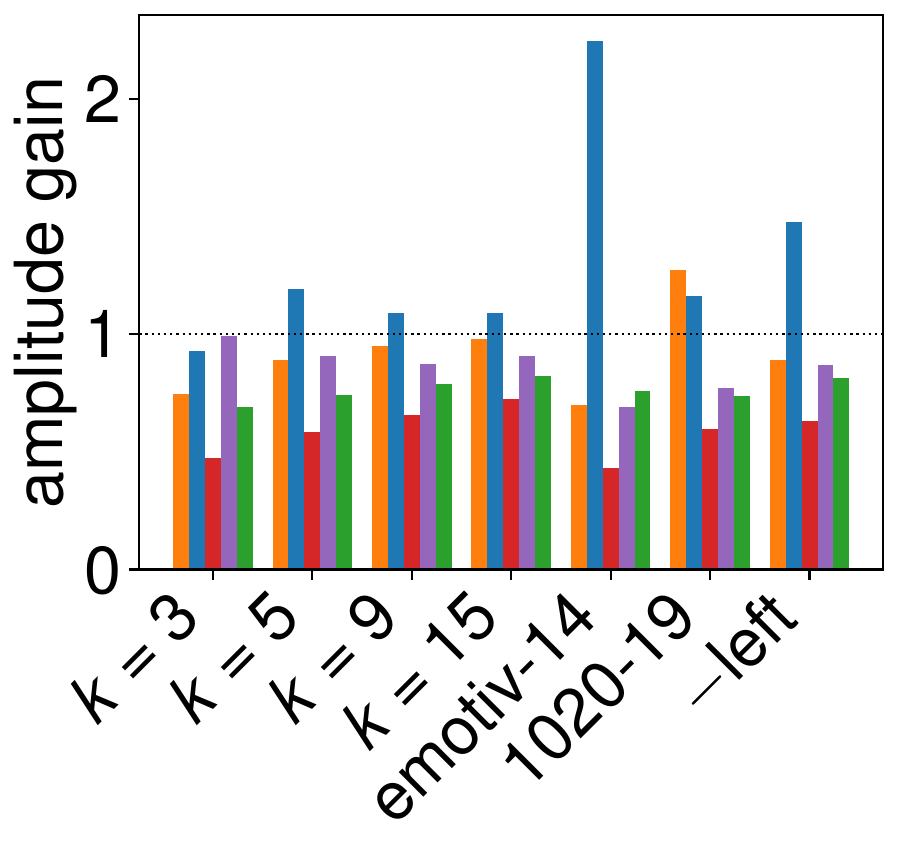}}\\[2mm]
\includegraphics[width=0.34\textwidth,trim=12 115 17 109,clip]{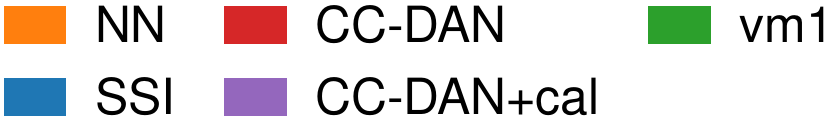}
\caption{Amplitude gain at missing electrodes (median over trials and missing electrodes of predicted over true standard deviation; dotted line: 1, unbiased) for NN, SSI, \ccdan{} (Tier-1), \ccdan{} with the post-hoc gain table (\ccdan{}+cal), and \ccdan{} pretrained with the variance-matching loss (vm1, $\lambda_{\mathrm{var}}=1$).
The corresponding MSE (cost of calibration) is given in Table~S7.}
\label{fig:calib}
\end{figure}

\begin{figure}[tbp]
\centering
\subfloat[MI, BCI IV-2a]{\includegraphics[width=0.245\textwidth]{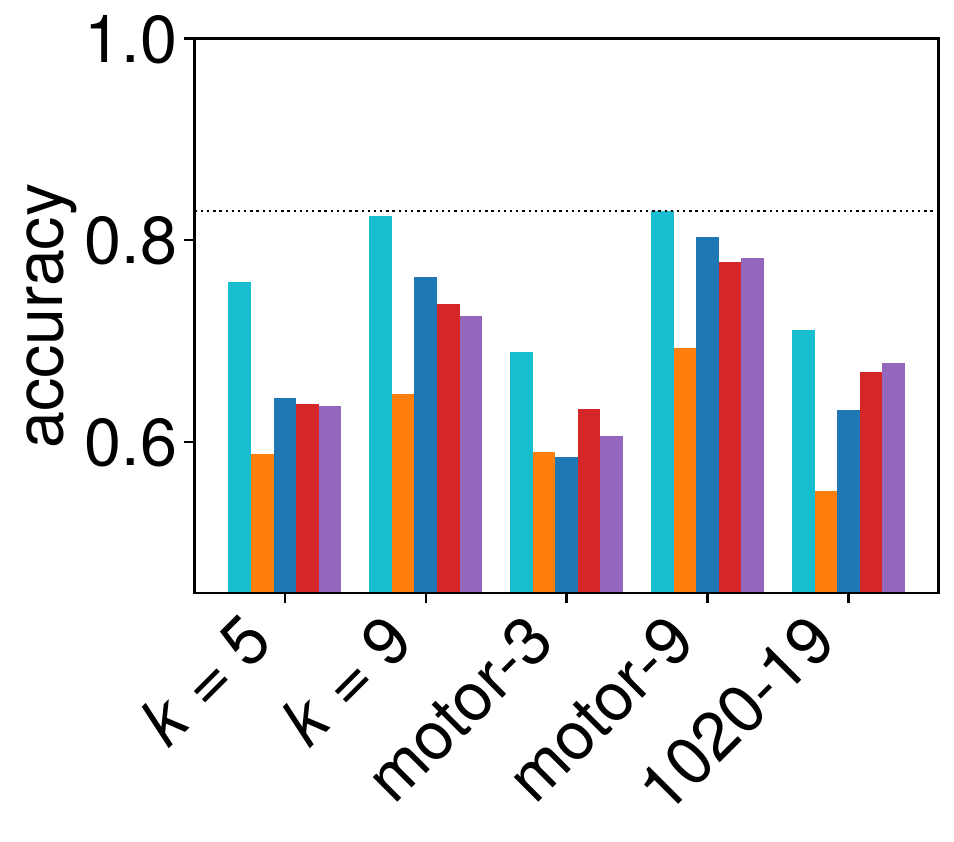}}\hfill
\subfloat[MI, Lee2019]{\includegraphics[width=0.245\textwidth]{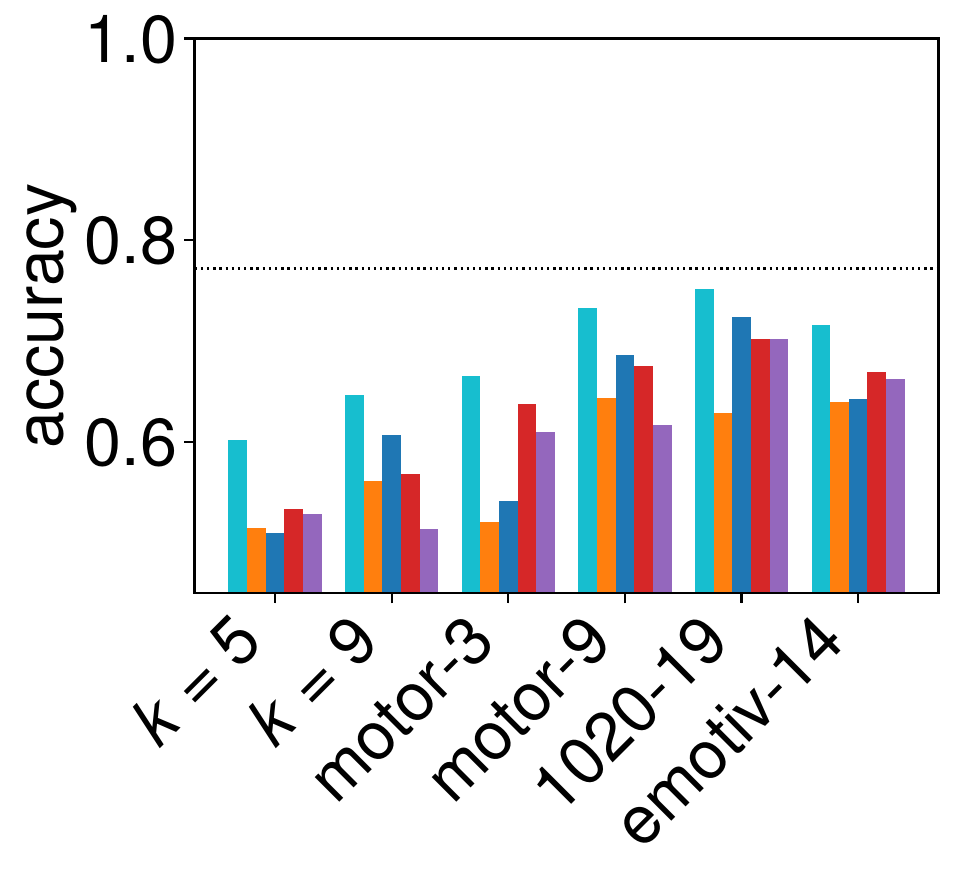}}\hfill
\subfloat[SSVEP, Lee2019]{\includegraphics[width=0.245\textwidth]{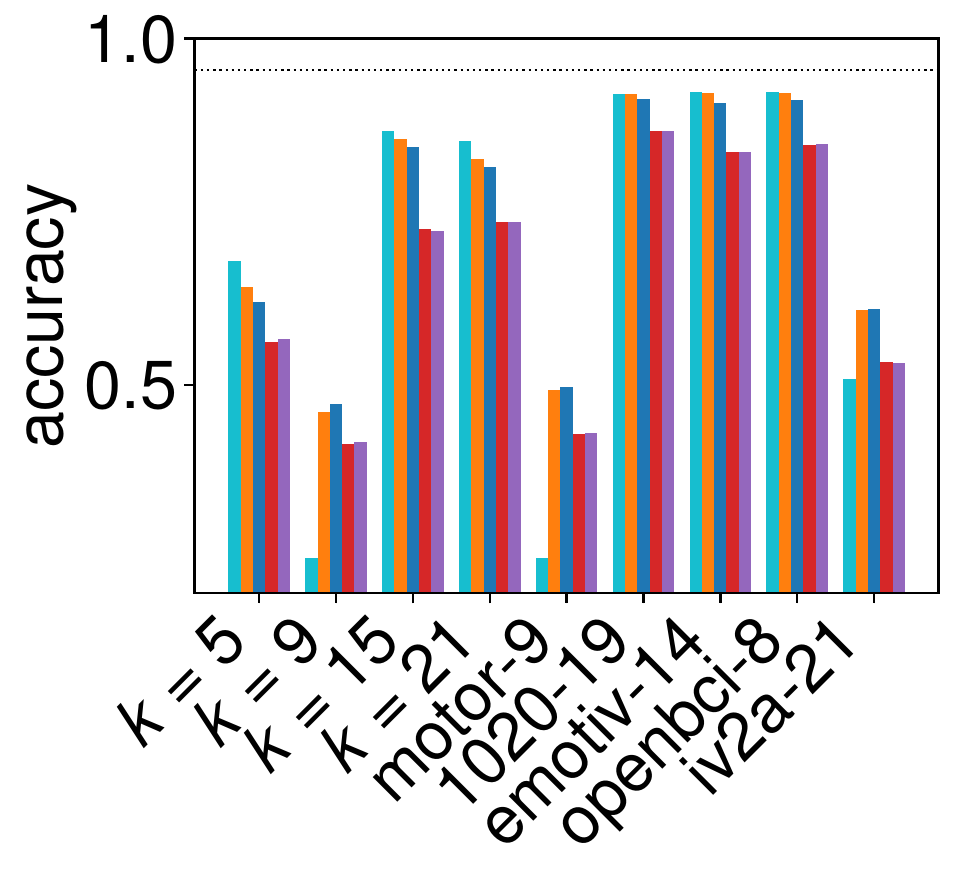}}\hfill
\subfloat[$\alpha$ topography, HBN]{\includegraphics[width=0.245\textwidth]{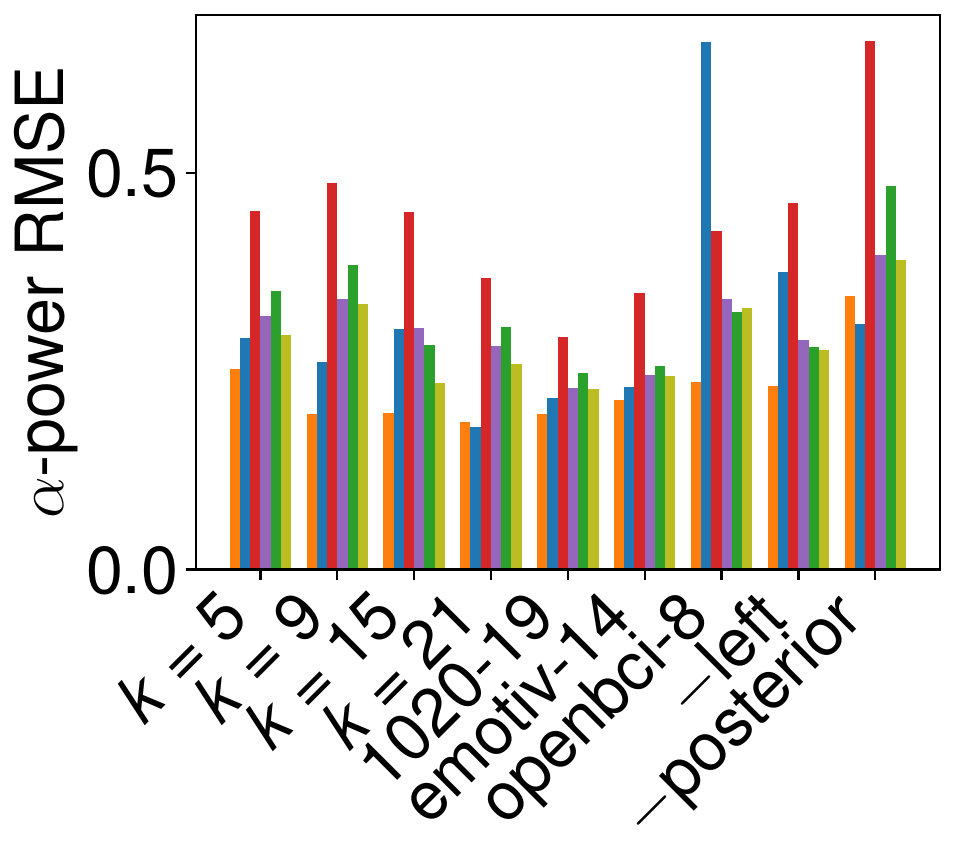}}\\[2mm]
\includegraphics[width=0.46\textwidth,trim=12 98 14 92,clip]{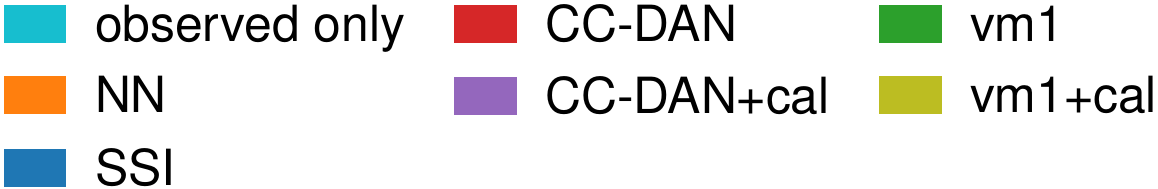}
\caption{Downstream tasks.
(a) BCI IV-2a (22 ch, LOSO over nine subjects) and (b) Lee2019 MI (62 ch, 44 training / 10 test subjects): accuracy of an EEGNet trained on real full-montage data and applied to completed low-density trials (dotted line: accuracy on real full-montage data, 0.829 / 0.772); bars: classifier retrained on the observed electrodes only (observed only), NN, SSI, \ccdan{}, and \ccdan{} with the post-hoc gain table (\ccdan{}+cal).
(c) Lee2019 SSVEP (54 subjects): CCA accuracy on 13 parieto-occipital electrodes after completion (dotted line: full montage, 0.954).
(d) HBN resting (EGI 128 ch, 40 subjects): RMS error of $\log_{10}\alpha$ power at missing electrodes for NN, SSI, \ccdan{}, \ccdan{}+cal, vm1 (pretrained with the variance-matching loss, $\lambda_{\mathrm{var}}=1$), and vm1+cal.}
\label{fig:downstream}
\end{figure}

\emph{Effect and cost of amplitude calibration.}
Fig.~\ref{fig:calib} and Table~S7 show amplitude gain and MSE on all evaluation sets.
(1) The variance-matching loss with $\lambda_{\mathrm{var}}=1$ (vm1) improved the gain by $+0.1$ to $0.25$ at almost unchanged MSE (BCI IV-1 $k$=9: MSE 20.6 $\to$ 20.0, gain 0.79 $\to$ 0.89; BCI IV-2a $k$=9: 8.4 $\to$ 8.2, 0.73 $\to$ 0.82); per subject, MSE improved by $-3$ to $-8\%$ for $k\ge9$ and hemispheric dropout on the 10-05 systems (BCI IV-2a: 9/9, $p$=0.004) but worsened by $+3$ to $12\%$ on the EGI net and low-channel devices (HBN $k$=3: 350 $\to$ 394).
(2) The post-hoc gain table restored the gain to nearly one (0.96--1.06) at a large MSE cost: $+1$ to $24\%$ for $k\ge9$ (BCI IV-1: $+7$ to $12\%$), $+2$ to $49\%$ at $k$=5, and $+5$ to $99\%$ at $k$=3 (BCI IV-1: 50 $\to$ 78; HBN: 350 $\to$ 697).
This follows from the framework above: the error of the MSE-optimal prediction is $(1-\rho^2)\sigma^2$, whereas restoring the amplitude to $\sigma$ gives $2\sigma^2(1-\rho)$, and the gap widens as $\rho$ decreases.
The table depends on geometry only, so it does not fit data whose amplitude structure differs from the corpus: on Nakanishi2015 the gain of \ccdan{} was 0.18--0.23 and remained 0.25--0.30 after calibration, so the MSE advantage there (Table~\ref{tab:main}) mainly reflects low-amplitude predictions.
(3) vm1 with the gain table (vm1+cal) is a compromise with gain $\approx1$ at an MSE cost of $-2$ to $+29\%$ for $k\ge5$ ($+5$ to $10\%$ on BCI IV-1 and IV-2a).
(4) The RMS error of the $\alpha$ topography fell from 0.29--0.67 to 0.23--0.40 with post-hoc calibration and to 0.25--0.48 with vm1 (vm1+cal: 0.23--0.39), but none reached NN (0.19--0.35), and the correlation barely changed (Table~S6): calibration corrects the bias but not the error in relative power across electrodes.

In summary, (i) MSE gains barely carry over to classification accuracy (\ccdan{} $\approx$ SSI), (ii) amplitude shrinkage penalizes power-based measures, and although the variance-matching loss and post-hoc calibration remove the bias, shape errors remain and NN/SSI stay ahead, and (iii) when a downstream model can be retrained on the observed electrodes, completion is unnecessary.
The value of completion is confined to using a fixed-montage pipeline without retraining; we recommend the variance-matching loss as the default in pretraining and the gain table for power analysis only.

\subsection{Interpretability}\label{sec:interp}
\begin{figure}[tbp]
\centering
\subfloat[atom \#17 (22\% of activation energy; $\bigstar$ T7, 10 Hz)]{\begin{minipage}[c]{0.49\textwidth}\centering
\begin{minipage}[c]{0.45\linewidth}\centering\includegraphics[width=\linewidth]{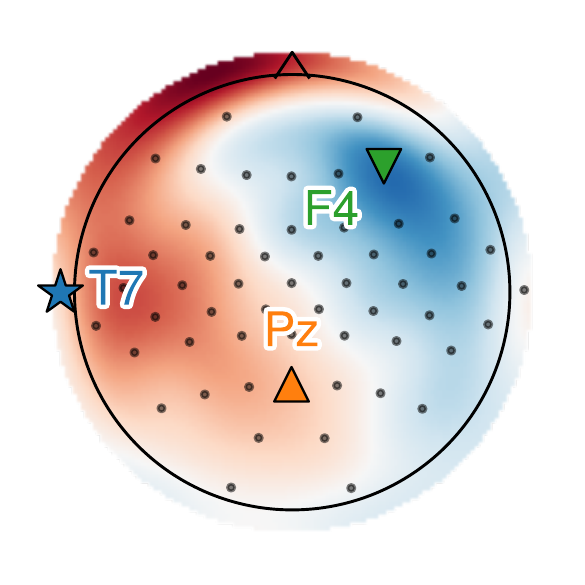}\end{minipage}\hfill
\begin{minipage}[c]{0.53\linewidth}\centering\includegraphics[width=\linewidth]{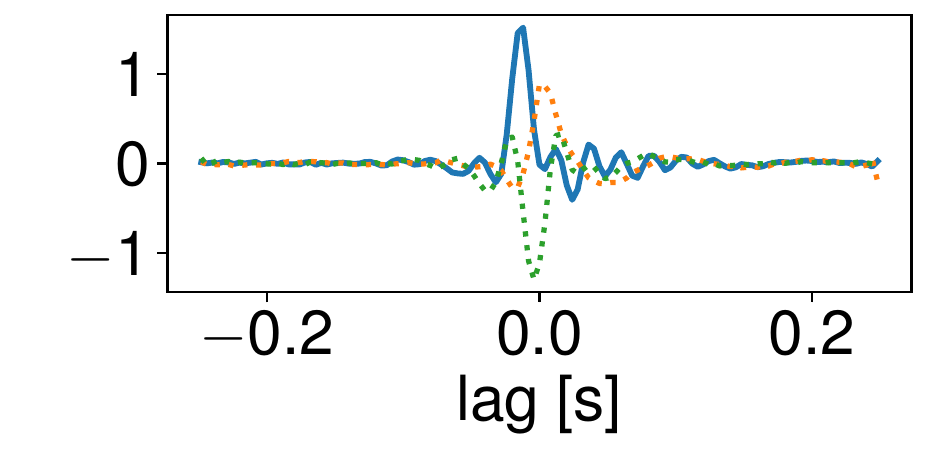}\\[0.5mm]\includegraphics[width=\linewidth]{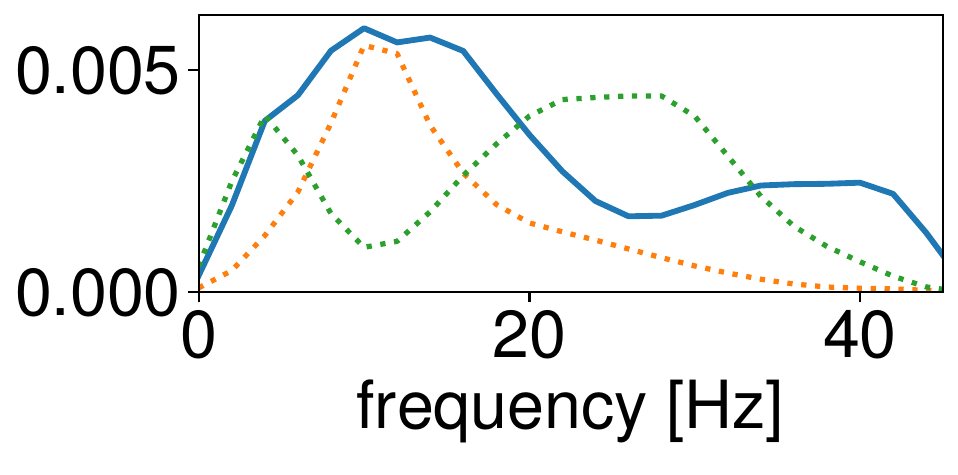}\end{minipage}
\end{minipage}}\hfill
\subfloat[atom \#1 (17\%; $\bigstar$ PO2, 10 Hz)]{\begin{minipage}[c]{0.49\textwidth}\centering
\begin{minipage}[c]{0.45\linewidth}\centering\includegraphics[width=\linewidth]{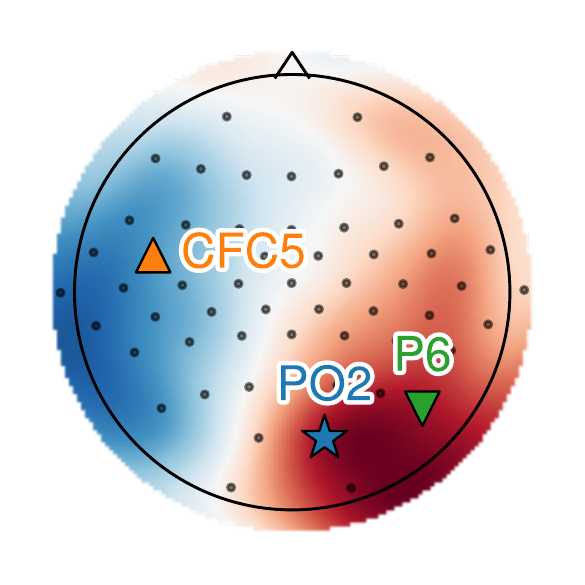}\end{minipage}\hfill
\begin{minipage}[c]{0.53\linewidth}\centering\includegraphics[width=\linewidth]{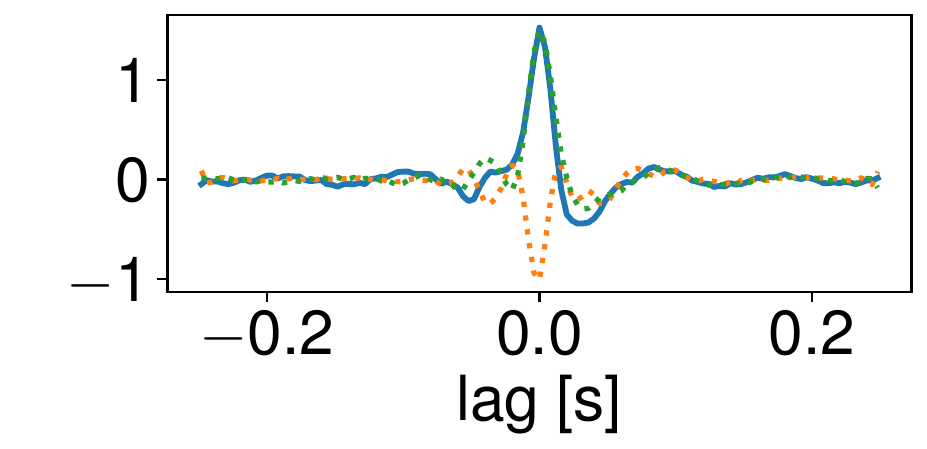}\\[0.5mm]\includegraphics[width=\linewidth]{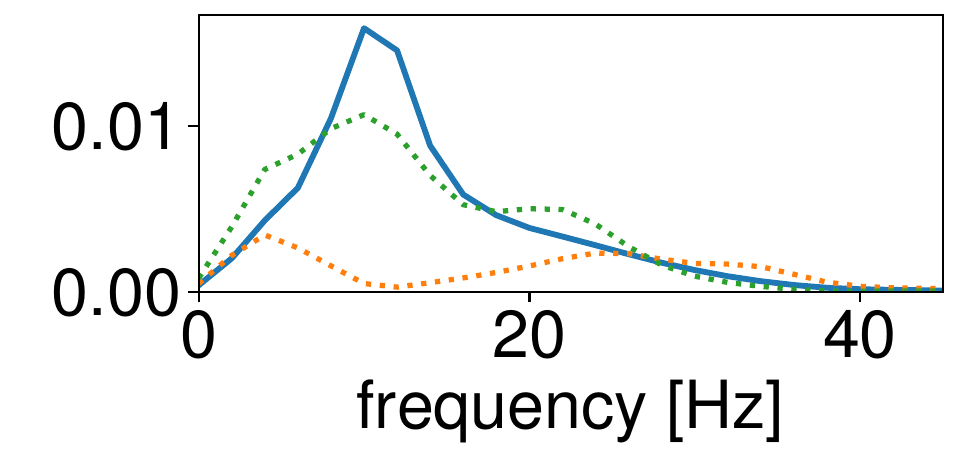}\end{minipage}
\end{minipage}}
\caption{The two most-used atoms of \ccdan{} (Tier-1) on 400 trials of BCI IV-1.
Left: scalp field at the peak lag (from the spherical-harmonic coefficients; RdBu, blue negative, red positive, symmetric scale per atom; dots: electrodes of BCI IV-1; star: electrode of maximum atom energy; triangles: two random electrodes; electrode names in the marker color).
Top right: waveform at these three electrodes (solid: star; dotted: triangles; colors match the markers).
Bottom right: power spectra of the same waveforms (Welch, linear scale); amplitude and power are in arbitrary units, and the frequency in each subcaption is the spectral peak at the starred electrode.}
\label{fig:atoms}
\end{figure}

\begin{figure}[tbp]
\centering
\begin{minipage}[c]{0.55\textwidth}\centering
\subfloat[NN]{\includegraphics[width=0.28\linewidth]{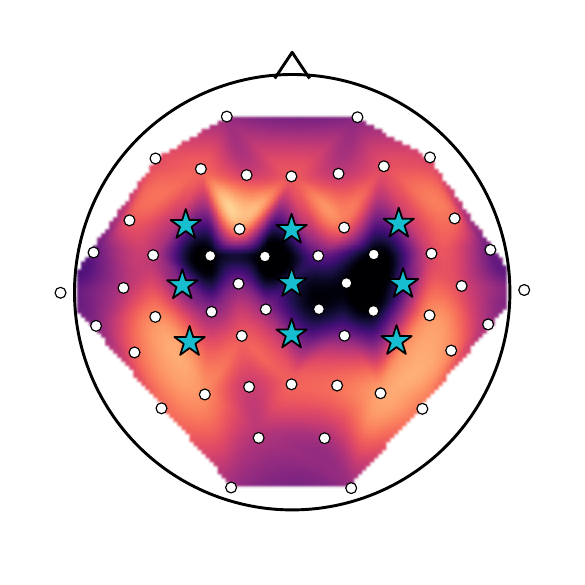}}\hfill
\subfloat[SSI]{\includegraphics[width=0.28\linewidth]{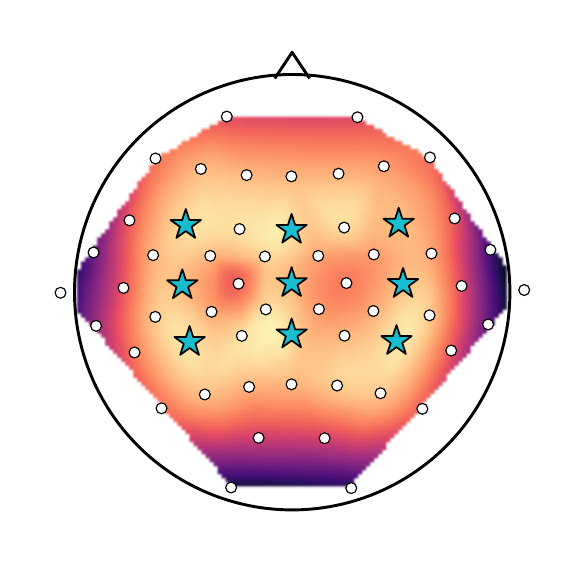}}\hfill
\subfloat[\ccdan{}]{\includegraphics[width=0.28\linewidth]{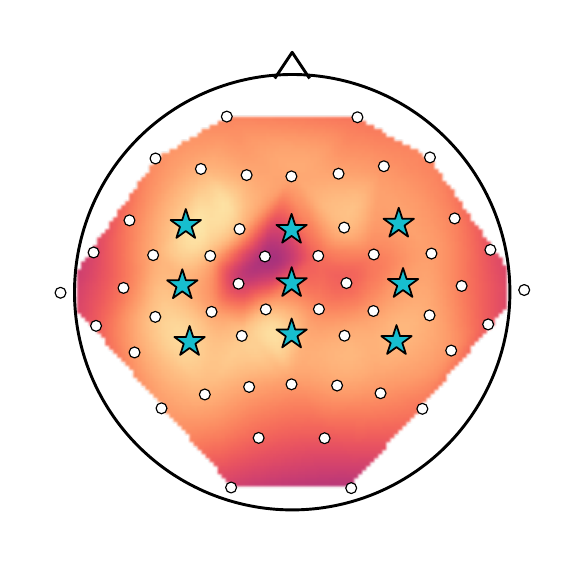}}\hfill
\begin{minipage}[c]{0.12\linewidth}\centering\includegraphics[width=\linewidth,trim=183 12 6 5,clip]{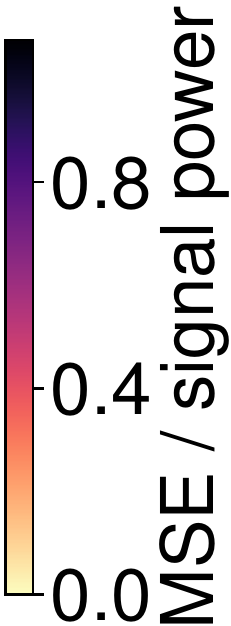}\end{minipage}
\end{minipage}\hfill
\begin{minipage}[c]{0.42\textwidth}\centering
\subfloat[Waveforms at CCP1 (top) and O2 (bottom)]{\begin{minipage}[c]{\linewidth}\centering\includegraphics[width=\linewidth]{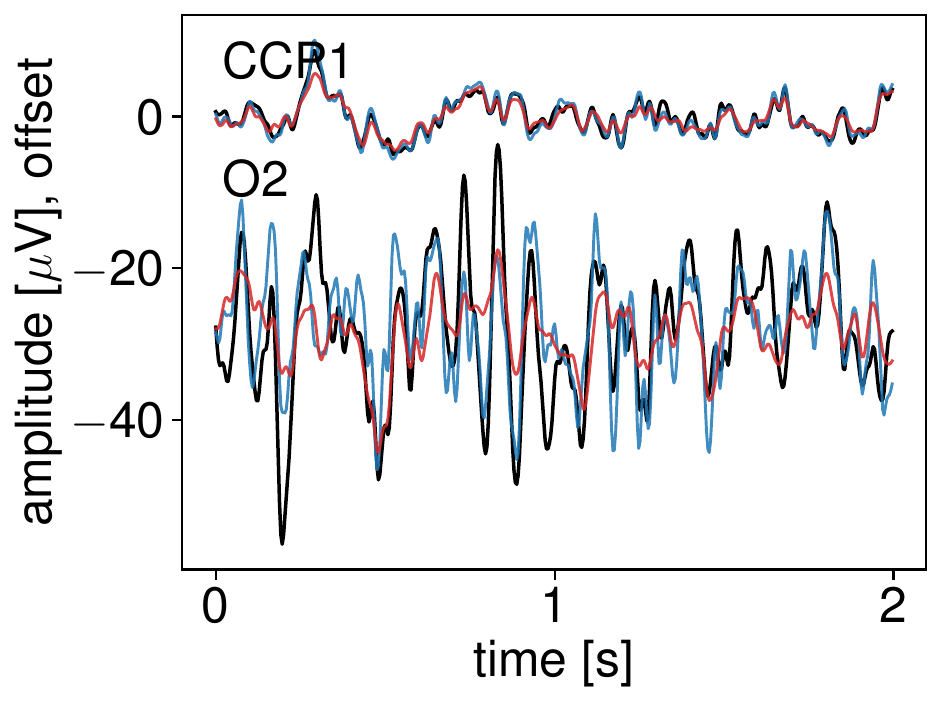}\\[0.5mm]\includegraphics[width=0.85\linewidth,trim=7 121 7 118,clip]{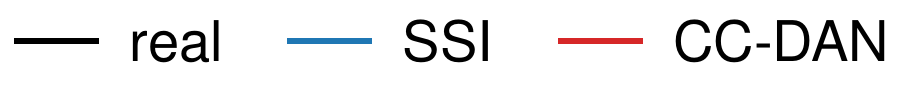}\end{minipage}}
\end{minipage}
\caption{Example on BCI IV-1 (subject f, motor 9 ch; stars: observed electrodes).
(a)--(c) MSE over signal power at each missing electrode (mean over missing electrodes: NN 0.55, SSI 0.33, \ccdan{} 0.28).
(d) Truth (black), SSI, and \ccdan{} at two missing electrodes of one trial (top: CCP1, where the SSI error is small; bottom: O2, where it is large).}
\label{fig:example}
\end{figure}

Fig.~\ref{fig:atoms} shows the two most-used atoms of the Tier-1 model (all eight in Fig.~S3).
Atoms have smooth, dipole-like continuous fields, and the waveforms share their temporal structure across electrodes with only amplitude and polarity changing, as implied by \eqref{eq:model}.
At the electrode of maximum energy, the two most-used atoms (temporal T7, parieto-occipital PO2) peak in $\alpha$ (10 Hz), two atoms in low $\beta$ (14 and 16 Hz), and the remaining four in slow waves (4 Hz), and the peak of one atom differs across electrodes (e.g., atom \#17 peaks at 10 Hz at T7 whereas low frequencies dominate at Pz).
On 400 trials of BCI IV-1, 74.5\% of the activations were zero and the eight most-used atoms carried 90\% of the activation energy.
The coordinate-dependent pooling weights of the detector formed fields selective to frontal, posterior, and temporal regions (Fig.~S4).
Fig.~\ref{fig:example} shows one subject with the motor 9-ch layout (an Emotiv 14-ch example is in Fig.~S5): NN leaves large errors at the electrodes farthest from an observed one, SSI extrapolates poorly toward the posterior edge, and \ccdan{} spreads a smaller error evenly (relative MSE 0.55, 0.33, and 0.28).
At the posterior electrode O2, SSI deviates strongly from the truth whereas \ccdan{} follows the waveform with the amplitude shrinkage discussed in Section~\ref{sec:downstream}.

\subsection{Issues in Prior Evaluation Protocols}\label{sec:protocol}
Subjects c, d, and e of BCI Competition IV-1 are artificially generated \cite{tangermann2012}; their channel variance (3176 \muVsq{}) and SSI error (51--489 \muVsq{}) are orders of magnitude larger than those of real subjects (2--6 \muVsq{}), and the tables of K M \emph{et al.} \cite{km26}, which average over seven subjects, are dominated by them.
Our SSI is 1/3 to 1/10 of the values reported there (118--131 vs 305--1156 \muVsq{} over all seven subjects; our GCN reimplementation, 303--348, agrees with their 260--431), so the SSI baseline was most likely misconfigured.
The reported 10-point gain in classification accuracy from combining real and completed data was not reproduced (EEGNet, LOSO over seven subjects: real 53.9\%, GCN combined 54.1\%, DAN combined 56.1\%; differences within the between-subject standard deviation), the task is described as left versus right hand although subjects a and f performed left hand versus foot, and five-fold cross-validation was used there whereas we use between-subject LOSO.
Finally, using the observed-set average as the target reference makes the evaluation reference-invariant and leakage-free and gave consistently better transfer than a CAR target (Section~\ref{sec:ablation}).

\section{Discussion}
\subsection{Summary of Findings}
Replacing electrode-name correspondence by coordinate conditioning did not cost accuracy (Section~\ref{sec:regression}).
A single model pretrained on five datasets and 132 subjects transferred without calibration to nine held-out evaluation sets; with five or fewer observed electrodes, sparse real layouts, and hemispheric dropout it beat $\minb$ in nearly every subject and approached the within-subject oracle (up to $k$=15 on the EGI net and 8- to 16-channel devices), whereas SSI was superior with 15 or more electrodes on 10-05 montages (Sections~\ref{sec:zeroshot}--\ref{sec:regime}).
\ccdan{} was the most robust to electrode-position error and outperformed in-dataset SRGDiff and zero-shot LUNA (Sections~\ref{sec:ablation} and \ref{sec:comparison}), but MSE gains barely propagated to downstream accuracy, and amplitude shrinkage degraded power-based measures (Section~\ref{sec:downstream}).

\subsection{Where to Use Learned Completion}
The gain of learned methods concentrates on sparse observations ($\le 5$ electrodes on 10-05 montages, $\le 15$ on the EGI net and low-channel devices), hemispheric dropout, and high-frequency bands; on dense 10-05 montages SSI is best.
At low density \ccdan{} is near the ceiling given by the within-subject oracle, whereas at high density a large margin remains between the ceiling and SSI.

\subsection{What Makes Plug-and-Play Work}
Transfer was enabled by defining the reference (observed-set average) and scale from the target's observed electrodes only, by a pretraining corpus harmonized in reference, scale, and dead channels, and by a structured mask distribution.
In preliminary experiments the reference convention was the largest domain difference between datasets, and transfer broke down before it was harmonized.

\subsection{MSE Versus Downstream Performance}
Amplitude shrinkage is a general property of MSE-optimal estimators, not a defect specific to learned methods (Section~\ref{sec:downstream}); it is strongest at the low densities where completion is most useful, so calibration is mandatory whenever completed signals feed power analysis or topography.
The variance-matching loss reduces the amplitude bias at nearly unchanged MSE on 10-05 systems, and the post-hoc gain table almost removes it but worsens MSE by up to 99\% at $k$=3; both reduced the $\alpha$-topography error (post-hoc 20--40\%, vm1 15--39\%) without reaching NN, so MSE and amplitude fidelity are hard to reconcile in a single point estimate.
Where a downstream model can be retrained on the observed electrodes, completion is unnecessary; its value lies in connecting an unseen device to a fixed-montage pipeline.
A natural target is EEG-based brain-age or pathology models trained on high-density clinical montages, but we did not test this, and because MSE gains did not carry over to classification here, such use requires task-level validation.
Finally, artificial subjects and weak SSI baselines inflate the apparent gain of learned methods (Section~\ref{sec:protocol}), which should be evaluated against $\minb$ with per-subject statistics.

\subsection{Limitations}
(1) MSE superiority does not translate directly into downstream accuracy, and full calibration to unit gain worsens MSE substantially at low density.
(2) Lee2019 cross-paradigm transfer shares subjects and device; true zero-shot is limited to Nakanishi2015 and BI2013a (33 subjects).
(3) Digitized coordinates have not been tested.
(4) The pretraining corpus has 132 subjects; extending it to the full HBN and to older adults is future work.
(5) The SRGDiff comparison is limited to BCI IV-1 with an inactive diffusion branch, so its potential may be underestimated, and downstream evaluations used a single seed without statistical tests.
(6) The detector does not model data-dependent interactions between electrodes, such as local interpolation from neighboring observed electrodes; adding such a path is future work.

\section{Conclusion}
With the coordinate-conditioned self-supervised model \ccdan{} and a plug-and-play protocol, a single pretrained model completes EEG at arbitrary positions from any observed electrode set without calibration and transfers zero-shot to unseen datasets, nets, and paradigms.
Nine evaluation sets quantified where learning beats classical methods and where SSI is superior, and showed robustness to coordinate error and an advantage over fixed-layout super-resolution and foundation-model reconstruction.
MSE gains, however, do not translate directly into downstream tasks, and amplitude shrinkage penalizes power-based measures; amplitude calibration is the key open issue.
Code and pretrained weights: \url{https://github.com/hgshrs/ccdan-eeg-completion}.

\section*{Acknowledgments}
The author thanks the maintainers of MOABB and MNE-Python and the providers of the public datasets used in this study.

\section*{Funding}
This work was supported in part by the Japan Society for the Promotion of Science (JSPS) KAKENHI under Grants 22H05163 and 24K15047, and in part by the Japan Science and Technology Agency (JST) Advanced International Collaborative Research Program (AdCORP) under Grant JPMJKB2307.

\section*{Data availability}
All datasets analyzed in this study are publicly available from the sources listed in Table~\ref{tab:datasets}.
Code, pretrained weights, and the scripts that reproduce the figures and tables are available at \url{https://github.com/hgshrs/ccdan-eeg-completion}.

\section*{Supplementary material}
The supplementary material, appended after the references, contains details of the corpus and coordinates, the layout templates and the gain table, cross-paradigm transfer within Lee2019, all eight atoms and the pooling fields of the detector, an Emotiv 14 ch example, and the downstream-task and amplitude-calibration tables.

\section*{Ethical statement}
This study analyzed publicly available, de-identified datasets only; no new human-subject experiments were performed, and each dataset was acquired under the ethics approval of the originating institution as described in the cited publications.

\appendix
\section{Notes on running the comparison methods}\label{app:comparison}
\subsection{SRGDiff}
The public code (\url{https://github.com/DhrLhj/ICLR2026SRGDiff}) has hard-coded paths, missing imports, and missing configuration files; about 40 lines in six files were patched to run the five-stage pipeline (VAE $\to$ latent encoder $\to$ latent DDIM $\to$ decoder fine-tuning $\to$ decoding).
The DDIM is conditioned on z-scored latents whereas the encoder is trained on raw latents, so the diffusion samples were unusable (300--900 \muVsq{}) and the final gated fusion reduced to the output of the convolutional encoder.
The VAE was trained for 1000 epochs instead of the configured 20000.
The reconstruction error of the VAE with all electrodes as input (5.9--7.2 \muVsq{}) shows that the bottleneck is the low-to-high-resolution latent mapping, not the autoencoder.

\subsection{TGSD}
The public code lacks the \texttt{model.mamba} module, has several tensor-shape inconsistencies, and is hard-wired to the 62-channel SEED montage.
A reimplementation from the paper would be required, and the computational cost of the selective scan is impractical, so the method was excluded from the comparison.

\subsection{REVE and LUNA}
The released REVE weights (69.2 M parameters) contain only the encoder, not the MAE decoder, and are gated on Hugging Face.
LUNA (base 6.7 M / large 40.7 M) includes a reconstruction decoder and can structurally be queried at the positional embedding of an unobserved channel, but a sinusoidal sweep showed that it reconstructs only frequencies below 4 Hz (below 15 Hz for the large model), so on 0.5--40 Hz EEG it was equivalent to predicting zero.

\bibliographystyle{iopart-num}
\bibliography{refs}

\clearpage
\setcounter{section}{0}\setcounter{figure}{0}\setcounter{table}{0}\setcounter{equation}{0}
\renewcommand{\thesection}{S\arabic{section}}
\renewcommand{\thesubsection}{S\arabic{section}.\arabic{subsection}}
\renewcommand{\thefigure}{S\arabic{figure}}
\renewcommand{\thetable}{S\arabic{table}}
\renewcommand{\theequation}{S\arabic{equation}}
\begin{center}
{\LARGE\bfseries Supplementary material}\\[6pt]
{\large for ``Coordinate-conditioned detector-atom network for montage-agnostic EEG channel completion: zero-shot transfer and a regime map''}
\end{center}
\suppressfloats[t]

This supplementary material contains details of the corpus and coordinates (Section~\ref{app:corpus}) and supplementary figures and tables (Section~\ref{app:supp}).
Section, figure, and table numbers without the prefix S (e.g., ``Section~2.3'', ``Fig.~3'', ``Table~4'') refer to the main text.
Abbreviations and condition names are listed in Section~\ref{app:abbr}.

\section{Corpus and Coordinate Details}\label{app:corpus}
\subsection{Dataset-Specific Processing}
\begin{figure}[tbp]
\centering
\subfloat[Schirrmeister2017 (127 of 128 ch, pretraining)]{\includegraphics[width=0.19\textwidth]{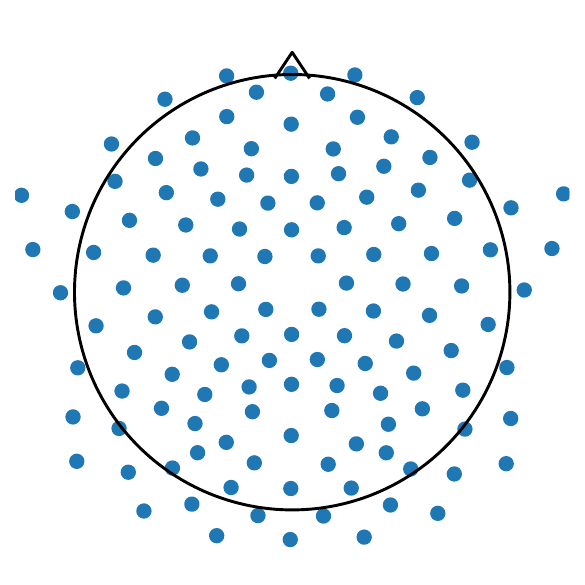}}\hfill
\subfloat[GrosseWentrup2009 (128 ch, pretraining)]{\includegraphics[width=0.19\textwidth]{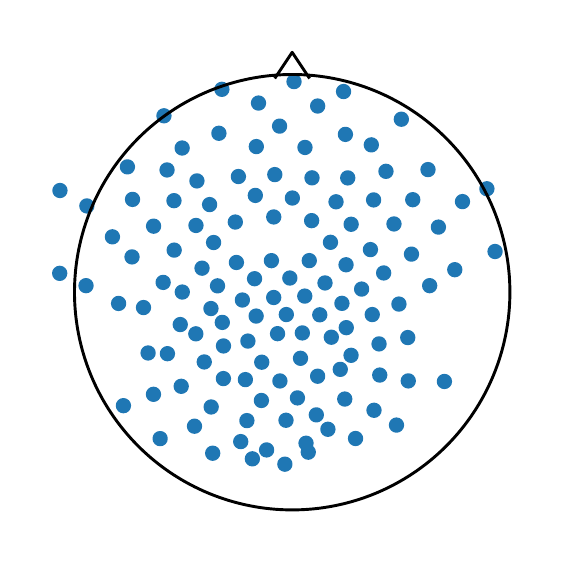}}\hfill
\subfloat[Lee2019 (62 ch, pretraining)]{\includegraphics[width=0.19\textwidth]{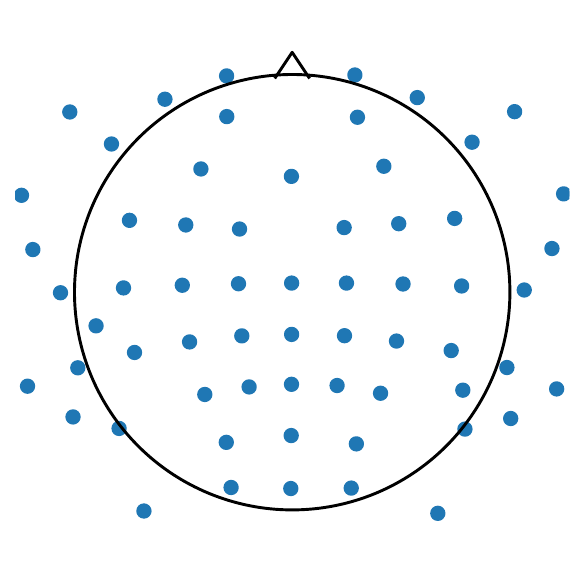}}\hfill
\subfloat[BCI IV-1 (59 ch, evaluation)]{\includegraphics[width=0.19\textwidth]{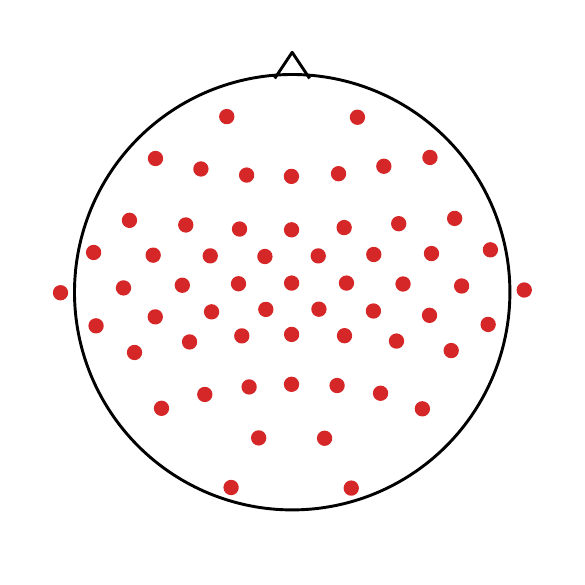}}\hfill
\subfloat[BCI III-IIIa grid (60 ch, evaluation)]{\includegraphics[width=0.19\textwidth]{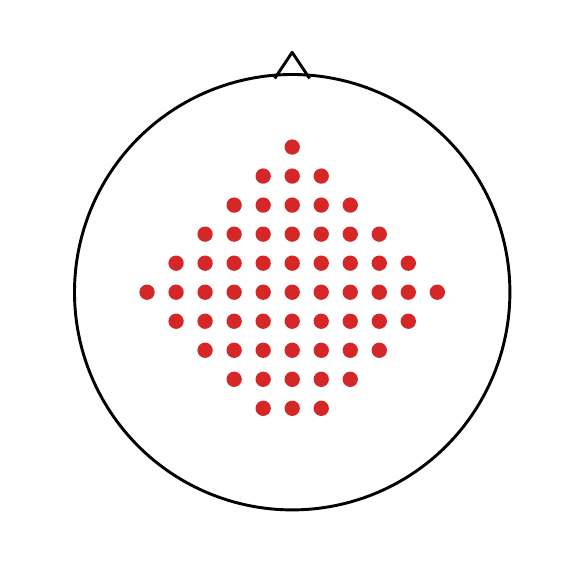}}\\[1mm]
\subfloat[BCI IV-2a (22 ch, evaluation)]{\includegraphics[width=0.19\textwidth]{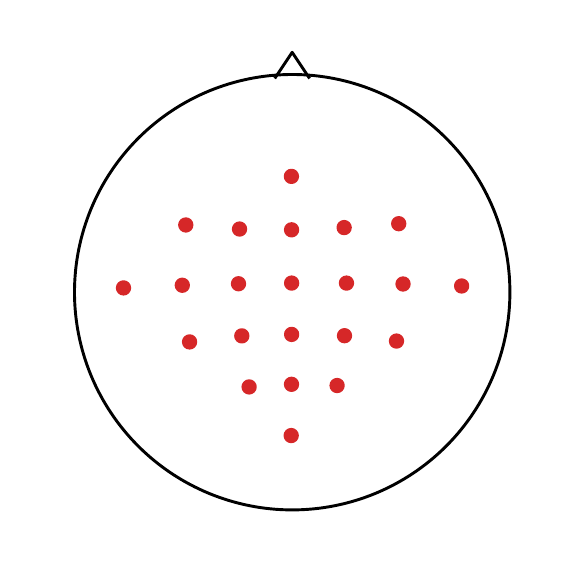}}\hfill
\subfloat[HBN: EGI GSN-128 (evaluation)]{\includegraphics[width=0.19\textwidth]{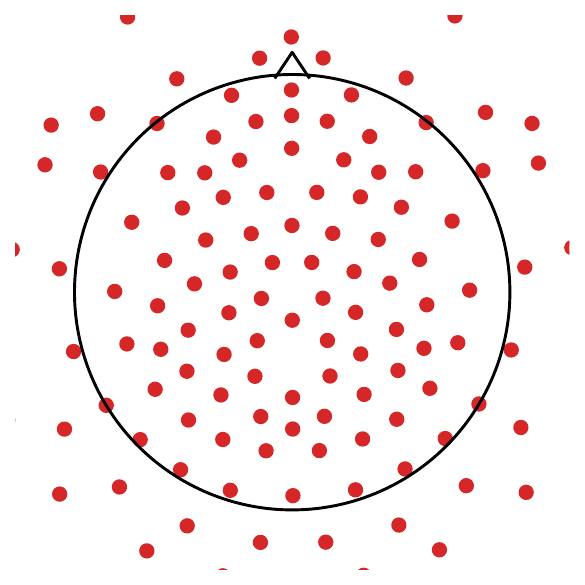}}\hfill
\subfloat[Nakanishi2015 (8 ch, evaluation)]{\includegraphics[width=0.19\textwidth]{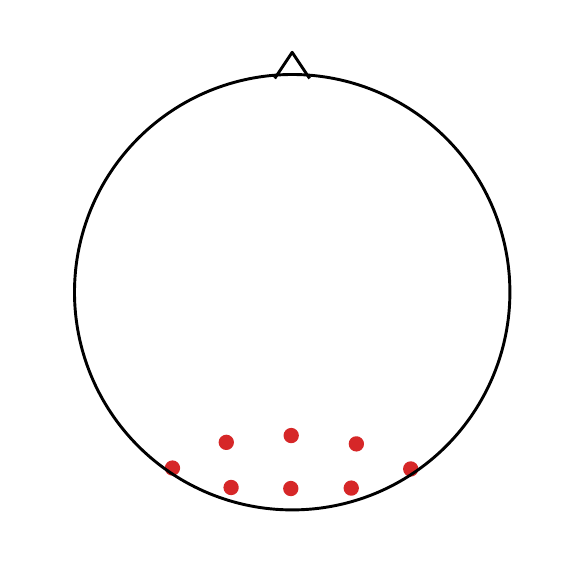}}\hfill
\subfloat[BrainInvaders 2013a (16 ch, evaluation)]{\includegraphics[width=0.19\textwidth]{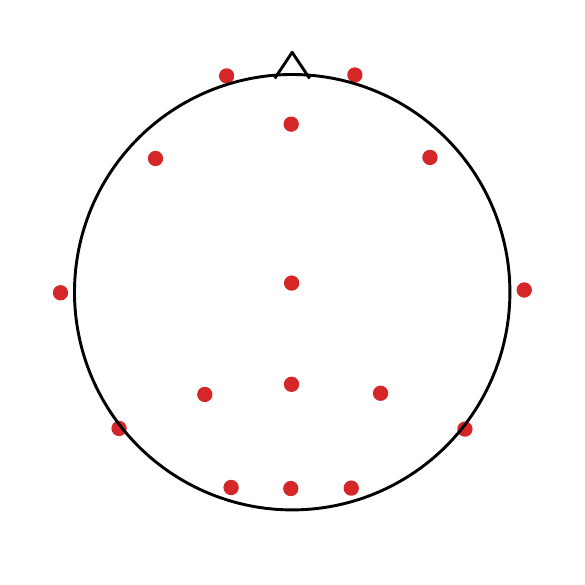}}\hfill
\subfloat[Low-density layout templates]{\includegraphics[width=0.19\textwidth]{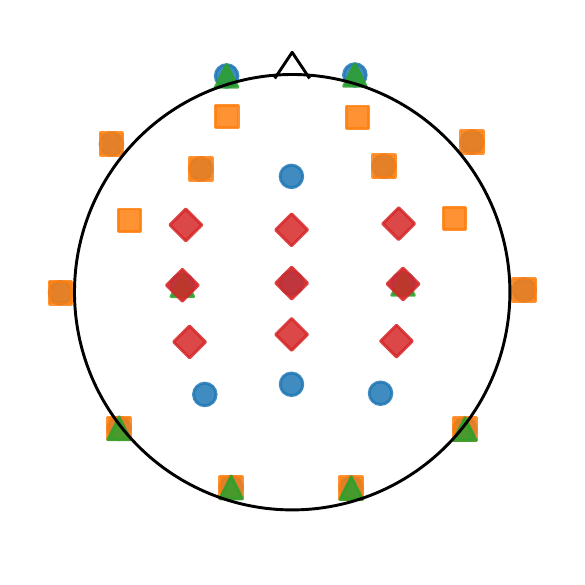}}
\caption{Electrode layouts in the canonical frame (top view, azimuthal equidistant projection; front is up; the circle is the head radius at $90^\circ$ from the vertex).
Blue: pretraining sets; red: evaluation sets.
The peripheral electrodes of HBN (EGI) lie outside the head circle, where spherical-spline extrapolation breaks down.
(j) Low-density templates used for coordinate matching (dots: 10-20 with 19 ch; squares: Emotiv 14 ch; triangles: OpenBCI 8 ch; diamonds: motor 9 ch).}
\label{fig:montages}
\end{figure}

The continuous signals of BCI Competition IV-1 are int16 in units of 0.1 \muV{} and were multiplied by 0.1 to obtain \muV{}.
Its electrode names were mapped to the 10-05 system (CFC1--6 $\to$ FCC1h--6h, CFC7/8 $\to$ FTT7h/8h, CCP1--6 $\to$ CCP1h--6h, CCP7/8 $\to$ TTP7h/8h).
BCI Competition III-IIIa provides no electrode names, only a grid diagram; because two electrodes lie between C3 and Cz, the grid was converted to 3-D coordinates as an azimuthal equidistant grid with $12^\circ$ spacing.
The HBN coordinates in the BIDS electrodes.tsv were the same EGI template for all subjects, so the MNE GSN-HydroCel-129 template was aligned to the canonical frame via the fiducials, and the 128 channels excluding Cz (the recording reference) were used.
Table~1 of the main text lists all datasets, and Fig.~\ref{fig:montages} shows the electrode layouts.

\subsection{Coordinate Error of the GSN-128 to 10-20 Correspondence Table}
Table~\ref{tab:gsn} lists the angular error, after fiducial alignment to the canonical frame, of the correspondence between GSN-HydroCel-128 electrodes and 10-20 electrodes (19 pairs) given in EGI's technical documentation.
Measured from the coordinate origin, the mean angle to the tabulated 10-20 electrode is $9.8^\circ$; measured from the sphere center of each montage it is $7.2^\circ$ (main text, Section~2.2).
Many GSN electrodes are closer to another 10-05 electrode than to the tabulated 10-20 electrode (e.g., E11 to AFFz rather than Fz, E45 to TP7 rather than T7).

\begin{table}[tbp]
\centering\footnotesize\setlength{\tabcolsep}{3pt}
\caption{Angular Error (Degrees, From the Coordinate Origin) of the GSN-128 to 10-20 Correspondence Table in EGI's Technical Documentation.
``Nearest'' Is the Closest 10-05 Electrode in the Canonical Frame and Its Angular Distance.}
\label{tab:gsn}
\begin{tabular}{llrlr}
\toprule
GSN & 10-20 (table) & Angle & Nearest 10-05 & Angle\\
\midrule
E22 & Fp1 & 6.3 & AFp5 & 2.1\\
E9 & Fp2 & 7.4 & AFp6 & 2.4\\
E33 & F7 & 17.3 & FT7 & 4.2\\
E24 & F3 & 6.1 & FFC3 & 5.9\\
E11 & Fz & 12.2 & AFFz & 1.0\\
E124 & F4 & 7.4 & FFC4 & 4.9\\
E122 & F8 & 17.7 & FT8 & 3.1\\
E45 & T7 & 21.3 & TP7 & 5.0\\
E36 & C3 & 7.8 & CCP3 & 3.9\\
E104 & C4 & 7.5 & CCP4 & 3.3\\
E108 & T8 & 20.5 & TP8 & 4.0\\
E58 & P7 & 9.6 & PPO7 & 2.4\\
E52 & P3 & 5.8 & P5h & 3.9\\
E62 & Pz & 11.0 & PPOz & 1.7\\
E92 & P4 & 6.5 & P6h & 3.8\\
E96 & P8 & 8.1 & PPO8 & 1.5\\
E70 & O1 & 4.6 & O1 & 4.6\\
E75 & Oz & 5.9 & OIz & 3.5\\
E83 & O2 & 4.2 & O2 & 4.2\\
\bottomrule
\end{tabular}
\end{table}

\subsection{Layout Templates}
Table~\ref{tab:layouts} lists the templates of real layouts (\texttt{standard\_1005} names) used in the training mask distribution (main text, Section~2.5) and in the evaluation (Section~3.4).
Templates are matched to each dataset by coordinates, not by names; a template electrode is observed only if the dataset has an electrode within $8^\circ$ of it (the number of matched electrodes is given in parentheses in the main text).

\begin{table}[tbp]
\centering\footnotesize
\caption{Layout Templates Used for Coordinate Matching (11 Templates).}
\label{tab:layouts}
\begin{tabular}{lp{5.6cm}}
\toprule
Template & Electrodes (\texttt{standard\_1005} names)\\
\midrule
1020\_19 & Fp1, Fp2, F7, F3, Fz, F4, F8, T7, C3, Cz, C4, T8, P7, P3, Pz, P4, P8, O1, O2\\
emotiv\_14 & AF3, F7, F3, FC5, T7, P7, O1, O2, P8, T8, FC6, F4, F8, AF4\\
openbci\_8 & Fp1, Fp2, C3, C4, P7, P8, O1, O2\\
openbci\_16 & Fp1, Fp2, C3, C4, P7, P8, O1, O2, F7, F8, F3, F4, T7, T8, P3, P4\\
muse\_4 & AF7, AF8, TP9, TP10\\
enobio\_8 & Fp1, Fp2, F3, F4, C3, C4, P3, P4\\
gtec\_16\_motor & FC3, FC1, FCz, FC2, FC4, C5, C3, C1, Cz, C2, C4, C6, CP3, CP1, CPz, CP2\\
iv2a\_22 & Fz, FC3, FC1, FCz, FC2, FC4, C5, C3, C1, Cz, C2, C4, C6, CP3, CP1, CPz, CP2, CP4, P1, Pz, P2, POz\\
iv2a\_21 & iv2a\_22 without POz\\
motor\_9 & FC3, FCz, FC4, C3, Cz, C4, CP3, CPz, CP4\\
motor\_3 & C3, Cz, C4\\
\bottomrule
\end{tabular}
\end{table}

\subsection{Gain Table for Amplitude Calibration}
Table~\ref{tab:gaintable} gives the post-hoc gain table of the Tier-1 model.
Predictions were made on the pretraining corpus (five sets, 600 trials each) under the training mask distribution, and the ratio $g=\mathrm{std}(x)/\mathrm{std}(\hat x)$ of each missing electrode was binned by the angular distance $d$ to the nearest observed electrode and by the number of observed electrodes $k$ (median per cell; 181{,}665 samples, overall median 1.28).
Cells with fewer than 50 samples were filled with the median of neighboring cells in the same $d$ bin (this applies to the $d<5^\circ$ row).

\begin{table}[tbp]
\centering\footnotesize\setlength{\tabcolsep}{4pt}
\caption{Post-Hoc Gain Table (Tier-1 \ccdan{}).
Rows: Angular Distance $d$ to the Nearest Observed Electrode (Degrees); Columns: Number of Observed Electrodes $k$.
Each Cell Gives the Median Amplitude Ratio and (in Parentheses) the Number of Samples.}
\label{tab:gaintable}
\resizebox{\textwidth}{!}{%
\begin{tabular}{lrrrrrrrr}
\toprule
$d$ [deg] $\backslash$ $k$ & 3 & 4--5 & 6--8 & 9--13 & 14--20 & 21--31 & 32--63 & $\ge$64\\
\midrule
0--5 & 1.02 (2) & 1.02 (5) & 1.02 (2) & 1.02 (6) & 1.02 (4) & 1.02 (27) & 1.02 (28) & 1.02 (1)\\
5--10 & 1.46 (70) & 1.12 (101) & 1.17 (229) & 1.16 (345) & 1.17 (583) & 1.16 (638) & 1.15 (880) & 1.26 (242)\\
10--15 & 1.39 (587) & 1.20 (1155) & 1.17 (3544) & 1.14 (3500) & 1.14 (5471) & 1.14 (4610) & 1.14 (5997) & 1.20 (1770)\\
15--20 & 1.41 (626) & 1.22 (2101) & 1.16 (4788) & 1.15 (4495) & 1.14 (8117) & 1.15 (4583) & 1.17 (5031) & 1.24 (1730)\\
20--30 & 1.69 (2242) & 1.30 (4586) & 1.23 (9715) & 1.22 (7439) & 1.20 (10792) & 1.19 (4929) & 1.21 (4661) & 1.35 (1813)\\
30--45 & 1.95 (3346) & 1.46 (6365) & 1.36 (10013) & 1.35 (5699) & 1.30 (6292) & 1.29 (2061) & 1.42 (2307) & 1.46 (2351)\\
$>$45 & 2.35 (6067) & 1.92 (9391) & 1.70 (8853) & 1.64 (2740) & 1.46 (3582) & 1.32 (851) & 2.01 (2394) & 1.71 (1908)\\
\bottomrule
\end{tabular}}
\end{table}

\subsection{Verification of the Spherical-Harmonic Implementation}
Our implementation of the real spherical harmonics ($L$=8) agrees with SciPy's \texttt{sph\_harm\_y} to within $3\times10^{-15}$, and the Monte Carlo orthogonality error on uniformly random points on the sphere was 0.006.

\section{Supplementary Figures and Tables}\label{app:supp}
\begin{figure}[tbp]
\centering
\subfloat[Lee2019 MI (62 ch)]{\includegraphics[width=0.32\textwidth]{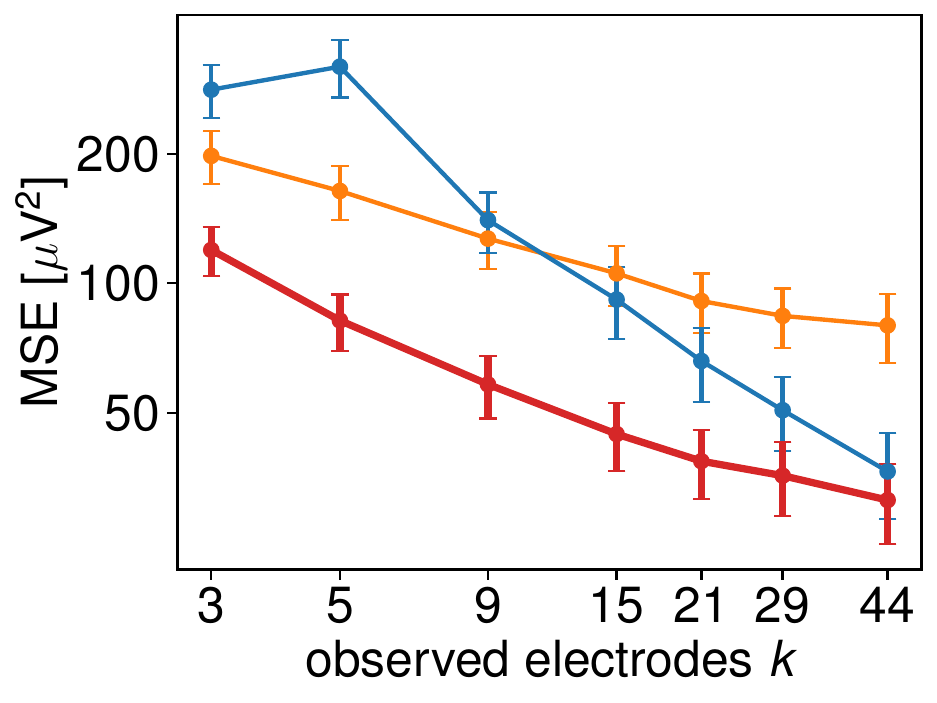}}\hfill
\subfloat[Lee2019 SSVEP (62 ch)]{\includegraphics[width=0.32\textwidth]{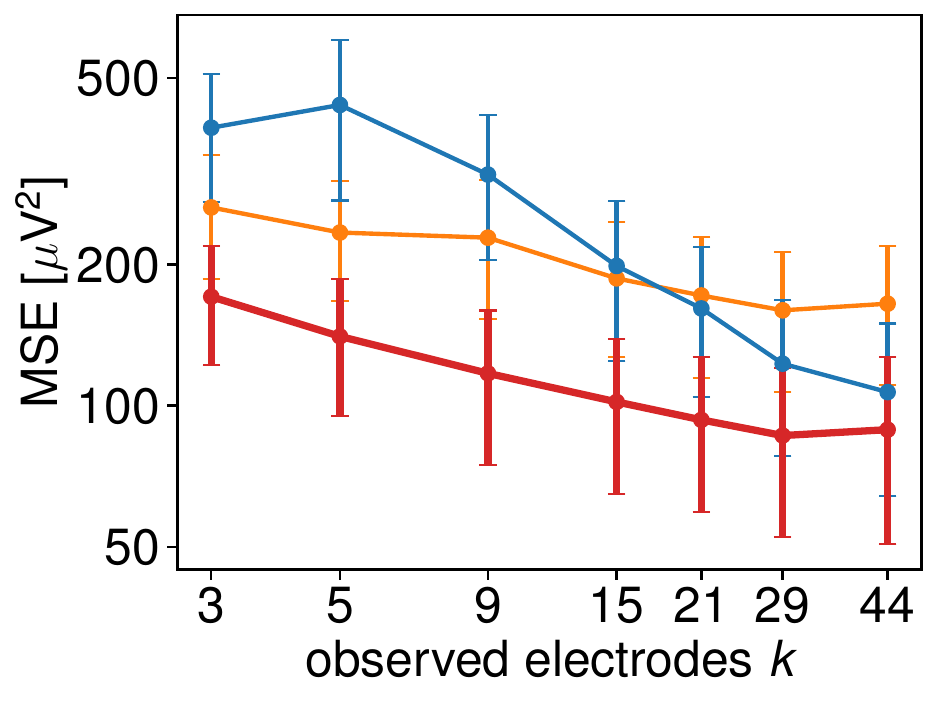}}\hfill
\subfloat[Lee2019 ERP (62 ch)]{\includegraphics[width=0.32\textwidth]{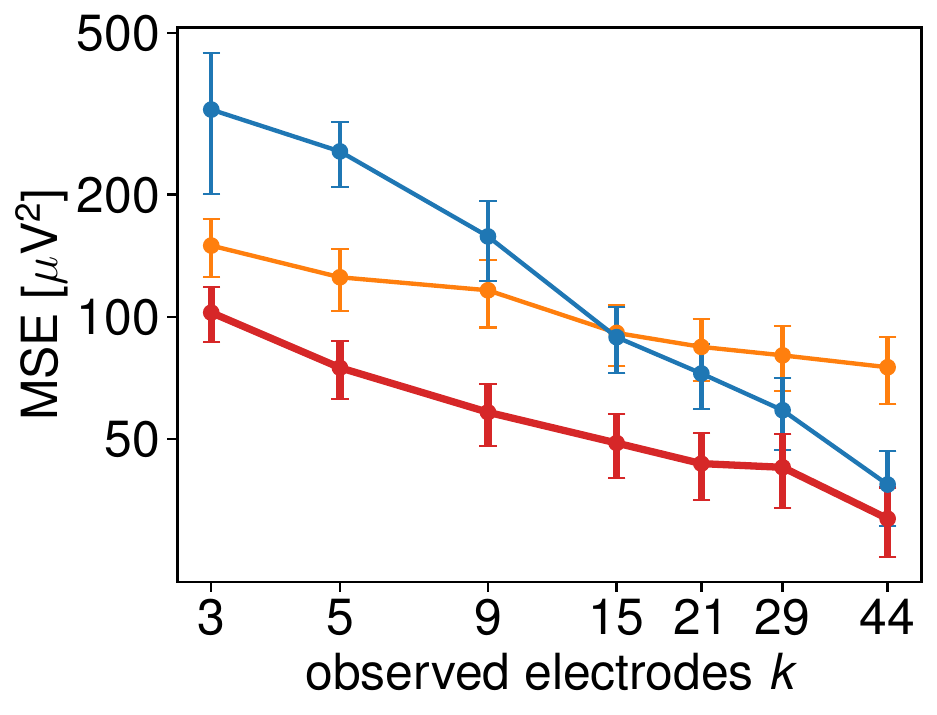}}\\[2mm]
\includegraphics[width=0.65\textwidth,trim=10 129 13 128,clip]{legend_methods_row.pdf}
\caption{Cross-paradigm transfer within Lee2019: MSE on missing electrodes versus the number of observed electrodes $k$ (mean over subjects $\pm$ SEM, log-log) when each paradigm is evaluated with a model trained on Tier-1 without that paradigm.
The three paradigms share the same 54 subjects, device, and montage, so this is within-laboratory, within-subject transfer.
Colors as in Fig.~3 of the main text.}
\label{fig:lee}
\end{figure}

\begin{figure}[p]
\centering
\subfloat[atom \#17 (22\% of activation energy; $\bigstar$ T7, 10 Hz)]{\begin{minipage}[c]{0.49\textwidth}\centering
\begin{minipage}[c]{0.45\linewidth}\centering\includegraphics[width=\linewidth]{atom1_topo_T1}\end{minipage}\hfill
\begin{minipage}[c]{0.53\linewidth}\centering\includegraphics[width=\linewidth]{atom1_wave_T1}\\[0.5mm]\includegraphics[width=\linewidth]{atom1_spec_T1}\end{minipage}
\end{minipage}}\hfill
\subfloat[atom \#1 (17\%; $\bigstar$ PO2, 10 Hz)]{\begin{minipage}[c]{0.49\textwidth}\centering
\begin{minipage}[c]{0.45\linewidth}\centering\includegraphics[width=\linewidth]{atom2_topo_T1}\end{minipage}\hfill
\begin{minipage}[c]{0.53\linewidth}\centering\includegraphics[width=\linewidth]{atom2_wave_T1}\\[0.5mm]\includegraphics[width=\linewidth]{atom2_spec_T1}\end{minipage}
\end{minipage}}\\[2mm]
\subfloat[atom \#2 (11\%; $\bigstar$ PO2, 4 Hz)]{\begin{minipage}[c]{0.49\textwidth}\centering
\begin{minipage}[c]{0.45\linewidth}\centering\includegraphics[width=\linewidth]{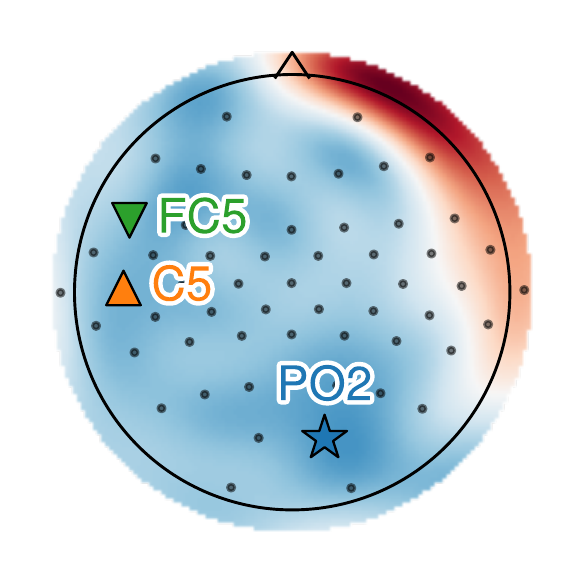}\end{minipage}\hfill
\begin{minipage}[c]{0.53\linewidth}\centering\includegraphics[width=\linewidth]{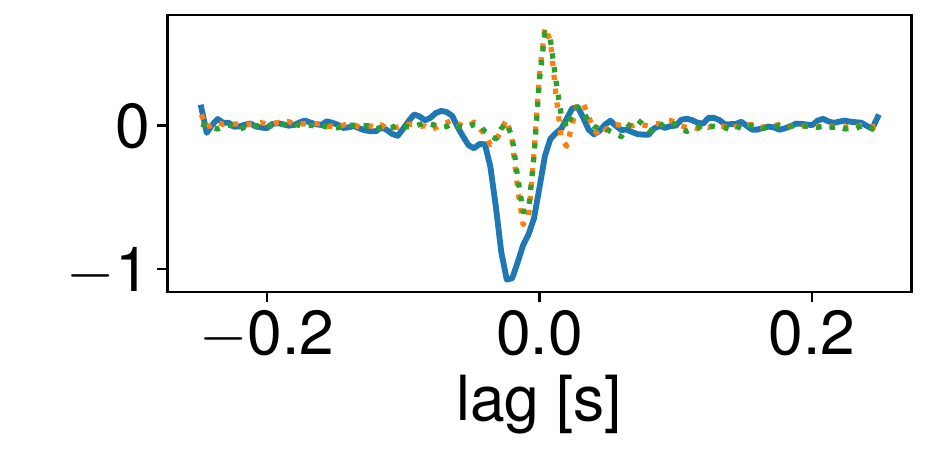}\\[0.5mm]\includegraphics[width=\linewidth]{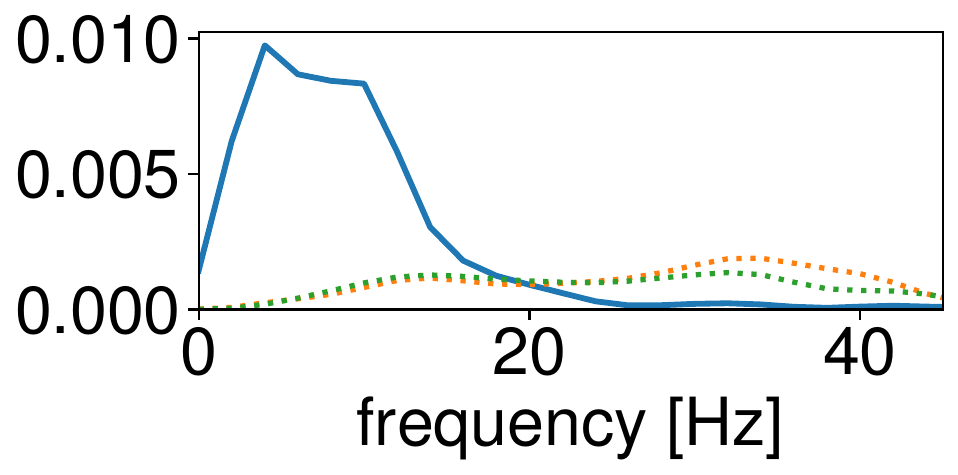}\end{minipage}
\end{minipage}}\hfill
\subfloat[atom \#20 (10\%; $\bigstar$ T8, 14 Hz)]{\begin{minipage}[c]{0.49\textwidth}\centering
\begin{minipage}[c]{0.45\linewidth}\centering\includegraphics[width=\linewidth]{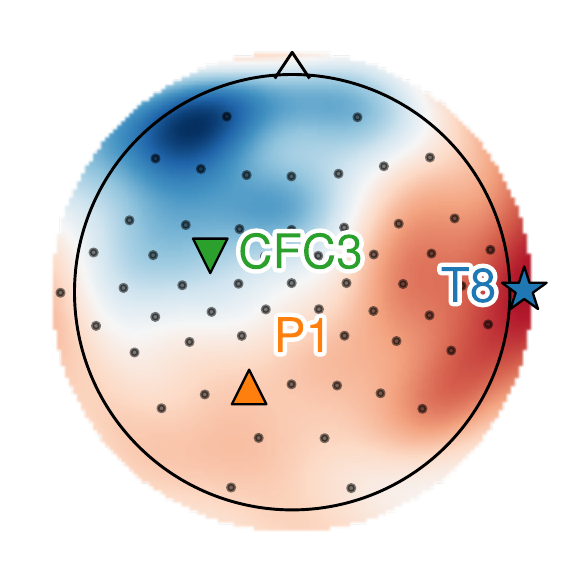}\end{minipage}\hfill
\begin{minipage}[c]{0.53\linewidth}\centering\includegraphics[width=\linewidth]{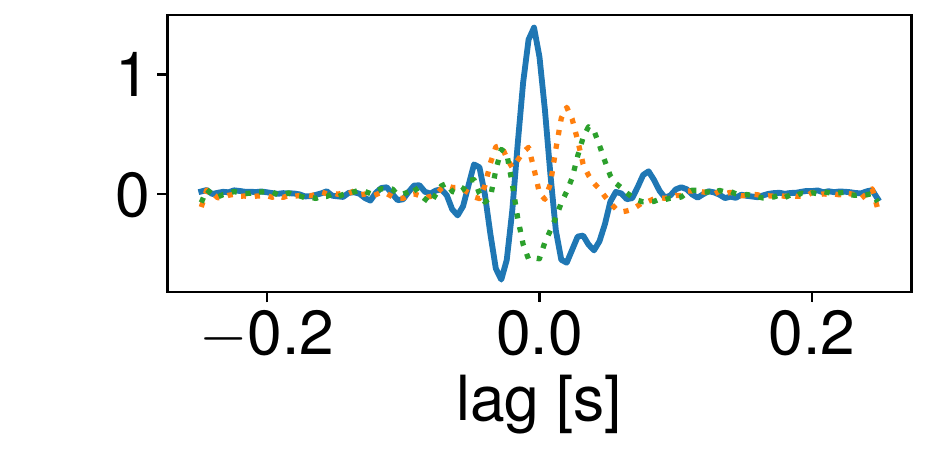}\\[0.5mm]\includegraphics[width=\linewidth]{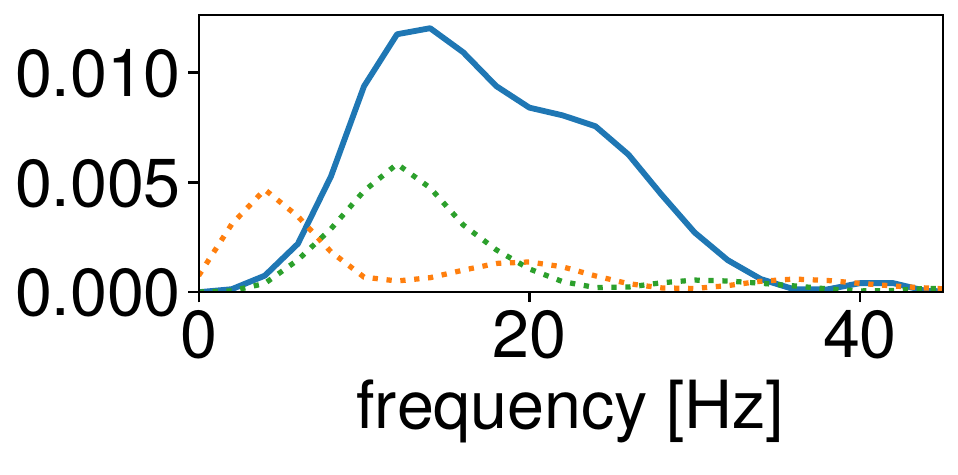}\end{minipage}
\end{minipage}}\\[2mm]
\subfloat[atom \#12 (10\%; $\bigstar$ AF3, 4 Hz)]{\begin{minipage}[c]{0.49\textwidth}\centering
\begin{minipage}[c]{0.45\linewidth}\centering\includegraphics[width=\linewidth]{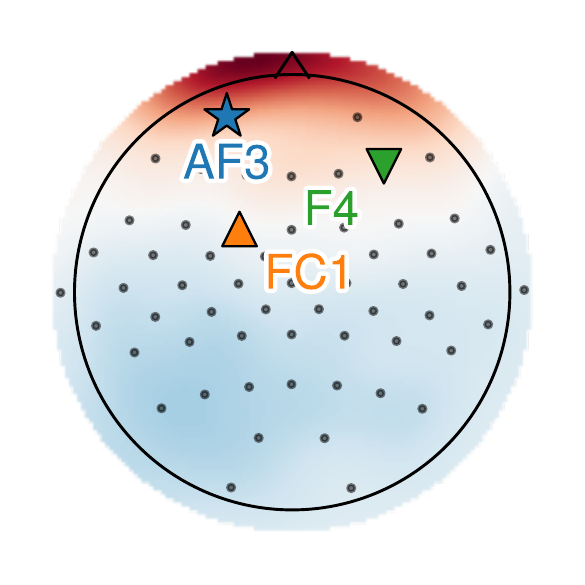}\end{minipage}\hfill
\begin{minipage}[c]{0.53\linewidth}\centering\includegraphics[width=\linewidth]{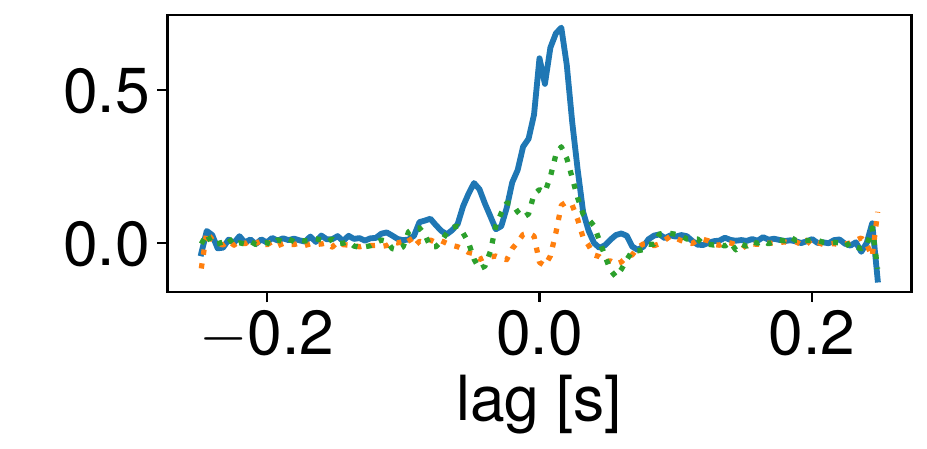}\\[0.5mm]\includegraphics[width=\linewidth]{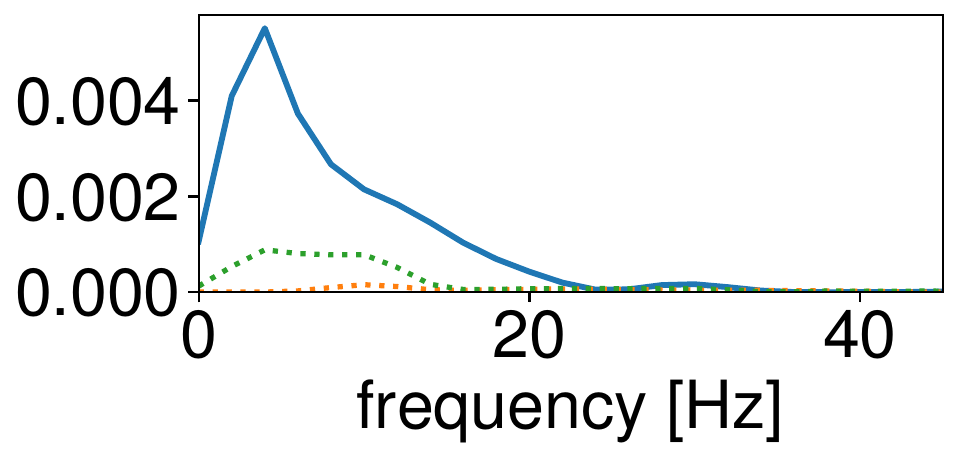}\end{minipage}
\end{minipage}}\hfill
\subfloat[atom \#21 (9\%; $\bigstar$ AF4, 4 Hz)]{\begin{minipage}[c]{0.49\textwidth}\centering
\begin{minipage}[c]{0.45\linewidth}\centering\includegraphics[width=\linewidth]{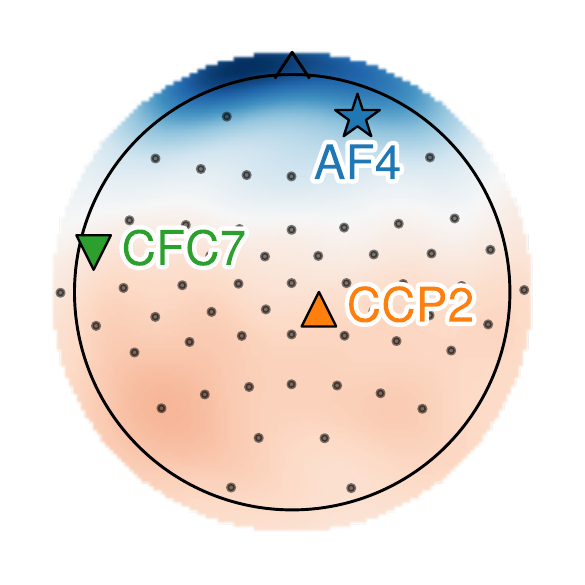}\end{minipage}\hfill
\begin{minipage}[c]{0.53\linewidth}\centering\includegraphics[width=\linewidth]{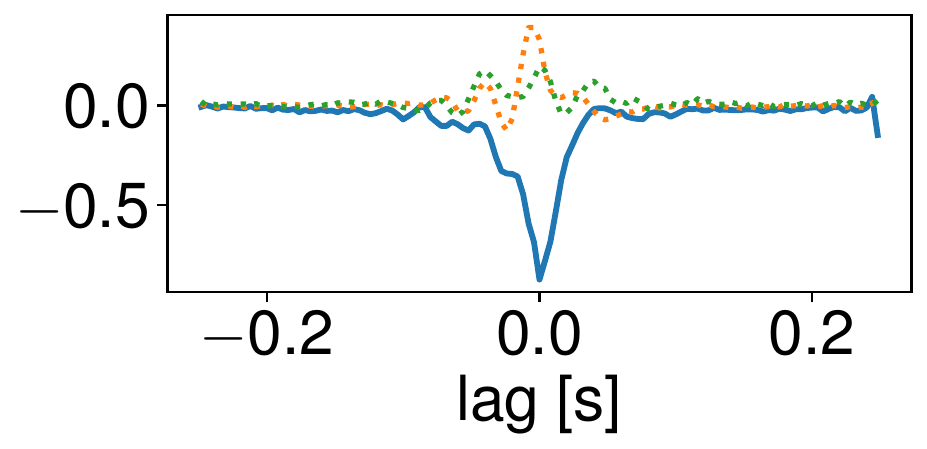}\\[0.5mm]\includegraphics[width=\linewidth]{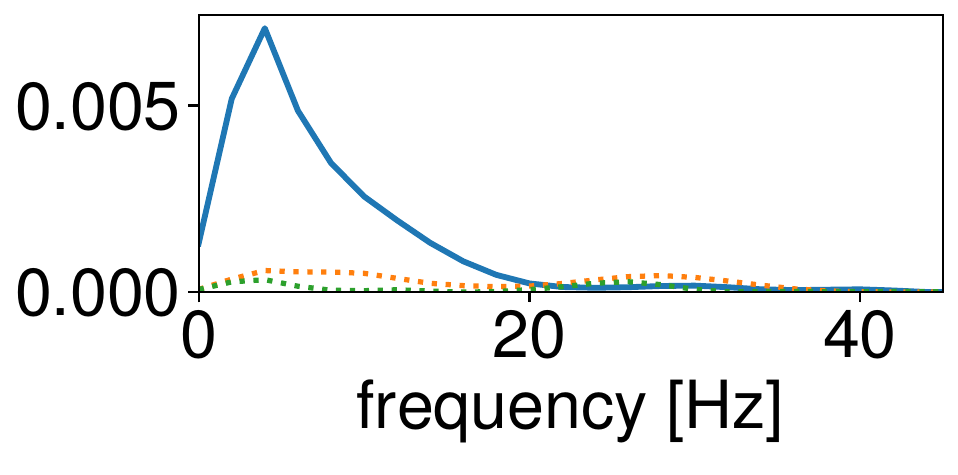}\end{minipage}
\end{minipage}}\\[2mm]
\subfloat[atom \#6 (6\%; $\bigstar$ T7, 4 Hz)]{\begin{minipage}[c]{0.49\textwidth}\centering
\begin{minipage}[c]{0.45\linewidth}\centering\includegraphics[width=\linewidth]{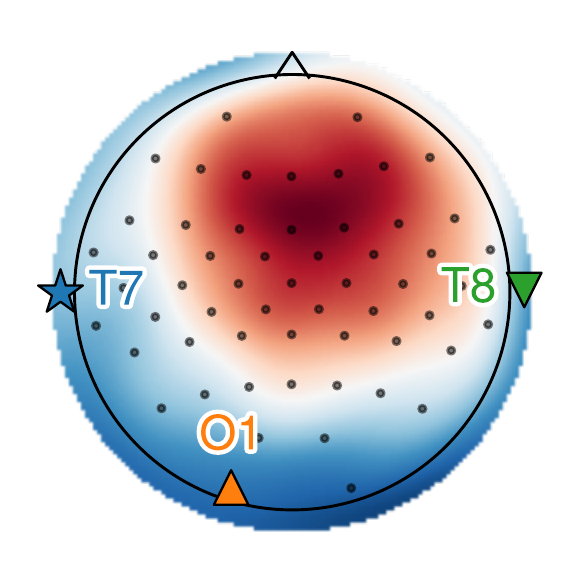}\end{minipage}\hfill
\begin{minipage}[c]{0.53\linewidth}\centering\includegraphics[width=\linewidth]{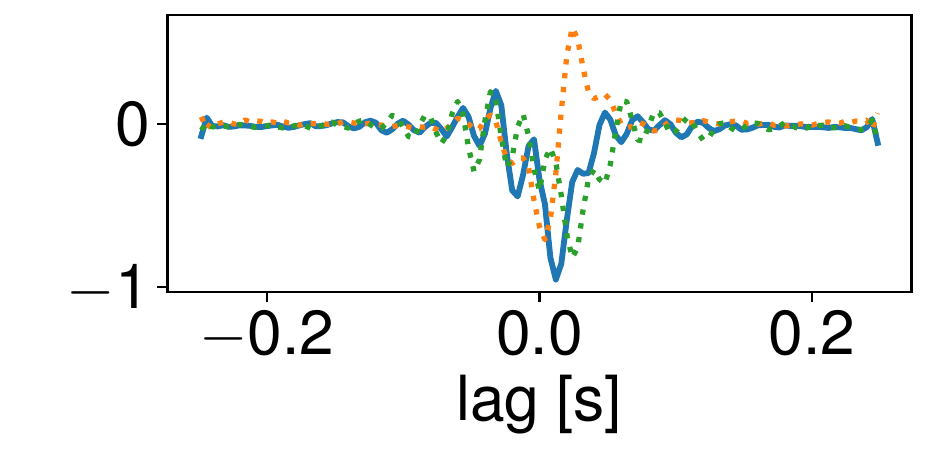}\\[0.5mm]\includegraphics[width=\linewidth]{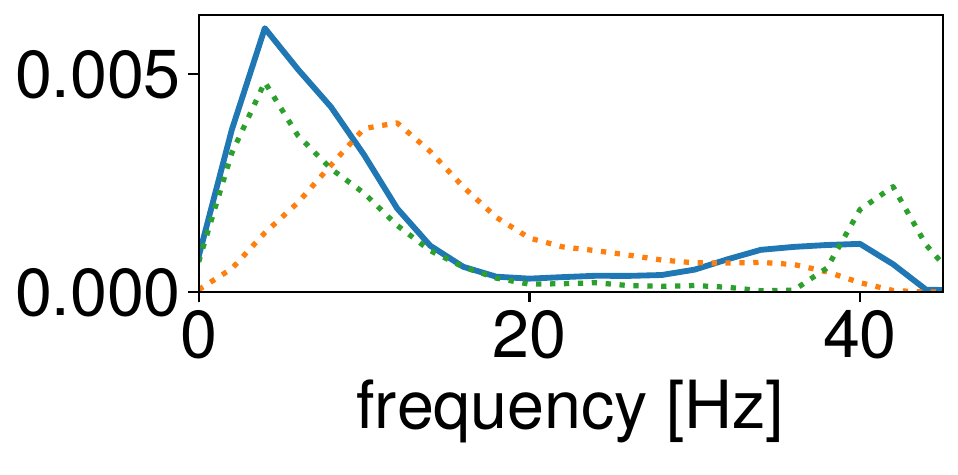}\end{minipage}
\end{minipage}}\hfill
\subfloat[atom \#16 (5\%; $\bigstar$ F3, 16 Hz)]{\begin{minipage}[c]{0.49\textwidth}\centering
\begin{minipage}[c]{0.45\linewidth}\centering\includegraphics[width=\linewidth]{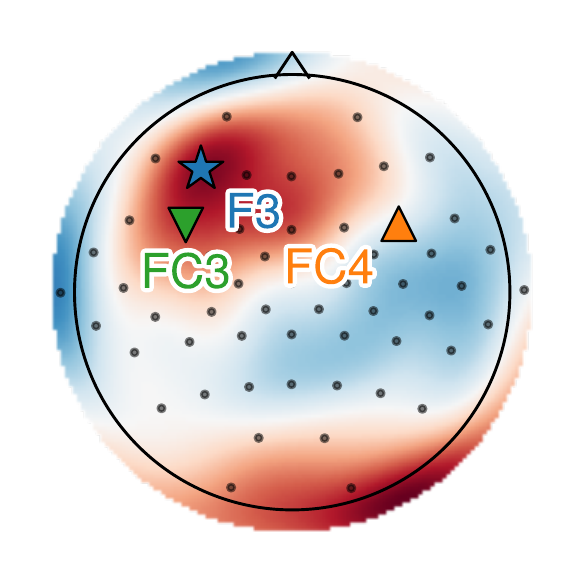}\end{minipage}\hfill
\begin{minipage}[c]{0.53\linewidth}\centering\includegraphics[width=\linewidth]{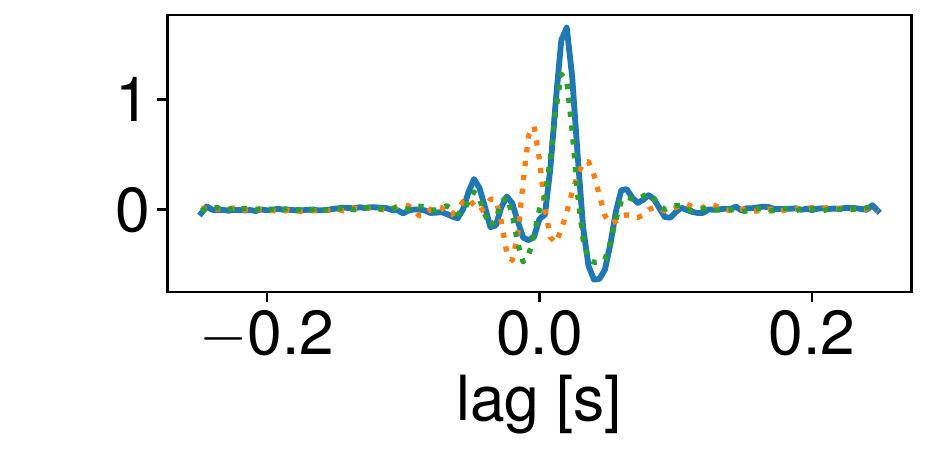}\\[0.5mm]\includegraphics[width=\linewidth]{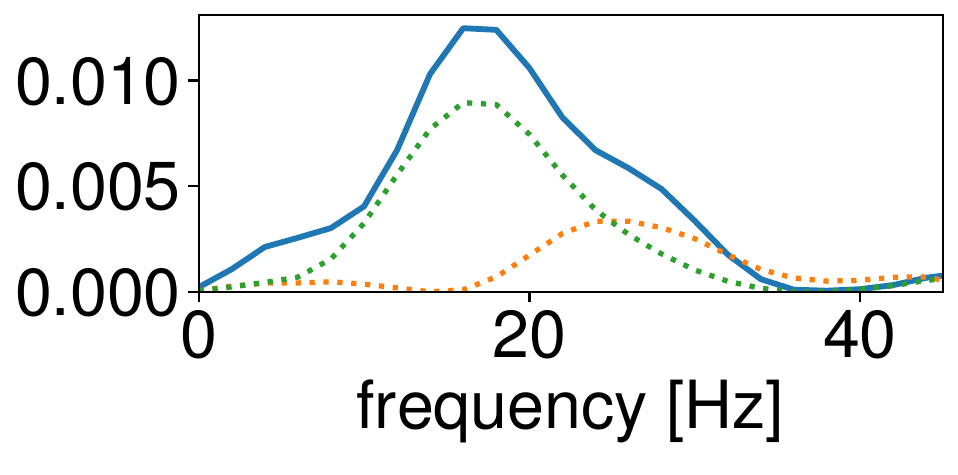}\end{minipage}
\end{minipage}}
\caption{The eight most-used atoms of \ccdan{} (Tier-1), evaluated on 400 trials of BCI IV-1.
For each atom, left: scalp field at the peak lag (from the spherical-harmonic coefficients; RdBu, blue negative, red positive, symmetric scale per atom; dots: electrodes of BCI IV-1; star: electrode of maximum atom energy; triangles: two random electrodes; electrode names in the marker color), top right: waveform at these three electrodes (solid: star; dotted: triangles; colors match the markers), bottom right: power spectra of the same waveforms (Welch, linear scale).
Amplitude and power are in arbitrary units; the frequency in each subcaption is the spectral peak at the starred electrode.}
\label{fig:atoms_supp}
\end{figure}

\begin{figure}[tbp]
\centering
\subfloat[feature $h$=24]{\includegraphics[width=0.235\textwidth]{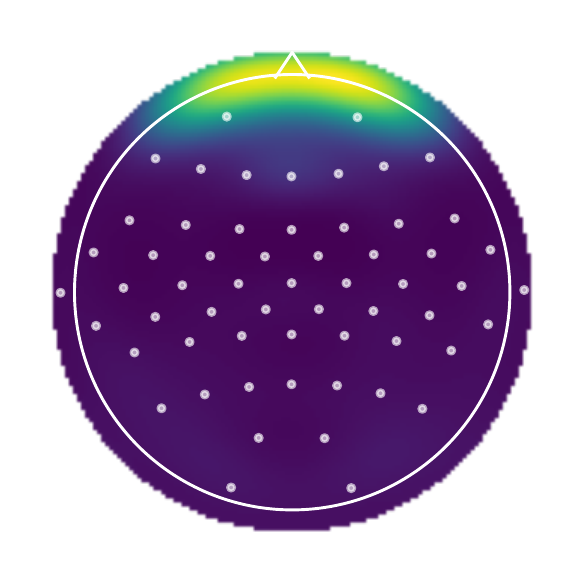}}\hfill
\subfloat[feature $h$=36]{\includegraphics[width=0.235\textwidth]{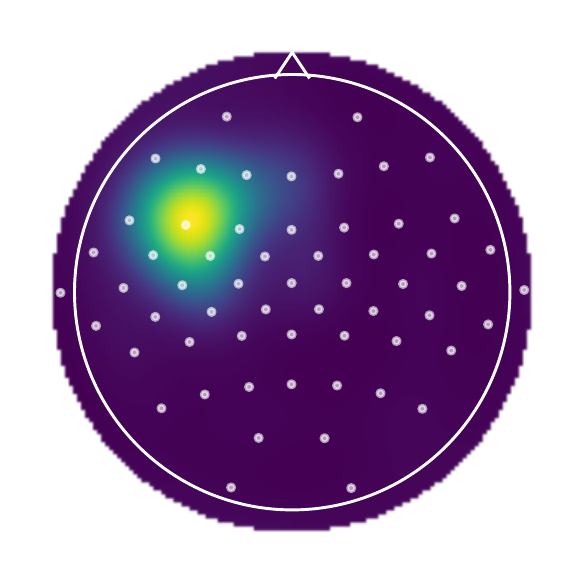}}\hfill
\subfloat[feature $h$=4]{\includegraphics[width=0.235\textwidth]{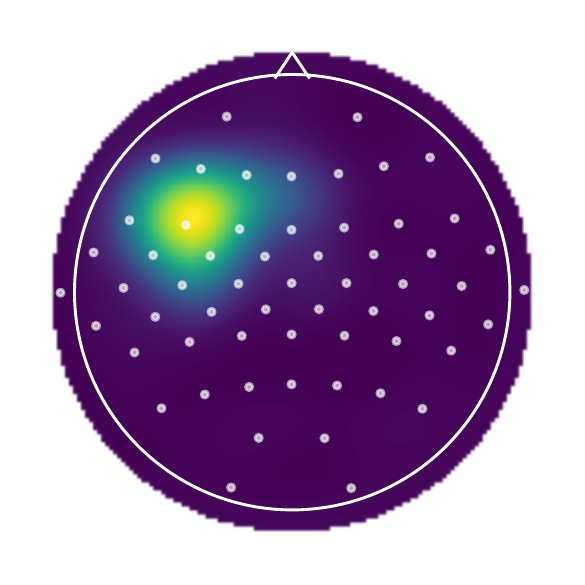}}\hfill
\subfloat[feature $h$=39]{\includegraphics[width=0.235\textwidth]{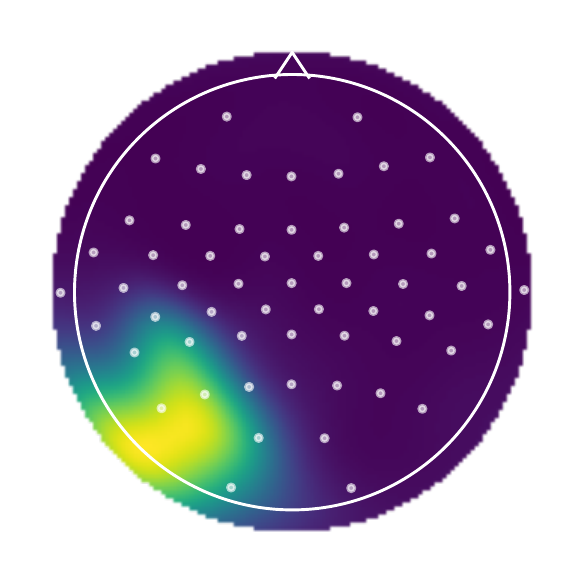}}\\[1mm]
\subfloat[feature $h$=32]{\includegraphics[width=0.235\textwidth]{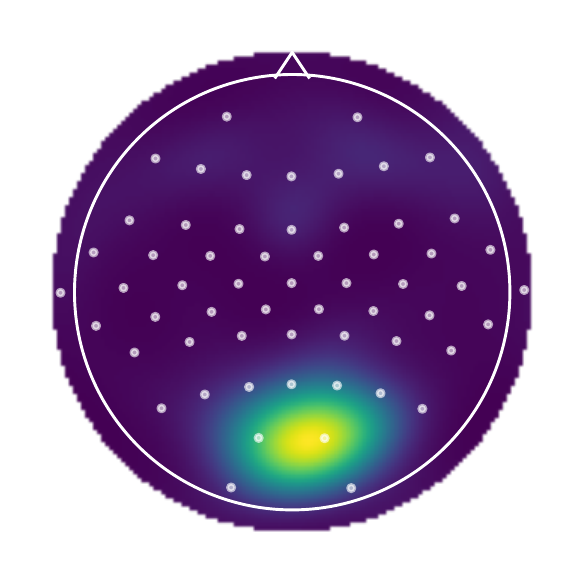}}\hfill
\subfloat[feature $h$=3]{\includegraphics[width=0.235\textwidth]{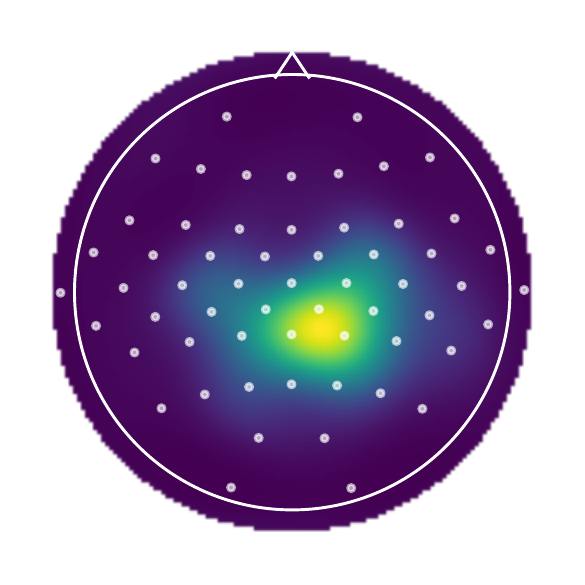}}\hfill
\subfloat[feature $h$=0]{\includegraphics[width=0.235\textwidth]{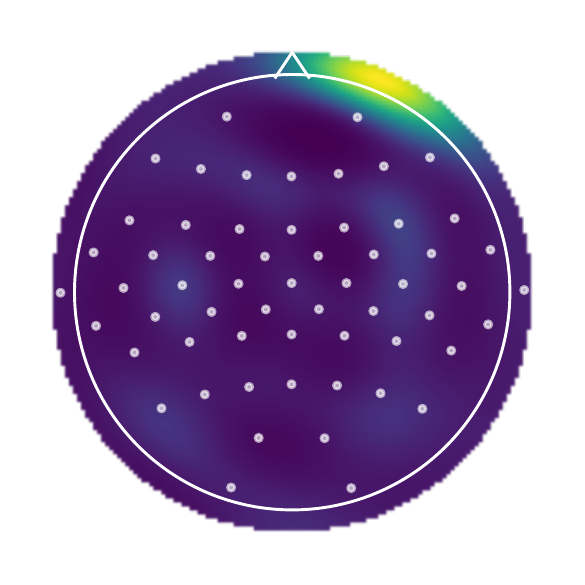}}\hfill
\subfloat[feature $h$=49]{\includegraphics[width=0.235\textwidth]{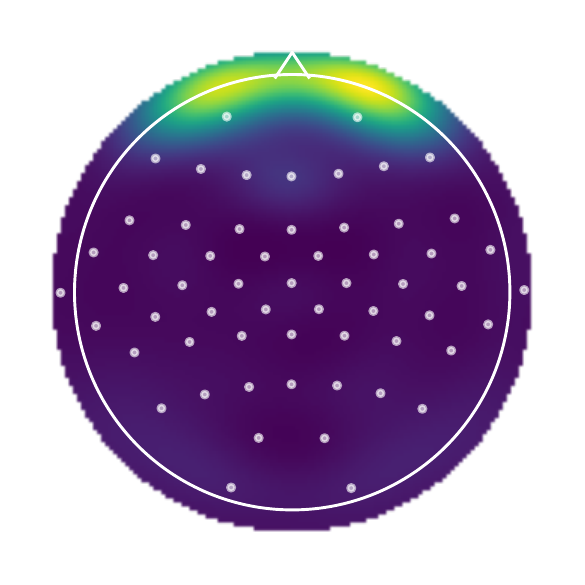}}
\caption{Coordinate-dependent pooling weights $w_h(\bm p)$ of the detector (the eight features with the largest spatial variance; viridis, dark: small, bright: large).
Fields selective to frontal, posterior, and temporal regions have formed.}
\label{fig:pooling}
\end{figure}

\begin{figure}[tbp]
\centering
\begin{minipage}[c]{0.55\textwidth}\centering
\subfloat[NN]{\includegraphics[width=0.28\linewidth]{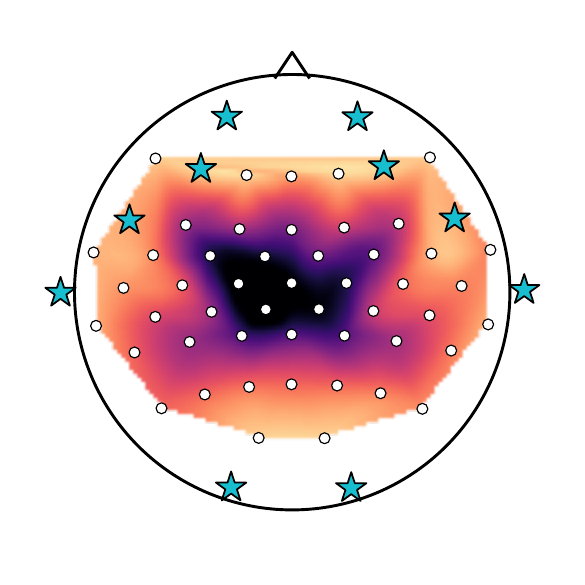}}\hfill
\subfloat[SSI]{\includegraphics[width=0.28\linewidth]{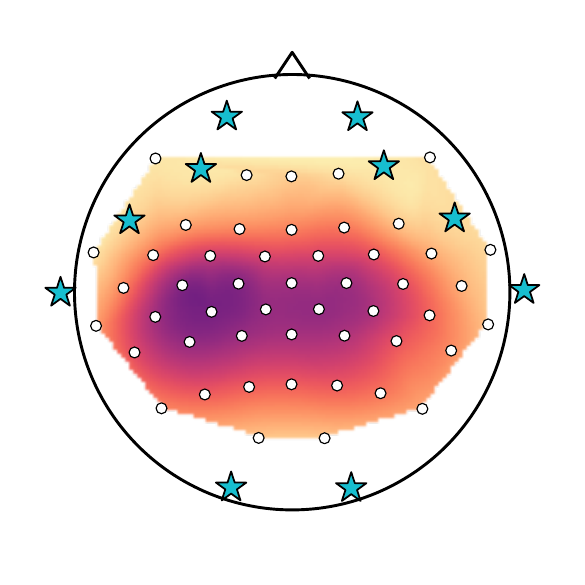}}\hfill
\subfloat[\ccdan{}]{\includegraphics[width=0.28\linewidth]{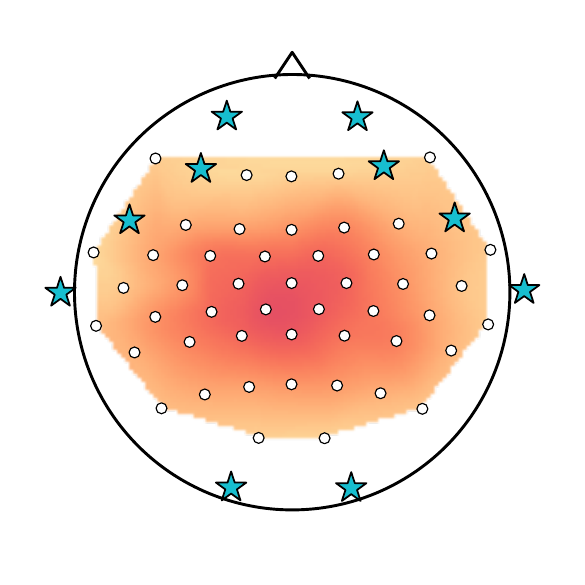}}\hfill
\begin{minipage}[c]{0.12\linewidth}\centering\includegraphics[width=\linewidth,trim=183 12 6 5,clip]{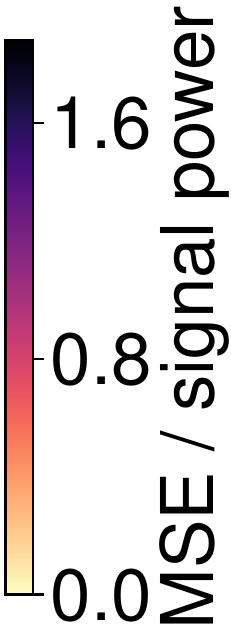}\end{minipage}
\end{minipage}\hfill
\begin{minipage}[c]{0.42\textwidth}\centering
\subfloat[Waveforms at CFC7 (top) and P2 (bottom)]{\begin{minipage}[c]{\linewidth}\centering\includegraphics[width=\linewidth]{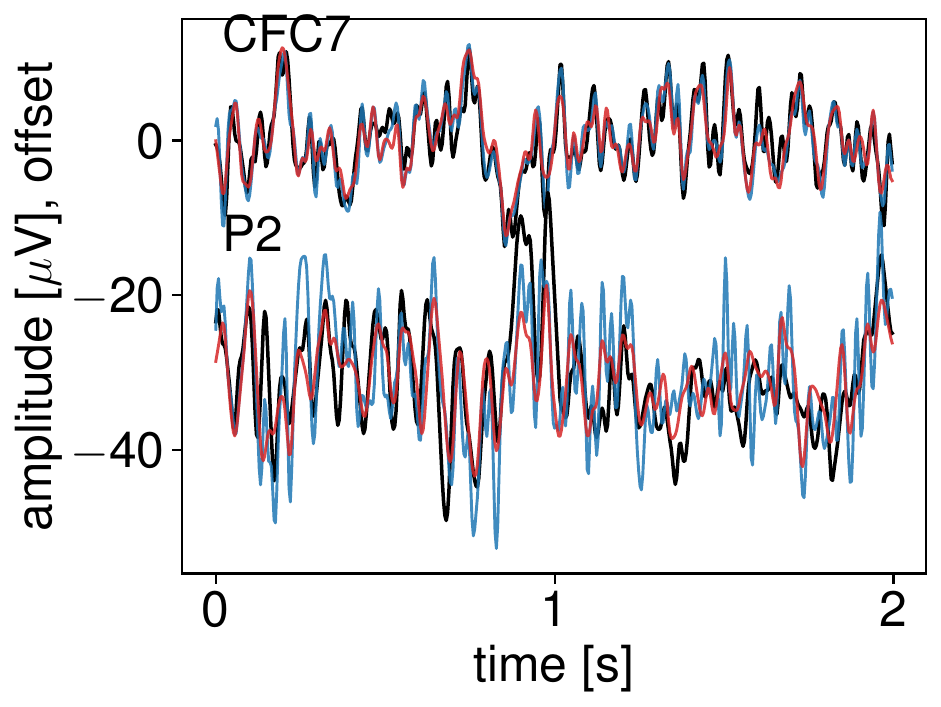}\\[0.5mm]\includegraphics[width=0.85\linewidth,trim=7 121 7 118,clip]{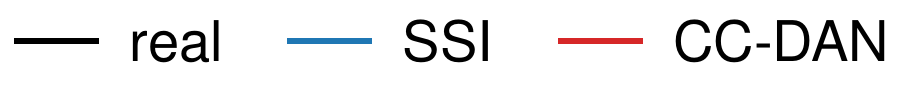}\end{minipage}}
\end{minipage}
\caption{Error topographies and waveform examples (BCI IV-1, subject a, Emotiv 14 ch with 10 matched electrodes; stars: observed electrodes; the motor 9 ch example is Fig.~10 of the main text).
(a)--(c) MSE over signal power at missing electrodes (mean over missing electrodes: NN 0.85, SSI 0.61, \ccdan{} 0.39).
(d) Truth (black), SSI, and \ccdan{} at two missing electrodes of one trial (top: CFC7, where the SSI error is small; bottom: P2, where it is large).}
\label{fig:examples}
\end{figure}

Fig.~\ref{fig:lee} shows the cross-paradigm transfer within Lee2019 (main text, Section~4.3).
Tables~\ref{tab:downstream}--\ref{tab:alpha} give the downstream-task numbers (Section~4.7, Fig.~8 of the main text), and Table~\ref{tab:gain} the amplitude-calibration benchmark on all evaluation sets (Fig.~7).
Fig.~\ref{fig:atoms_supp} shows the eight most-used atoms, Fig.~\ref{fig:pooling} the coordinate-dependent pooling fields of the detector, and Fig.~\ref{fig:examples} error topographies and waveform examples (Section~4.8).

\begin{table}[tbp]
\centering\footnotesize
\caption{Downstream Task: Motor-Imagery Classification Accuracy (EEGNet Trained on Real Full-Montage Data and Applied to Completed Trials).
``Observed only'': Classifier Retrained on the Observed Electrodes.
``+cal'': \ccdan{} Calibrated With the Post-Hoc Gain Table.
Lee2019 Values Are From the Rerun Including Calibration (Zero-Fill Only From the First Run).}
\label{tab:downstream}
\resizebox{\textwidth}{!}{%
\begin{tabular}{llrrrrrrr}
\toprule
Data & Condition & Full montage & Observed only & Zero-fill & NN & SSI & \ccdan{} & +cal\\
\midrule
BCI IV-2a (22 ch, LOSO, 9 subjects) & motor\_9 & 0.829 & 0.829 & 0.669 & 0.693 & \textbf{0.803} & 0.779 & 0.782\\
 & motor\_3 & & 0.689 & 0.525 & 0.590 & 0.585 & \textbf{0.632} & 0.605\\
 & 1020\_19 (5) & & 0.711 & -- & 0.552 & 0.631 & 0.669 & \textbf{0.679}\\
 & $k$=9 & & 0.824 & 0.582 & 0.647 & \textbf{0.764} & 0.736 & 0.725\\
 & $k$=5 & & 0.758 & 0.546 & 0.588 & \textbf{0.643} & 0.637 & 0.636\\
Lee2019 MI (62 ch, 44 training / 10 test subjects) & motor\_9 (8) & 0.772 & 0.733 & 0.674 & 0.644 & \textbf{0.686} & 0.676 & 0.616\\
 & motor\_3 & & 0.666 & 0.645 & 0.520 & 0.541 & \textbf{0.638} & 0.610\\
 & 1020\_19 & & 0.752 & 0.644 & 0.629 & \textbf{0.724} & 0.702 & 0.702\\
 & emotiv\_14 & & 0.716 & 0.534 & 0.640 & 0.643 & \textbf{0.669} & 0.662\\
 & $k$=9 & & 0.646 & 0.518 & 0.561 & \textbf{0.606} & 0.568 & 0.514\\
 & $k$=5 & & 0.601 & 0.499 & 0.515 & 0.510 & \textbf{0.533} & 0.528\\
\bottomrule
\end{tabular}}
\end{table}

\begin{table}[tbp]
\centering\footnotesize\setlength{\tabcolsep}{3pt}
\caption{Downstream Task: SSVEP Detection Accuracy (Lee2019 SSVEP, 54 Subjects, CCA on 13 Parieto-Occipital Electrodes; Full Montage 0.954, Chance 0.25).}
\label{tab:ssvep}
\resizebox{\textwidth}{!}{%
\begin{tabular}{lrrrrr}
\toprule
Condition & Observed only & NN & SSI & \ccdan{} & +cal\\
\midrule
1020\_19 (19) & 0.920 & 0.920 & 0.913 & 0.866 & 0.866\\
emotiv\_14 (14) & 0.923 & 0.922 & 0.907 & 0.836 & 0.837\\
openbci\_8 (8) & 0.923 & 0.922 & 0.911 & 0.847 & 0.848\\
iv2a\_21 (20) & 0.509 & 0.608 & 0.610 & 0.534 & 0.532\\
$k$=21 / 15 / 5 & 0.853 / 0.866 / 0.679 & 0.826 / 0.855 / 0.641 & 0.814 / 0.843 / 0.621 & 0.735 / 0.725 / 0.562 & 0.735 / 0.723 / 0.566\\
motor\_9 (8) / $k$=9 (no occipital) & 0.250 / 0.250 & 0.493 / 0.461 & 0.497 / 0.473 & 0.429 / 0.416 & 0.431 / 0.418\\
\bottomrule
\end{tabular}}
\end{table}

\begin{table}[tbp]
\centering\footnotesize\setlength{\tabcolsep}{2.5pt}
\caption{Downstream Task: $\alpha$ Topography (HBN Resting 128 ch, 40 Subjects).
RMS Error of $\log_{10}\alpha$ Power at Missing Electrodes and Topography Correlation $r$ Over Missing Electrodes.
vm1: Pretrained With the Variance-Matching Loss ($\lambda_{\mathrm{var}}=1$); ``+cal'': Post-Hoc Gain Table.}
\label{tab:alpha}
\resizebox{\textwidth}{!}{%
\begin{tabular}{lrrrrrrrrrrrr}
\toprule
 & \multicolumn{6}{c}{RMS error} & \multicolumn{6}{c}{Correlation $r$}\\
\cmidrule(lr){2-7}\cmidrule(lr){8-13}
Condition & NN & SSI & \ccdan{} & +cal & vm1 & vm1+cal & NN & SSI & \ccdan{} & +cal & vm1 & vm1+cal\\
\midrule
$k$=5 & 0.253 & 0.292 & 0.452 & 0.319 & 0.351 & 0.296 & 0.26 & 0.18 & 0.25 & 0.26 & 0.23 & 0.23\\
$k$=9 & 0.196 & 0.262 & 0.487 & 0.341 & 0.384 & 0.335 & 0.56 & 0.41 & 0.20 & 0.22 & 0.15 & 0.16\\
$k$=15 & 0.197 & 0.303 & 0.451 & 0.304 & 0.283 & 0.235 & 0.59 & 0.44 & 0.37 & 0.42 & 0.42 & 0.44\\
$k$=21 & 0.187 & 0.180 & 0.368 & 0.282 & 0.307 & 0.259 & 0.60 & 0.65 & 0.34 & 0.33 & 0.35 & 0.35\\
1020\_19 (14) & 0.196 & 0.217 & 0.293 & 0.228 & 0.248 & 0.227 & 0.55 & 0.52 & 0.37 & 0.44 & 0.37 & 0.41\\
emotiv\_14 (12) & 0.214 & 0.230 & 0.349 & 0.245 & 0.256 & 0.244 & 0.52 & 0.51 & 0.32 & 0.36 & 0.33 & 0.35\\
openbci\_8 (6) & 0.237 & 0.665 & 0.427 & 0.341 & 0.325 & 0.330 & 0.37 & 0.30 & 0.04 & 0.10 & 0.06 & 0.11\\
left-hemisphere dropout (59) & 0.231 & 0.375 & 0.462 & 0.290 & 0.281 & 0.277 & 0.52 & 0.22 & 0.45 & 0.36 & 0.35 & 0.28\\
posterior dropout (57) & 0.345 & 0.309 & 0.666 & 0.397 & 0.484 & 0.390 & 0.23 & 0.45 & $-0.22$ & 0.07 & $-0.12$ & $-0.03$\\
\bottomrule
\end{tabular}}
\end{table}

\begin{table}[tbp]
\centering\footnotesize\setlength{\tabcolsep}{3pt}
\caption{Amplitude-Calibration Benchmark: MSE on Missing Electrodes (\muVsq{}) and, in Parentheses, Amplitude Gain (Median of Predicted Over True Standard Deviation; 1 Is Unbiased).
Tier-1 \ccdan{}, Post-Hoc Gain-Table Calibration (+cal), Variance-Matching Loss $\lambda_{\mathrm{var}}=1$ (vm1), Calibrated vm1, and $\lambda_{\mathrm{var}}=3$ (vm3).
Same Corpus, Settings, and 40 Epochs.
This Benchmark Was Run Independently; Its NN, SSI, and \ccdan{} Values Coincide With Table~I of the Main Text.}
\label{tab:gain}
\resizebox{\textwidth}{!}{%
\begin{tabular}{llrrrrrrr}
\toprule
Data & Condition & NN & SSI & \ccdan{} & +cal & vm1 & vm1+cal & vm3\\
\midrule
BCI IV-1 & $k$=3 & 64.7 (0.73) & 62.7 (0.87) & 50.2 (0.54) & 77.6 (1.06) & 52.9 (0.76) & 66.2 (1.00) & 58.0 (0.85)\\
 & $k$=5 & 48.3 (0.85) & 48.0 (1.05) & 32.6 (0.69) & 40.7 (0.99) & 32.8 (0.85) & 36.9 (0.97) & 35.7 (0.91)\\
 & $k$=9 & 33.0 (0.91) & 21.0 (1.02) & 20.6 (0.79) & 23.1 (0.98) & 20.0 (0.89) & 21.6 (0.99) & 21.5 (0.94)\\
 & $k$=15 & 24.1 (0.94) & 11.5 (1.00) & 14.6 (0.84) & 15.7 (0.99) & 14.0 (0.92) & 14.8 (1.00) & 15.2 (0.95)\\
 & motor\_9 & 32.5 (0.75) & 34.2 (1.04) & 19.0 (0.79) & 23.1 (0.98) & 19.3 (0.89) & 21.5 (0.98) & 21.1 (0.93)\\
 & left-hemi. dropout & 48.6 (0.95) & 64.2 (1.27) & 35.3 (0.65) & 40.3 (0.92) & 33.0 (0.82) & 37.3 (0.99) & 39.4 (0.84)\\
\midrule
BCI III-IIIa & $k$=3 & 14.2 (0.75) & 17.9 (0.98) & 12.2 (0.47) & 16.1 (0.83) & 12.8 (0.69) & 14.8 (0.85) & 14.9 (0.82)\\
 & $k$=5 & 10.3 (0.86) & 11.6 (1.04) & 8.6 (0.59) & 9.3 (0.77) & 8.6 (0.74) & 9.1 (0.81) & 9.8 (0.86)\\
 & $k$=9 & 6.5 (0.94) & 4.0 (0.99) & 5.4 (0.69) & 5.6 (0.81) & 5.3 (0.80) & 5.5 (0.86) & 5.9 (0.88)\\
 & $k$=15 & 4.6 (0.97) & 2.1 (0.98) & 4.2 (0.73) & 4.2 (0.84) & 4.1 (0.82) & 4.2 (0.88) & 4.4 (0.88)\\
 & motor\_9 & 6.2 (0.99) & 4.1 (1.00) & 4.6 (0.73) & 4.6 (0.87) & 5.2 (0.87) & 5.4 (0.93) & 5.3 (0.94)\\
 & left-hemi. dropout & 9.4 (0.91) & 12.0 (1.22) & 9.2 (0.58) & 9.6 (0.73) & 8.4 (0.65) & 8.7 (0.74) & 9.7 (0.77)\\
\midrule
BCI IV-2a & $k$=3 & 23.7 (0.72) & 24.3 (0.92) & 19.7 (0.48) & 26.3 (0.90) & 20.5 (0.68) & 24.1 (0.88) & 23.0 (0.79)\\
 & $k$=5 & 17.4 (0.83) & 18.0 (1.05) & 13.3 (0.61) & 14.6 (0.84) & 13.3 (0.76) & 14.1 (0.85) & 14.2 (0.84)\\
 & $k$=9 & 12.6 (0.91) & 7.4 (1.01) & 8.4 (0.73) & 8.8 (0.90) & 8.2 (0.82) & 8.5 (0.90) & 8.7 (0.87)\\
 & $k$=15 & 10.3 (0.94) & 3.8 (1.00) & 5.6 (0.81) & 5.8 (0.96) & 5.4 (0.85) & 5.5 (0.93) & 5.8 (0.88)\\
 & motor\_9 & 12.7 (0.85) & 6.7 (1.02) & 7.2 (0.80) & 7.9 (0.98) & 7.5 (0.90) & 8.0 (0.99) & 8.2 (0.94)\\
 & left-hemi. dropout & 18.9 (0.82) & 21.8 (1.18) & 16.6 (0.60) & 16.8 (0.73) & 15.3 (0.70) & 15.9 (0.79) & 16.5 (0.76)\\
\midrule
HBN & $k$=3 & 382 (0.77) & 1035 (0.89) & 350 (0.49) & 697 (1.02) & 394 (0.70) & 532 (0.95) & 432 (0.79)\\
 & $k$=5 & 300 (0.89) & 720 (1.16) & 256 (0.62) & 380 (0.96) & 278 (0.78) & 331 (0.94) & 290 (0.85)\\
 & $k$=9 & 259 (0.96) & 437 (1.12) & 200 (0.70) & 245 (0.90) & 212 (0.81) & 236 (0.92) & 219 (0.86)\\
 & $k$=15 & 219 (0.98) & 261 (1.07) & 173 (0.72) & 190 (0.86) & 181 (0.81) & 194 (0.90) & 186 (0.85)\\
 & motor\_9 & 246 (0.68) & 1176 (1.34) & 264 (0.70) & 423 (0.96) & 278 (0.83) & 351 (0.96) & 275 (0.90)\\
 & left-hemi. dropout & 376 (1.00) & 822 (1.55) & 220 (0.56) & 243 (0.80) & 229 (0.71) & 245 (0.85) & 233 (0.74)\\
\midrule
BI2013a & $k$=3 & 135 (0.75) & 264 (0.93) & 100 (0.47) & 177 (0.99) & 107 (0.69) & 133 (0.95) & 115 (0.78)\\
 & $k$=5 & 126 (0.89) & 297 (1.19) & 75.4 (0.59) & 101 (0.91) & 78.2 (0.74) & 88.8 (0.90) & 86.6 (0.81)\\
 & $k$=9 & 141 (0.95) & 218 (1.09) & 63.1 (0.66) & 78.6 (0.87) & 65.0 (0.79) & 72.6 (0.90) & 70.6 (0.84)\\
 & $k$=15 & 82.3 (0.98) & 83.1 (1.09) & 31.5 (0.72) & 37.4 (0.91) & 32.2 (0.82) & 35.5 (0.93) & 35.6 (0.85)\\
 & left-hemi. dropout & 94.6 (0.89) & 248 (1.48) & 80.8 (0.63) & 91.7 (0.87) & 82.6 (0.82) & 89.2 (0.95) & 85.1 (0.77)\\
\midrule
Nakanishi2015 & $k$=3 & 53.4 (0.71) & 120 (1.14) & 37.6 (0.18) & 39.5 (0.28) & 40.1 (0.37) & 41.7 (0.42) & 42.0 (0.47)\\
 & $k$=5 & 57.0 (0.83) & 103 (1.29) & 32.5 (0.20) & 33.2 (0.25) & 34.6 (0.37) & 35.0 (0.39) & 35.6 (0.46)\\
 & left-hemi. dropout & 52.5 (0.72) & 322 (2.79) & 33.4 (0.23) & 34.0 (0.30) & 36.0 (0.43) & 36.8 (0.47) & 38.0 (0.54)\\
\bottomrule
\end{tabular}}
\end{table}

\section{Abbreviations and Condition Names}\label{app:abbr}
Abbreviations follow the main text: \ccdan{}, coordinate-conditioned detector-atom network; DAN, detector-atom network (D-1 / D-2: detector designs of the name-based multichannel DAN, Section~2.3); SH, spherical harmonics of degree $L$; RBF, radial basis functions; rank-4 / rank-0, rank of the coordinate-dependent detector filters (rank-0: shared filters only); NN, nearest-neighbor interpolation; SSI, spherical spline interpolation; oracle, within-subject ridge regression from observed to missing electrodes trained on other trials of the same subject; MSE, mean squared error on missing electrodes (\muVsq{}); SEM, standard error of the mean; LOSO, leave-one-subject-out; CAR, common average reference; Tier-1, the full pretraining corpus (five datasets, 132 subjects); vm1 / vm3, models pretrained with the variance-matching loss ($\lambda_{\mathrm{var}}$ = 1 / 3); +cal, post-hoc gain calibration; gain, median ratio of predicted to true standard deviation at missing electrodes; MI, motor imagery; SSVEP, steady-state visual evoked potential; ERP, event-related potential; CCA, canonical correlation analysis; RMS, root mean square; EGI GSN, Electrical Geodesics Geodesic Sensor Net; HBN, Healthy Brain Network; BI2013a, BrainInvaders 2013a; MAE, masked autoencoder; VAE, variational autoencoder; DDIM, denoising diffusion implicit model; LB, Laplace--Beltrami; $k$, number of observed electrodes.
Condition names such as motor\_9, 1020\_19, emotiv\_14, openbci\_8, iv2a\_21, and drop\_left denote the coordinate-matched layout templates and region-dropout conditions of Table~\ref{tab:layouts} (Section~3.4 of the main text); the number in parentheses after a layout is the number of matched (observed) electrodes.

\end{document}